\documentclass[a4paper,11pt]{article}

\usepackage{jheppub}
\allowdisplaybreaks
\usepackage{graphicx}
\usepackage{epstopdf}
\usepackage{simplewick}
\usepackage[utf8]{inputenc}

\newcommand{\be}{\begin{equation}}
\newcommand{\ee}{\end{equation}}
\newcommand{\bea}{\begin{eqnarray}}
\newcommand{\eea}{\end{eqnarray}}
\newcommand{\nn}{\nonumber}

\title{Effective Field Theories of Post-Newtonian Gravity
:\\A comprehensive review}

\author[a,b]{Mich\`ele Levi}

\affiliation[a]{Institut de Physique Th\'eorique, 
CEA \& CNRS, Universit\'e Paris-Saclay,
\\ 91191 Gif-sur-Yvette, France}
\affiliation[b]{Niels Bohr International Academy, Niels Bohr Institute, 
University of Copenhagen, 
\\Blegdamsvej 17, 2100 Copenhagen, Denmark}

\emailAdd{michelelevi@nbi.ku.dk}

\abstract{
This review article presents the progress made over the last decade, since 
the introduction of effective field theories (EFTs) into post-Newtonian (PN) 
gravity. These have been put forward in the context of gravitational waves 
(GWs) from the compact binary inspiral. The mature development of this 
interdisciplinary field has resulted in significant advances of wide interest 
to physics at several levels serving various purposes. The field has firmly 
demonstrated, that seemingly disparate physical domains, such as quantum 
field theory and classical gravity, are related, and that the EFT framework 
is a universal one, where it has been proven to supply a robust methodology 
to boost progress in the development of PN theory. In this review emphasis was 
put on an accessible pedagogic presentation of the field theoretic aspects of 
the subject, with the view, that these are in fact common across the whole of 
theoretical physics, rather than in their original narrow quantum context. 
The review is aimed at a broad audience, from general readers new to the 
field, to specialists and experts in related subjects. 

The review begins with an overview of the introduction of EFTs into classical 
gravity and their development. Then, the basic ideas, which form the 
conceptual foundation of EFTs, are provided, and the strategy of a 
multi-stage EFT framework, which is utilized for the PN binary inspiral 
problem, is outlined. The main body of the review is then dedicated to 
presenting in detail the study of each of the effective theories at each of 
the intermediate scales in the problem, up to the actual GW observables. 
First, the EFT for a single compact object is considered, from which one 
proceeds to the EFT of a compact binary system, viewed as a composite 
particle with internal binding interactions. Finally, one arrives at the 
effective theory of the time-dependent multipole moments of the radiating 
system. The review is concluded with the multiple prospects of building on 
the progress in the field, and using further modern field theory insights 
and tools, to specifically address the study of GWs, as well as to broadly 
expand our fundamental understanding of gauge and gravity theories across 
the classical and quantum regimes.}


\dedicated{
To my loving father, \\sister, and old friend, \\who is a brother to me.}

\begin{document}

\maketitle

\flushbottom

\section{EFTs in classical gravity}
\label{intro}

Gravitational radiation is a probing prediction of any candidate complete 
theory of gravity. As a relativistic field theory of gravity, the general 
theory of relativity (GR), which was formulated in 1915 by Einstein, indeed 
led him soon after to predict gravitational waves (GWs) within GR in 1916. 
However, the first evidence of GWs arrived only with the discovery of the first 
binary pulsar by Hulse and Taylor in 1974 \cite{Hulse:1974eb}, where in the 
analysis, done in the following years, the decay of the orbital period 
fit with the loss of energy and angular momentum due to gravitational 
radiation, as expected from GR \cite{Taylor:1982zz,Weisberg:1984zz,
Damour:1986}. In the following decades laser interferometers were developed 
for the purpose of observing GWs directly and making GW astronomy a reality, 
which led to the construction of a few ground-based GW detectors, all 
at a frequency range of about $10-10^4$ Hz, appropriate for the detection of 
comparable mass compact binary coalescence (CBC) events: TAMA in Japan, the 
twin LIGO detectors in the USA \cite{Ligo}, and GEO600 \cite{GEO600} and 
Virgo \cite{Virgo} in Europe. TAMA started operating in 1999, with both LIGO 
and GEO600 starting in 2002. Later, TAMA was closed and Virgo was put online in 
2007. For a decade none of these detectors has achieved detection, and 
subsequently, LIGO and Virgo were shut down to undergo major upgrading to 
second-generation GW detectors in order to make them more sensitive.

The first observation of GWs, labeled `GW150914' \cite{Abbott:2016blz}, was 
indeed realized in 2015 with the Advanced LIGO detectors, only 4 days 
before the operation of LIGO's advanced version officially commenced. Three 
further detections - GW151226 \cite{Abbott:2016nmj}, GW170104 
\cite{Abbott:2017vtc}, and GW170608 \cite{Abbott:2017gyy} - by Advanced LIGO 
followed, before the Advanced Virgo detector joined the observations in 2017. 
The first joint detection by these three advanced detectors in the USA and 
Europe, GW170814 \cite{Abbott:2017oio}, arrived shortly after, allowing for 
the localization of the source and testing of GW polarizations for the first time. 
Moreover, a second joint detection, GW170817 \cite{TheLIGOScientific:2017qsa}, 
was attained just after, this time involving a neutron star (NS) binary merger, 
after all five previous GW events featured only black hole (BH) binaries. 
Thus, this was the first multi-messenger detection of a NS merger, with 
electromagnetic counterparts, in particular the gamma ray burst (GRB) 
170817A, which confirmed the longstanding conjecture, that short GRBs 
originate from NS mergers \cite{Abbott:2018wiz,Abbott:2018exr}.

These first earthshaking GW detections proved to be even more telling 
than expected (see, e.g.~also \cite{TheLIGOScientific:2016src, 
TheLIGOScientific:2016wfe}), and paved the way for further plans of advanced 
GW detectors to multiply worldwide. In particular, second-generation 
ground-based GW detectors, such as KAGRA, the successor of TAMA in Japan 
\cite{Kagra}, and IndiGO in India \cite{IndiGO}, are expected to start 
operation in the coming years \cite{Aasi:2013wya}, complementing the 
distribution of the worldwide network\footnote{We note however, that it would 
be desirable to further extend the distribution of the worldwide network to 
the Southern Hemisphere in the future.}. Furthermore, there are already 
specific plans for third-generation ground-based detectors, such as the 
Einstein Telescope in Europe \cite{Punturo:2010zz}, and the Cosmic Explorer 
in the USA \cite{Evans:2016mbw}. Additionally, the space-based GW detector LISA 
\cite{Lisa} is designed to operate in a complementary frequency range of 
about $10^{-5}-1$ Hz, appropriate for the detection of extreme mass ratio 
inspirals (EMRIs). Moreover, further GW detectors, such as the ground-based 
SOGRO \cite{PAIK:2016yxn}, and the space-based Japanese DECIGO 
\cite{Sato:2017dkf} and Chinese TianQin \cite{Luo:2015ght}, in complementary 
intermediate frequency ranges, have also been proposed. 

The GW detections made so far have all involved the CBC signal, where 
binary constituents of comparable mass are concerned. The evolution of such a 
compact binary is comprised of three consecutive yet distinguishable phases 
\cite{Buonanno:2014aza}: 1.~The inspiral, where the components of the binary 
are still moving at non-relativistic (NR) velocities, and their orbital 
separation is slowly decaying; 2.~The merger, when the separation between the 
components falls roughly below the innermost stable orbit of a BH, and the 
objects reach relativistic velocities, and merge into a single compact 
object; 3.~The ringdown phase for BH binaries, where spacetime 
settles down to that of a rotating Kerr BH via quasinormal mode oscillations 
\cite{Vishveshwara:1970zz, Vishveshwara:1970cc}\footnote{For a binary of NSs, 
there is likely an intermediate stage of a hypermassive metastable NS as the 
merger product.}. Each of these phases has been properly studied with a 
distinct methodology, as the physical conditions around the peak amplitude of 
the CBC event rapidly switch. The initial inspiral phase, where 
the NR approximation holds, is ideally treated with post-Newtonian (PN) theory 
\cite{Blanchet:2013haa,Schafer:2018kuf}, whereas in entering the merger phase 
and the strong field regime, PN theory, in principle, breaks down. The merger 
and ringdown phases are currently studied in detail via numerical 
simulations, which solve for the fully exact relativistic evolution of the 
binary \cite{Pretorius:2005gq}. Finally, the ringdown phase can also be 
analytically treated with BH perturbation theory and self-force formalism, 
where the BH spacetime is considered as slightly deformed by a mass much 
smaller than the mass of the remnant compact object. 

\begin{figure}[t]
\centering
\includegraphics[width=0.5\linewidth]{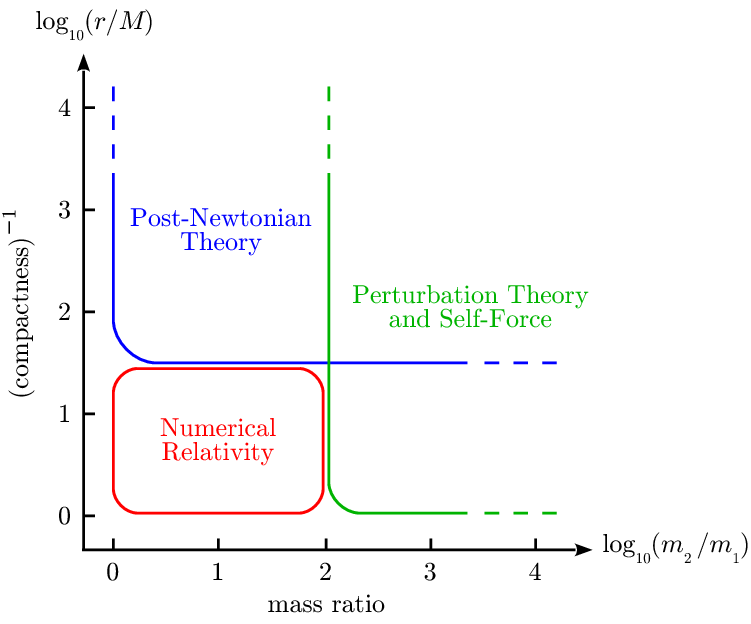}
\caption{The various elementary analytical and numerical methodologies used to 
study CBCs, depending on the mass ratio, $0<m_1/m_2\le1$, of the components 
of the binary, and the compactness of the system, evaluated by the parameter 
$M/r$, where $M=m_1+m_2$ is the total mass, and $r$ the typical size of the 
system. However, in order to model the complete CBC signal, one needs 
to resort to the comprehensive EOB framework, which enables their integration  
into the full theoretical GW templates, currently used in the GW detectors. 
Note that BH perturbation theory also accounts for the case of EMRIs, 
which will be targeted by the future space-based GW detector LISA.
Reproduced with permission from Le Tiec 
\cite{Tiec:2014lba}.}
\label{methods} 
\end{figure}

It is important to stress, that while numerical simulations currently exclusively 
treat the strong field regime in full detail, covering the merger and ringdown phases of 
the CBC signal, they are inherently not suitable for tackling the inspiral phase. 
The long inspiral phase makes up the major portion of the CBC signal if observed in its 
entirety, and depending on the masses in the binary, it can be the only observed 
signal falling within the frequency band of the detectors. The inspiral phase 
is also uniquely telling, providing exclusive input on gravity 
theories in the PN regime, and on the individual components of the binary, yet 
it can not be treated numerically due to the intrinsic long timescale, which 
characterizes the evolution at this stage. The regimes of validity 
of the different and complementary elementary physical methodologies used to study 
the CBC event, and their overlap, are illustrated in figure \ref{methods}, 
depending on the mass ratio of the components of the binary, and the 
compactness of the system \cite{Tiec:2014lba}. The latter is evaluated by the 
ratio of the total mass of the system to its typical size, i.e.~$M/r$, where 
$M=m_1+m_2$ is the total mass of the binary, and $r$ is its orbital 
separation.

Remarkably, even prior to the breakthrough in numerical relativity, which 
enabled the first exact treatment of the CBC event in the strong field regime 
\cite{Pretorius:2005gq}, an inclusive theoretical framework, which tackles 
and models the complete evolution of the compact binary, has already been devised 
by Buonanno and Damour \cite{Buonanno:1998gg,Buonanno:2000ef,Damour:2012mv}. 
This framework, dubbed the `effective one-body' (EOB) approach, provided 
the first complete theoretical waveform templates for the CBC signal. The 
idea behind the EOB approach was to relate the first and last phases of 
evolution, where analytical control has been feasible, via a mapping of the 
two-body problem to an effective one-body problem, in the spirit of Newtonian 
physics. This mapping, which thus enables fully tracking the evolution of the 
system, as it continuously transforms from having two compact objects to a 
single one, is non-trivially constructed in curved spacetime 
\cite{Damour:2012mv}. Therefore, the EOB framework currently enables 
extraction of the desirable elements from all the abovementioned elementary 
methodologies, and to integrate them into the full theoretical CBC template 
models, that are currently used in the GW detectors. There are additional 
phenomenological models, which similarly to EOB ones, also combine analytical 
and numerical results into complete inspiral-merger-ringdown waveform models 
\cite{Hannam:2013oca,Khan:2015jqa}.
An elaborate discussion of the various methodologies used to study and 
model the CBC signal, and their multiple interfaces, can be found in 
\cite{Tiec:2014lba}.

As the CBC signal is weak with respect to the sensitivities and various noise 
sources of current ground-based GW detectors, matched filtering is utilized 
for the detection of GW events, which places high demands on the accuracy of 
theoretical waveform templates. Moreover, with the expected increasing influx 
of even more detailed and telling GW data, we have truly entered a new era of 
high precision gravity. To that end, the framework of effective field 
theories (EFTs) is as a powerful means to advance theoretical 
progress, as put forward by Goldberger et al.~\cite{Goldberger:2004jt, 
Goldberger:2006bd,Goldberger:2007hy}. In particular, EFTs, which are designed 
for high precision computation, can be applied in the inspiral and ringdown 
phases of the CBC, where multiple widely separated physical scales are 
involved, and the evolution is amenable to analytical perturbative treatment. 
However, the concept of EFTs is inextricably tied to that of renormalization, 
historically devised in the realm of quantum field theory (QFT) and particle 
physics, where these and classical gravity are considered as disparate 
branches of physics. Despite this, the novel EFT approach for gravitational 
radiation presented in \cite{Goldberger:2004jt,Goldberger:2006bd, 
Goldberger:2007hy} soon inspired the use of the EFT framework in some modern 
problems in classical gravity, where a clear hierarchy of scales also exists. 
Notably, EFTs were applied to higher-dimensional gravity with compact 
dimensions, inspired by string theory \cite{Chu:2006ce,Kol:2007rx, 
Emparan:2009cs,Gilmore:2009ea,Emparan:2009at}, or to weak ultra-relativistic 
scattering \cite{Kol:2011pj}.

Following the introduction of the EFT formulation for the compact binary 
inspiral, where the small parameter, controlling the hierarchy of scales, is 
the NR orbital velocity, $v\ll1$\footnote{Throughout this review the speed of 
light, $c$, and the reduced Planck constant, $\hbar$, are taken as 
$c=\hbar=1$, such that the Planck mass, $m_p$, and the gravitational 
constant, $G$, are normalized via $m_p^2\equiv\frac{1}{32\pi G}$.}, 
consistent progress was made along this research program in PN theory. After 
the first PN (1PN) order\footnote{The $n$PN order is the 
$\left(v/c\right)^{2n}$ order correction in GR to Newtonian gravity.} 
correction in the conservative orbital dynamics was reproduced\footnote{A 
comprehensive account of PN theory methodologies and results can be found in 
the seminal review by Blanchet \cite{Blanchet:2013haa}, as well as the recent 
review by Sch{\"a}fer and Jaranowski \cite{Schafer:2018kuf}, whose focus was 
solely on Hamiltonian formulations and results. Though the progress made 
in the EFT field presented here belongs to the advances in PN theory, the 
latter constitutes a monumental body of work, which is beyond the limited 
scope of this review.} \cite{Goldberger:2004jt,Goldberger:2007hy}, and was 
simplified with a useful Kaluza-Klein (KK) reduction of the spacetime metric 
\cite{Kol:2007bc,Kol:2010si,Kol:2010ze}, higher PN order corrections in the 
point-mass sector were reproduced: The 2PN order \cite{Gilmore:2008gq, 
Chu:2008xm}, where in \cite{Chu:2008xm} it was first considered within a 
generic $n$-body problem, and an automatization of the computations was put 
forward, as well as the 3PN order \cite{Foffa:2011ub}. Subsequently, the 
current state of the art at the 4PN order was approached in 
\cite{Foffa:2012rn}, and in parallel with the progress and completion of this 
sector, attained via other PN methodologies \cite{Blanchet:2013haa}, it 
was partly computed in \cite{Foffa:2016rgu}, making use of the analogy with 
amplitudes of massless gauge theory. Moreover, after the introduction of the 
effective worldline mass-induced action of a single compact object in 
\cite{Goldberger:2004jt,Goldberger:2006bd,Goldberger:2007hy}, its non-minimal 
coupling part has been extended in \cite{Bini:2012gu}. Furthermore, the 
Wilson coefficients of this part of the effective action were studied for BHs 
in an arbitrary dimension in \cite{Kol:2011vg}, where a classical 
renormalization group (RG) flow was found at higher dimensions.

However, the knowledge of spin effects in the compact binary inspiral is also of 
crucial importance to gravity, quantum chromodynamics (QCD), and astrophysics 
(see, e.g.~\cite{Abbott:2016izl,Gerosa:2017xik}). Spin is one of 
the two unique features of BHs in nature, along with the mass, whereas spin 
effects in NSs are required in order to learn about their internal structure 
\cite{Harry:2018hke}, which in turn is also expected to reveal information on QCD and 
strong gravity. Indeed, spin effects in the orbital dynamics were first 
tackled within the EFT approach in \cite{Porto:2005ac}, where leading order 
(LO) corrections at the 1.5PN and 2PN orders\footnote{The explicit PN order 
of spin effects is evaluated, in accordance with the standard practice found in 
the templates for the LIGO/Virgo detectors, for rapidly rotating compact objects, 
particularly for near extremal Kerr BHs.} were reproduced. Subsequently, 
next-to-leading order (NLO) corrections at the 3PN order were approached in 
\cite{Porto:2006bt,Porto:2008tb,Levi:2008nh,Porto:2008jj}, and reproduced in 
\cite{Perrodin:2010dy,Porto:2010tr,Levi:2010zu} at the 2.5PN order, where in 
\cite{Porto:2008jj} finite size effects were tackled, and in 
\cite{Levi:2008nh,Levi:2010zu} the explicit relations to corresponding 
Hamiltonians were provided. Moreover, next-to-next-to-leading order (NNLO) 
corrections \cite{Levi:2011eq,Levi:2014sba,Levi:2015ixa, Levi:2016ofk} as 
well as LO ones \cite{Levi:2014gsa} at the 4PN order were obtained. Further, 
NNLO corrections \cite{Levi:2015uxa} and LO ones \cite{Levi:2014gsa} at the 
3.5PN order were reproduced, where in \cite{Levi:2014gsa,Levi:2015ixa, 
Levi:2016ofk} finite size effects were considered. This state-of-the-art 
PN accuracy in the spin sector was based on the EFT formulation for spinning 
objects presented in \cite{Levi:2015msa,Levi:2017oqx}, building on 
\cite{Goldberger:2004jt,Goldberger:2007hy,Levi:2008nh,Levi:2010zu,
Levi:2014sba}, where in \cite{Levi:2015msa} all spin effects up to NLO were 
consistently reproduced. It should be noted, that a distinct EFT formulation 
for general gravitating rotating objects was provided in 
\cite{Delacretaz:2014oxa}, which is formally and phenomenologically 
complementary to that in \cite{Levi:2015msa}, and was not specified in the PN 
context.

Dissipative effects of the components of the inspiraling compact binary, such 
as the absorption of gravitational energy by the horizon of BHs, or the 
dissipative tidal deformations of NSs, also have an effect on GW 
emission. For astrophysical objects both finite size and dissipative effects 
provide an indirect probe of the microphysics of their internal structure. 
Thus, the absorption by BH horizons was considered and reproduced in the EFT 
framework in \cite{Goldberger:2005cd,Porto:2007qi,Kol:2008hc}, where the 
leading PN dissipative effect is superradiance of rotating BHs at the 5PN 
order. More general dissipative effects of arbitrary rotating gravitating 
objects were considered in \cite{Endlich:2015mke,Endlich:2016jgc}, based on 
the EFT formulation in \cite{Delacretaz:2014oxa}, which are expected to 
contribute at higher PN orders. Tidal deformations of non-rotating 
gravitating objects were also treated in \cite{Bini:2012gu} via an effective 
action approach. Finally, dynamical tidal deformations of non-rotating 
objects, in particular resonances in NSs, were treated in 
\cite{Chakrabarti:2013lua,Chakrabarti:2013xza,Steinhoff:2016rfi}.

The direct effects of the gravitational radiation, emitted from the binary 
system, were first formulated and treated in detail in the EFT framework in 
\cite{Goldberger:2009qd}, within an effective theory with a set of multipole 
moments of the system, which sources the radiation. Next, general expressions 
for the multipole moments, radiated power, and GW amplitude, were derived 
from a general multipole expansion of the action in \cite{Ross:2012fc}. The 
1PN order correction to an infinite subset of the multipoles was derived 
explicitly in \cite{Birnholtz:2013nta,Birnholtz:2014fwa}, and NLO multipole 
moments, that depend on the spins of the compact components of the binary, 
were tackled in \cite{Porto:2010zg,Porto:2012as}. Hereditary tail effects, 
where the radiation emitted from the binary is scattered by its self-generated 
gravitational background, were studied and reproduced in 
\cite{Goldberger:2009qd}, where a classical RG flow of the mass quadrupole of 
the binary was found. A subleading RG flow of the total mass of the system 
was uncovered in \cite{Goldberger:2012kf}. 

Subsequently, radiation reaction, which is the backreaction of the 
gravitational radiation, emitted from the system, on the system itself, 
should also be taken into account. In fact, the treatment of PN radiation 
reaction in compact binary inspirals is in close analogy with the self-force 
in EMRIs, where there is a small mass, $m$, moving in the gravitational 
background of a large mass, $M$ (see, e.g. a recent review in 
\cite{Barack:2018yvs}). EMRIs are studied perturbatively in terms of the small 
mass ratio, $m/M\ll1$, where one considers the self-forces acting on the 
small mass due to its own effect on the gravitational background, generated 
by the large mass. Moreover, examining the test particle limit is similar to 
considering EMRIs with BH perturbation theory and self-force, and it provides 
information on the ringdown phase of the comparable mass CBC in the strong 
field regime. Self-force in EMRIs was first studied in an EFT approach in 
\cite{Galley:2008ih}, where the LO correction was 
reproduced\footnote{However, the primary obstacle here is that the relevant 
Green's functions in the BH spacetime background are highly non-trivial, and 
require numerical computation, so one would have to find out how to 
implement a semi-analytic EFT.}. The self-force in EMRIs was further studied 
within the EFT framework at higher orders in the mass ratio in a non-linear 
scalar theory model in \cite{Galley:2010xn,Galley:2011te}. Moreover, the 
self-forces in the ultra-relativistic limit, and in non-vacuum spacetimes, 
were investigated within an EFT approach in \cite{Galley:2013eba}, and 
\cite{Zimmerman:2015rga}, respectively.

The radiation reaction in the compact binary inspiral with components of 
comparable mass was then tackled in an EFT approach in \cite{Galley:2009px}, 
where the LO effect at the 2.5PN order was reproduced. Next, the NLO 
correction at the 3.5PN order, an effect still linear in $G$, was reproduced 
in \cite{Galley:2012qs}. The leading non-linear tail effect, which yields a 
radiation reaction effect at the 4PN order, was first approached in 
\cite{Foffa:2011np}, and further studied in \cite{Galley:2015kus}, using a 
formulation for classical causal actions of generic dissipative systems, 
which was put forward in \cite{Galley:2012hx,Galley:2014wla}. An independent 
formulation of classical causal actions for the study of radiation and 
the radiation reaction, which can also be generally applied at the level of the 
equations of motion (EOMs), was devised in \cite{Birnholtz:2013nta, 
Birnholtz:2014fwa}, and extended for a general spacetime dimension in 
\cite{Birnholtz:2013ffa,Birnholtz:2015hua}. The leading non-linear radiation 
reaction effect contributes logarithmic corrections to the binding energy, 
which were first derived via the EFT approach in \cite{Goldberger:2012kf}, 
though not through an analysis of the radiation reaction. Lastly, the leading 
radiation reaction correction linear in the spin of the objects at the 4PN 
order was reproduced in \cite{Maia:2017gxn}, while the related LO spin-orbit 
tail effect was previously reproduced in \cite{Porto:2012as}. The leading 
radiation reaction correction quadratic in the spins at the 4.5PN order,  
including spin-squared finite size effects, was derived in 
\cite{Maia:2017yok}.

All in all, as outlined above, the mature development of this interdisciplinary 
field over the last decade, initiated in \cite{Goldberger:2004jt, 
Goldberger:2007hy}, has resulted in significant advances of wide interest to 
physics at several levels serving various purposes. First, the field has 
brought novel perspectives to such important concepts of theoretical physics 
as renormalization, and the RG. The most fundamental lesson, which this field 
has validated, is how much the EFT framework is in fact a universal one, apt 
at describing an unprecedented scope of physical phenomena. As long as a 
hierarchy of scales, widely separated by a small ratio parameter, is 
identified in a problem prone to non-linear perturbations, EFTs can be used, 
whether the non-linear coupling involves the quantum parameter, 
$\hbar$, or any other non-quantum parameter. The EFT scheme provides a systematic 
methodology for constructing an effective theory, which describes the physics at 
some relevant scale to arbitrarily high accuracy. It organizes the 
perturbative calculation efficiently by employing powerful tools, which are 
standard in QFT, such as Feynman diagrams and calculus. It also provides a 
natural framework to handle the regularization required at higher-order 
perturbative corrections within the standard renormalization scheme. Thus, 
the universality of the EFT framework is extremely useful, as multi-scale 
problems are abundant across all of physics. Consequently, the field has 
instigated the widespread use of the EFT framework in classical gravity in 
several modern problems, which are in fact motivated by high energy 
physics, and also address the attempt to arrive at a complete theory of 
gravity across all scales.

The impact of the field within the definite domain of PN theory and GWs has 
also been established and varied. Goldberger et al.~have put forward an 
original self-contained EFT program to tackle the intricate PN binary 
inspiral problem. It should be noted though that some ingredients, inherently 
included in the EFT methodology, have appeared in some form (which bears some 
resemblance) in past work \cite{Fokker:1929}, or specifically in the domain, 
e.g.~\cite{Damour:1995kt, Damour:1998jk}. In the EFT formulation a manifest 
power counting in the small PN expansion parameter was achieved by performing 
a decomposition of the gravitational field into modes with definite scaling 
properties, at the level of the action. Further, the use of action 
formalism and symmetries, intrinsic in the EFT framework, was invigorated in 
the various traditional PN methodologies. The tackling of several PN effects 
via the EFT framework propelled further progress in other PN methodologies in 
parallel. Notable within the progress made in the field is the improvement in 
the general understanding of classical spins in gravity, and consequently 
also several new results in the spinning sectors, some of which so far 
exclusively obtained via the EFT methodology. Some of the remarkable findings  
within the field include the classical RG flows, and related higher PN order 
logarithm corrections, which also constitute a unique prediction arising from 
the EFT framework. The final marked advance in the field is the `EFTofPNG' 
public code, which incorporates the EFT framework for high precision 
computation in PN gravity, including spin effects \cite{Levi:2017kzq}. Beyond 
its contribution at the strict level of computational and mathematical 
physics, it is an accessible tool for the broader community, which also 
practically fills in the current analytic gap for this significant part of the 
gravitational waveform models. Notably, this is the first comprehensive code 
in PN theory to be made public, and hence complements current public codes for GWs, 
which include numerical codes, codes for BH perturbation theory 
\cite{BHPToolkit}, and codes for data analysis. 

However, the field continues to develop. In particular, further progress is 
desirable in the non-conservative sectors of PN theory. Building on the 
progress in the field, the various aspects to be advanced within the context of 
GWs, in addition to the abundant potential of using further timely field theory 
advances to improve our understanding of gravity at all scales, are 
elaborately discussed in section \ref{future}. 

This review article is aimed at a broad audience, from general readers new to 
the field, to specialists and experts in related subjects. In this review 
emphasis was put on an accessible pedagogic presentation of the field 
theoretic aspects of the subject, with the view, that these are in fact 
common across the whole of theoretical physics, rather than in their original 
narrow quantum context. The review presents the progress made in the field 
since the introduction of the EFT approach to the PN binary inspiral, where 
works that served as mileposts are spelled out in more detail. 
Different reviews in the field of EFTs in PN gravity, highlighting various 
aspects of it from several perspectives, are found in 
\cite{Goldberger:2006bd,Goldberger:2007hy,Foffa:2013qca,Rothstein:2014sra, 
Porto:2016pyg}. 

The outline of this review is as follows. Section \ref{efts} provides the 
basic ideas, which form the conceptual foundation of EFTs, and outlines the 
strategy of a multi-stage EFT framework, which is deployed in the PN binary 
inspiral problem. This section actually serves as the formal complement of 
the introductory section. The main body of the review is then dedicated to 
presenting in detail the study of each of the effective theories at each of 
the intermediate scales in the problem, up to the actual GW observables: From 
the EFT of the single compact object in section \ref{oneeft}, on to that of 
the binary, viewed as a composite particle with internal binding 
interactions, in section \ref{compeft}, to the final effective theory of the 
time-dependent multipole moments of the radiating system in section 
\ref{togwobs}. The review is concluded in section \ref{future} with the 
multiple prospects of building on the progress in the field, and using further 
modern field theory insights and tools, to specifically address the study of GWs, 
as well as to expand our fundamental understanding of gauge and Gravity theories 
at all scales.

\section{Tower of EFTs}
\label{efts}

\begin{figure}[t]
\centering
\includegraphics[width=0.9\linewidth]{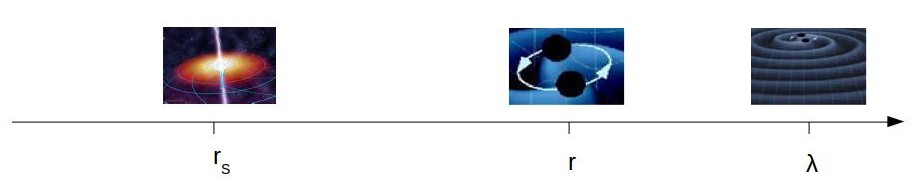}
\caption{The hierarchy of scales in the binary inspiral problem: $r_s$, the 
scale of the single compact object; $r$, the scale of the orbital separation 
between the components of the binary; $\lambda$, the wavelength of 
radiation, emitted from the inspiraling binary; it holds that $r_s\lll r\ll 
\lambda$. To eliminate the smallest scale of the single object, we construct 
the one-particle EFT. Next, we explicitly integrate out the field modes at 
the orbital separation scale, to be subsequently matched to the EFT of the 
composite particle, that is, the binary system. Finally, we eliminate the 
scale of radiation by explicitly integrating out the radiation field modes.} 
\label{efttower}
\end{figure}

Let us first establish the hierarchy of scales in the binary inspiral 
problem, as depicted in figure \ref{efttower} \cite{Goldberger:2004jt, 
Goldberger:2007hy}. Since each isolated object in the binary is compact, with 
a characteristic mass $m$, it holds that its gravitational radius, $Gm$, is 
of the order of the characteristic scale of its internal structure, $r_s$, 
i.e.~$r_s \sim Gm$. For a PN binary system, namely a gravitationally bound 
one, where the motion of the components has typical NR velocities, $v\ll1$, 
the virial theorem holds, such that $Gm/r \sim v^2$, where $r$ is the typical 
orbital separation of the bound system. The binary system emits gravitational 
radiation, e.g.~the quadrupole radiation at LO, and so the radiation 
frequency is fixed by the orbital frequency of the binary, $\omega$. Since 
the observed radiation consists of on-shell field modes, i.e.~such that their 
momentum, $(k_0$,$\vec{k})$, satisfies $k_0=|\vec{k}|\equiv k$, it holds that 
the radiation wavelength, $\lambda$, satisfies 
$\lambda^{-1}\sim k \sim \omega \sim v/r$. All in all, the three characteristic 
scales, $r_s$, $r$, and $\lambda$, satisfy the following hierarchy:
\be 
r_s\sim rv^2 \sim \lambda v^3,
\ee 
such that the mass, $m$, is the only scale in the full theory, where the NR 
velocity, $v$, is the small parameter, that controls the effective theories 
at different scales. It should also be noted, that once spin is also taken 
into account, there is an additional scalar, that is the spin length, $S^2$, 
and yet it holds that $S\lesssim m^2$.

To arrive at the orbital dynamics and GW emission observables, we eliminate each 
of the scales in our problem, one at a time, by constructing an effective 
theory, where each scale is integrated out, such that we go through three 
intermediate stages in a tower of effective theories. There are in fact two 
generic procedures to construct an EFT, referred to as the `bottom-up' and 
`top-down' approaches. They differ in how the effective action, which 
formally represents the EFT, is obtained. In the bottom-up approach the 
effective action is constructed from scratch as an infinite series of 
operators, consisting of the degrees of freedom (DOFs), and constrained by 
the symmetries, identified at the relevant scale. The top-down approach, on 
the other hand, arrives at the effective action by explicitly eliminating 
DOFs from the full action of the small scale (high energy) theory. 

In the binary inspiral problem we apply both of these procedures to construct 
the effective theories in our tower, as detailed in sections \ref{oneeft}, 
\ref{compeft}, and \ref{togwobs}. Indeed, in section \ref{towerefts} we 
outline the general strategy, which is deployed in this problem, to 
ultimately arrive at the orbital dynamics and GW emission observables through 
three intermediate stages. However, before we do that, let us open with a simple 
QFT prelude in section \ref{laymaneft}, which provides an idea of how a tower of 
EFTs works and on the fundamental concepts involved \cite{Peskin:1995ev}. We 
assume familiarity with the basics of QFT.

\subsection{EFTs for pedestrians}
\label{laymaneft}

Let us consider the interacting $\phi^4$ theory, given by the following 
Lagrangian:
\be
{\cal{L}}(\phi)=
\frac{1}{2}\left(\partial_\mu\phi\right)^2+\frac{1}{2}m^2\phi^2
+\frac{\lambda}{4!}\phi^4,
\ee
with some sharp ultraviolet (UV) momentum cutoff, $\Lambda$, and find the 
EFT, which corresponds to lowering the cutoff of the theory to $b\Lambda$, 
where $b$ is some parameter, such that $0<b<1$.

To this end, we proceed to apply Wilson's approach to renormalization to this 
simple theory, and integrate over the momentum shell. Let us first redefine 
the field according to its Fourier modes, $\phi(k)$, which will become the 
integration variable in a functional integral, in the following manner:
\be
\begin{array}{ll}  
\phi(k)\equiv\left\{\begin{array}{ll}  
\phi(k)   & \quad |k|<b\Lambda\\
0      & \quad |k|\ge b\Lambda
\end{array}\right.\qquad ,\qquad
\hat{\phi}(k)\equiv\left\{\begin{array}{ll}  
\phi(k)   & \quad b\Lambda\le |k| < \Lambda\\
0      & \quad \text{Otherwise}  
\end{array}\right.,
\end{array}
\ee
where we have also defined $\hat{\phi}$, which represents the high momentum 
DOFs, that we want to integrate out, and the original field, $\phi$, has been 
decomposed into two distinct orthogonal momentum components by transforming 
$\phi\to\phi+\hat{\phi}$. Let us then rewrite the theory in terms of the two 
distinct components of field modes: 
\begin{align}\label{lnew} 
{\cal{L}}(\phi,{\hat{\phi}})={\cal{L}}(\phi)
+\frac{1}{2}\left(\partial_\mu{\hat{\phi}}\right)^2
&+\frac{1}{2}m^2{\hat{\phi}}^2\nn\\
&+\lambda\left(\frac{1}{6}\phi^3 {\hat{\phi}}
+\frac{1}{4}\phi^2 {\hat{\phi}}^2
+\frac{1}{6}\phi {\hat{\phi}}^3
+\frac{1}{4!}{\hat{\phi}}^4\right).
\end{align} 
We can now carry out integration over the high momentum DOFs to arrive at 
an EFT with the cutoff $b\Lambda$, where ${\cal{L}}_{\text{eff}}(\phi)$ is 
actually defined by a functional integral, as follows:
\be\label{leffdef}
\int {\cal{D}}\phi \,\,\,\, \text{exp}\left(-\int d^dx \, 
{\cal{L}}_{\text{eff}}(\phi)\right)\equiv
\int {\cal{D}}\phi \int {\cal{D}}{\hat{\phi}}\,\,\,\, 
\text{exp}\left(-\int d^dx \, {\cal{L}}(\phi,{\hat{\phi}})\right),
\ee
where the functional integral is considered in the Euclidean form, i.e.~after 
Wick rotation.

\begin{figure}[t]
\centering
\includegraphics[width=0.5\linewidth]{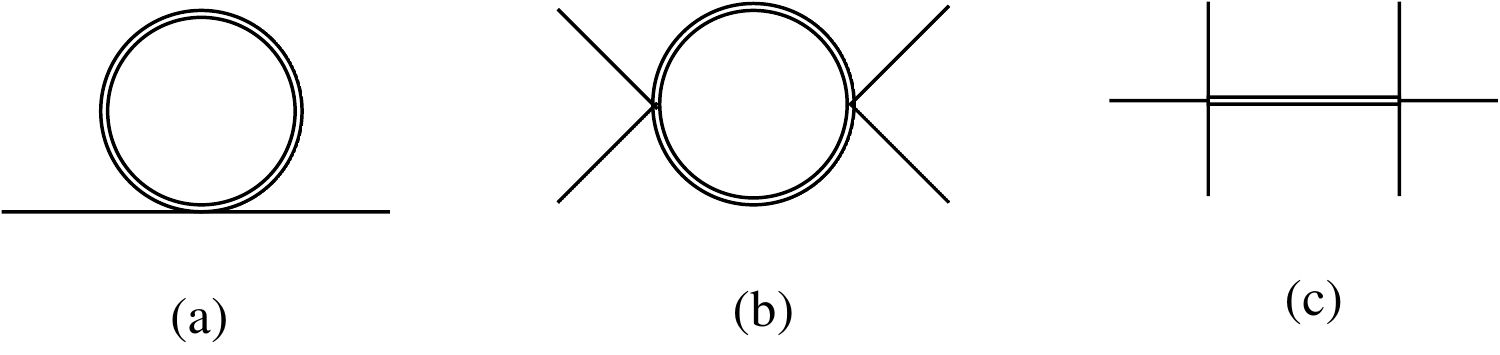}
\caption{A $\phi^4$ EFT through Wilson's approach: ${\hat{\phi}}$, the high 
momentum field component is represented by a double line, while $\phi$, the 
low momentum component, is represented by a single line. Note that the former 
will always appear as an internal line, whereas the latter will always appear 
as external. Diagram (a) is evaluated as $\frac{1}{2}\rho\phi^2$, where the 
new coefficient, $\rho$, gives a correction to the mass parameter, $m^2$. 
Diagram (b) is evaluated as $\frac{1}{4!}\zeta\phi^4$ at LO, where $\zeta$ is 
a correction to the coupling $\lambda$. Diagram (c) generates a new $\phi^6$ 
interaction with a new coupling constant, $\kappa$; and so on.}
\label{phi4eftgr}
\end{figure}

The integral over the Fourier component ${\hat{\phi}}$ is performed using the 
standard QFT perturbative method, involving the extraction of Feynman rules 
from the action, followed by a diagrammatic expansion of the exponential on 
the right hand side (RHS) of equation \eqref{leffdef}. Thus we read off the 
propagator $\langle {\hat{\phi}\,\hat{\phi}}\rangle$ from the kinetic term of 
$\hat{\phi}$ in equation \eqref{lnew}, and all the following terms in this 
equation are regarded as perturbations. Note that this also includes the mass 
term, since the situation considered here is $m^2\ll\Lambda^2$. Then, the 
Feynman graphs generated in the expansion are as depicted\footnote{All Feynman 
graphs in this manuscript were created using JaxoDraw \cite{Binosi:2003yf, 
Binosi:2008ig}, based on Axodraw \cite{Vermaseren:1994je}.} in figure 
\ref{phi4eftgr}. Integrating over $\hat{\phi}$ in such a manner leads to the 
following theory for $\phi$ with $|k|<b\Lambda$: 
\begin{align} \label{phi4new}
{\cal{L}}_\text{eff}(\phi)& = 
 \frac{1}{2}\left(\partial_\mu \phi\right)^2+\frac{1}{2}m^2\phi^2
+\frac{\lambda}{4!}\phi^4\nn\\
&\quad+\left(\text{sum of connected diagrams}\right)\nn\\
& = \frac{1}{2}\left(\partial_\mu\phi\right)^2
+\frac{1}{2}\left(m^2+\rho\right)\phi^2
+\frac{1}{4!}\left(\lambda+\zeta\right)\phi^4
+\frac{1}{4}\eta\phi^2\left(\partial_{\mu}\phi\right)^2+\kappa\phi^6+\ldots,
\end{align}
where the sum is only over connected diagrams, since those are exponentiated 
to also yield the disconnected ones in the expansion of equation \eqref{leffdef}.
The diagrammatic contributions include corrections to the original coupling 
constants, $m^2$ and $\lambda$, as well as operators of higher mass dimension 
with new interactions and coupling constants. It should be noted, that there 
are also derivative interactions from taking into account the external 
momenta of the diagrams via their Taylor expansion. Indeed, one finds, that 
the effective Lagrangian, ${\cal{L}}_{\text{eff}}(\phi)$, equals the original 
Lagrangian, ${\cal{L}}(\phi)$, plus new interaction terms, which compensate 
for the removal of the high momentum field DOFs. 

In general, the procedure of explicitly integrating out the high momentum 
component, ${\hat{\phi}}$, generates all possible interactions of the low 
momentum field modes, $\phi$, and their derivatives, allowed by the 
symmetries of the effective theory. This means, that one can arrive at the 
same result, i.e.~EFT, just by augmenting the form of the original Lagrangian 
through writing down an infinite series of all possible operators of higher 
mass dimension, which satisfy the symmetries, that apply below the new cutoff 
scale. The original and new operators will be preceded by new coupling 
constants, which encapsulate the UV information, that has been suppressed 
from the new effective theory, like the coefficients computed explicitly in 
figure \ref{phi4eftgr} of the operators $\phi^2$, $\phi^4$, $\phi^6$, and so 
on. In EFT terminology these coupling constants are 
referred to as \emph{Wilson coefficients}. What we have just described, in 
fact, constitutes the content of the \emph{decoupling theorem}, which states 
that physics at different scales factorizes, and that the UV physics can be 
safely suppressed and accounted for in a systematic manner 
\cite{Appelquist:1974tg}. 

One can now recognize the two approaches to tackle an EFT, that were 
alluded to in the introduction of this section: The top-down procedure, which 
corresponds to Wilson's approach, where one explicitly integrates out high 
momentum DOFs from a small scale theory, as outlined above, and the 
bottom-up approach, which relies on the universality of the decoupling 
theorem. The latter can always be deployed, whereas for the top-down 
approach, the full theory at the small scale should be known, and amenable to 
perturbative treatment. If possible, the two approaches can then be applied 
in parallel, as is actually demonstrated in our problem here, in what is 
described in section \ref{towerefts} as the second stage in the tower of 
EFTs, where we construct an EFT for the binary system as a composite 
particle. Using the two approaches can, in fact, be one way of fixing the 
Wilson coefficients, which encode the coupling to the UV scales, a procedure 
referred to as \emph{matching}. Otherwise, the matching of the Wilson 
coefficients can be done by using data from experiments. 

Let us further consider the new action in equation \eqref{phi4new} by rescaling 
distances and momenta according to $x'=xb$, $k'=k/b$, with the new cutoff 
parameter, $b$. Then the field, $\phi$, can also be rescaled, such that the 
unperturbed action returns to its initial form, while the various 
perturbations undergo a transformation, that changes the coupling parameters 
of the Lagrangian. Hence the combination of integrating out high momentum 
DOFs with the rescaling of distances and momenta can be rewritten as a 
transformation of the Lagrangian. One can then further lower the cutoff, and 
repeat a similar procedure iteratively. Such continuously generated 
transformations of Lagrangians are referred to as the \emph{renormalization 
group} (RG). They do not form a group in the strict mathematical sense, since 
the operation of integrating out DOFs is not invertible (resolution is lost). 
However, they renormalize the fields and coupling constants of the theory. There 
is a universal function, $\beta$, related to the shift in the field strength 
and coupling constants, that compensates for the shift in the renormalization 
scale of the theory, $\mu$. The $\beta$ function is just the rate of the 
\emph{RG flow} of the coupling constants. For the coupling parameter 
$\lambda$ in our example here, e.g., the $\beta$ function reads as follows: 
\be
\beta(\bar{\lambda}) = \mu\frac{d}{d\mu} \bar{\lambda }(\mu),
\ee
where $\bar{\lambda}$ is referred to as the \emph{running} coupling constant, 
and this is the \emph{RG equation }.

\subsection{Tower of EFTs for GWs}
\label{towerefts}

Equipped with the basic concepts of renormalization and EFTs, let us now get 
back to the compact binary inspiral problem. Having identified the problem as 
a multi-scale one, with three widely separated scales, we now realize that it 
is ideal to employ the EFT framework to remove each of these scales, one at 
a time: We can integrate them out in three stages, thereby constructing a 
tower of EFTs, that eventually leads to the orbital dynamics and GW emission 
observables. At the first stage we construct the one-particle EFT to remove 
the small scale of the single isolated compact object. At the next stage we 
integrate out the orbital scale in order to arrive at the EFT of a composite 
particle, namely the binary system. Finally, in order to obtain the effects, 
that involve the radiation directly, we integrate out the radiation scale, 
which yields an effective theory of dynamical multipole moments. Let us 
describe here briefly each of these three stages, on which each of the 
following detailed sections \ref{oneeft}, \ref{compeft}, and \ref{togwobs}, 
respectively, elaborate.

\subsubsection{One-particle EFT} 

As we noted, the first goal is to remove the scale of the single compact 
object, $r_s$, from the purely gravitational action of an isolated compact 
object (i.e.~outside of its interior). For the theory of GR this means 
starting from the following full theory, given by the Einstein-Hilbert action 
for the gravitational field, $g_{\mu\nu}(x)$:
\be \label{sgfull}
S[g_{\mu\nu}]= -\frac{1}{16\pi G} \int d^4x \sqrt{g}\,R[g_{\mu\nu}].
\ee

To this end we take the bottom-up approach in order to attain a one-particle 
EFT, where we decompose the metric into two distinct momentum components, 
$g_{\mu\nu}\equiv g^s_{\mu\nu}+\bar{g}_{\mu\nu}$, and $g^s_{\mu\nu}$ denotes 
the strong field modes to be removed from the theory. Thus, we construct the 
one-particle EFT to describe the single object at the orbital scale by 
introducing an infinite series of worldline operators, which contain new 
worldline DOFs, that depend on a worldline parameter, $\sigma$. These 
operators are added to the initial action form in equation \eqref{sgfull} as 
additional interactions of the new DOFs with the gravitational field 
component, $\bar{g}_{\mu\nu}$, such that the new action is of the form: 
\begin{align} \label{oneparteft}
S_{\text{eff}}[\bar{g}_{\mu\nu}(x),y^\mu(\sigma),e^\mu_A(\sigma)]
&=S[\bar{g}_{\mu\nu}] \,+\, 
S_{\text{pp}}[\bar{g}_{\mu\nu}(y),y^\mu(\sigma),e^\mu_A(\sigma)]\nn\\
&=-\frac{1}{16\pi G}\int d^4x\sqrt{\bar{g}}\,R[\,\bar{g}_{\mu\nu}]
+\sum_{i=1}^{\infty} C_i(r_s) \int d\sigma \,{\cal{O}}_i(\sigma).
\end{align}
In this action we noted the generic new worldline DOFs, $y^\mu$, and 
$e^\mu_A$, corresponding to the position, and rotation of the object, 
respectively, and we refer to the new part of the effective action 
localized on the worldline as the point-particle action, $S_{\text{pp}}$. 

In the point-particle action all the UV physics goes into the Wilson 
coefficients, $C_i$, which depend on the only scale in the full theory, 
$r_s\sim m$, and precede the worldline operators ${\cal{O}}_i$, which depend 
only on the scale of the effective theory, i.e.~$r$. The terms in the 
point-particle action in equation \eqref{oneparteft} are thus inserted in 
increasing orders of the ratio $r_s/r$, i.e.~according to their PN relevance. 
Hence, the crucial task at this stage is to correctly identify the pertinent 
DOFs, and the symmetries at this scale, both of spacetime and of the object, 
and to construct all possible worldline operators, which couple the DOFs in 
all ways allowed by the symmetries. However, we note that the initial theory in 
equation \eqref{sgfull} is the full theory for a BH, but for generic compact objects, 
one should actually add a matter action, i.e.~some fluid description of matter. 
This part of the initial action will also be integrated out, leading to a 
point-particle action augmented by harmonic oscillator matter DOFs, which are coupled 
to tidal forces \cite{Levi:2015msa}. Finally, the Wilson coefficients are to be 
matched from the full theory to the one-particle EFT, where the detailed 
construction of the point-particle action in its explicit form is presented in 
section \ref{oneeft}.

\subsubsection{EFT of a composite particle}

At this subsequent stage we construct the EFT of a composite particle, namely 
the binary system, and remove the orbital separation scale, $r$. To reach 
this goal, both the top-down and bottom-up approaches used to construct EFTs, 
which are mentioned in section \ref{laymaneft}, are deployed here. 
We can write the gravitational field, $\bar{g}_{\mu\nu}$, as an expansion on 
the asymptotic flat spacetime, 
$\bar{g}_{\mu\nu}=\eta_{\mu\nu}+H_{\mu\nu}+\widetilde{h}_{\mu\nu}$, where 
$H_{\mu\nu}$ denotes the orbital or `potential' modes, and 
$\widetilde{h}_{\mu\nu}$ denotes the radiation modes. The orbital modes are 
those that mediate the gravitational interaction between the two objects, 
whereas the radiation modes consist of the on-shell gravitons emitted from 
the system, and measured by the asymptotic observer, i.e.~the GW detectors. 
Let us note then the scale dependence of these two different components, 
which is given by
\be \label{eq:kmodes}
\partial_t H_{\mu\nu} \sim \frac{v}{r}H_{\mu\nu}, \quad
\partial_i H_{\mu\nu} \sim \frac{1}{r}H_{\mu\nu}, \quad  
\partial_\mu\widetilde{h}_{\mu\nu} \sim \frac{v}{r}\widetilde{h}_{\mu\nu}, 
\ee
so the orbital modes are hard momentum modes with respect to the radiation 
modes. 

We start from the full theory at small scales, which is at this stage the 
one-particle EFT in equation \eqref{oneparteft}, that we constructed in the 
previous stage. As we write the theory here for the two objects, we include a 
copy of the point-particle action in equation \eqref{oneparteft} for each of the 
objects in the initial action, namely a two-particle theory: 
\be \label{twoparteft}
S_{\text{eff}}
[\bar{g}_{\mu\nu},y_1^\mu,y_2^\mu,e_{1\,A}^{\,\,\mu},e_{2\,A}^{\,\,\mu}]=
S[\bar{g}_{\mu\nu}] \,+\, 
\sum_{a=1}^{2}
S_{pp}[\bar{g}_{\mu\nu}(y_a),(y_a)^\mu,(e_a)_{A}^{\,\mu}](\sigma_a).
\ee
We integrate out explicitly the orbital modes from the full two-particle  
theory in equation \eqref{twoparteft} by using standard perturbative methods in 
QFT, with a diagrammatic expansion. Then, the effective action of the
composite particle is defined, similar to equation \eqref{leffdef}, by the 
following functional integral: 
\be \label{seff2part}
e^{iS_{\text{eff}}[\widetilde{g}_{\mu\nu},(y_c)^{\mu},(e_c)_{A}^{\,\mu}]}
\equiv
\int {\cal{D}} H_{\mu\nu}\,e^{iS_{\text{eff}}
[\bar{g}_{\mu\nu}, y_1^\mu,y_2^\mu,e_{1\,A}^{\,\,\mu},e_{2\,A}^{\,\,\mu}]},
\ee
where $\widetilde{g}_{\mu\nu}\equiv\eta_{\mu\nu}+\widetilde{h}_{\mu\nu}$, and 
the subscript `c' denotes the generic worldline DOFs of the composite object, 
similar to the worldline DOFs of a single object in equation \eqref{oneparteft}. 
The functional integral is considered here only in the classical limit, where 
the relevant Feynman diagrams to evaluate in the expansion are graphs, which 
are tree level in the fields\footnote{A graph with a graviton loop is suppressed 
by a factor of the ratio 
$\hbar/L\sim\left(m_p/m_{\odot}\right)^2\approx10^{-76}$, where $L$ is the 
angular momentum of a compact object, and $m_{\odot}$ is the solar mass. 
Hence graviton loops can be safely neglected, see also figure \ref{exclude}. 
However, it should be stressed that the diagrammatic expansion involves loop 
integrals due to the non-linearity of GR in the gravitational constant, $G$, 
i.e.~the self-interaction of the field in classical gravity.}. 

At the same time a bottom-up approach is taken, and the effective action of 
a single composite particle, coupled to the gravitational radiation field,
$\widetilde{g}_{\mu\nu}$, is written as follows \cite{Goldberger:2009qd,
Ross:2012fc}:
\be \label{seffcomp}
S_{\text{eff}}[\widetilde{g}_{\mu\nu},(y_c)^\mu,(e_c)_{A}^{\,\mu}]
=-\frac{1}{16\pi G}\int d^4x
\sqrt{\widetilde{g}}\,R\left[\widetilde{g}_{\mu\nu}\right] 
\,+\,S_{\text{pp(comp)}}
[\widetilde{g}_{\mu\nu},(y_c)^\mu,(e_c)_{A}^{\,\mu}](\sigma_c),
\ee
where $S_{\text{pp(comp)}}$ is the effective worldline action at the 
radiation scale, which describes the composite object with a suitable `center 
of object' coordinate, $(y_c)^\mu$, and a tetrad $(e_c)_{A}^{\,\mu}$. The 
explicit form of the effective worldline action, $S_{\text{pp(comp)}}$, is 
given in section \ref{radrg}, similar to the generic structure in 
equation \eqref{oneparteft}, where the multipole moments of the composite object 
are regarded as the Wilson coefficients of the theory. 

The matching of the Wilson coefficients of the EFT of the composite particle 
in equation \eqref{seffcomp} is done by explicit perturbative computation from the 
full two-particle theory as in equation \eqref{seff2part}. This perturbative 
computation consists of two types of Feynman diagrams: Graphs, which contain 
only internal orbital field modes between the two objects, with no external 
radiation field modes, and account for the conservative sector of the theory, 
from which the orbital dynamics is derived, and which is presented in detail 
in section \ref{compeft}; and graphs, which contain a single external 
radiation field mode, that are matched onto the radiative sector of the 
theory, and are discussed in the beginning of section \ref{radrg}.

\subsubsection{Effective theory of dynamical multipoles}

At this final stage we are concerned with effects, involving the radiation 
directly. In the conservative sector, where no radiation modes, 
$\widetilde{h}$, are present, the EFT progression is actually done after the 
two previous stages: One just needs to further process the resulting set of 
interaction potentials, which enable derivation of the EOMs for the constituents 
of the binary, the Hamiltonian potentials, the binding energy, and other 
gauge invariant observables, as discussed in sections \ref{redacteomh} and
\ref{eftofpngcode}. Otherwise, where radiation modes are involved, explicit 
perturbative computations are carried out, essentially taking the top-down 
approach. Here one starts with the full theory from the previous stage in 
equation \eqref{seffcomp}, of a composite particle with a set of multipole moments 
at the radiation scale, $\lambda$. 

These computations involve Feynman graphs with a single worldline of the 
composite particle, where radiation modes are integrated out, such that 
eventually there are no field DOFs left in the theory. However, since at this 
stage the system is dissipative, and time reversal no longer holds, the 
action in equation \eqref{seffcomp} should in fact be adjusted in order to 
describe the non-conservative evolution. This is indeed attained via the 
closed time path (CTP) formalism, and the corresponding use of retarded 
propagators, which is presented in detail in section \ref{radrg}. With this 
developed formulation of the action, one can compute the GW energy flux, 
gravitational waveform, tail effects, radiation reaction forces, and RG flows 
of the time-dependent multipole moments. Thus, after all radiation modes are 
integrated out, one is left with an effective theory of the set of dynamical 
multipole moments of the binary system. This eventual stage of the tower is 
discussed in detail in section \ref{togwobs}.

\section{One-particle EFT}
\label{oneeft}

In this section we discuss in detail the construction of the one-particle 
EFT presented in equation \eqref{oneparteft}. This EFT removes the 
short scale of the internal structure of the isolated compact object, $r_s$, 
and provides a description to a single object coupled to gravity, that holds 
at the orbital scale, $r$. As we noted in section \ref{towerefts}, this task 
is tackled with the bottom-up approach, where the challenging aspect is to 
properly identify the relevant DOFs and symmetries at this scale. 
Accordingly, we start by considering in section \ref{pointmass} the addition 
of worldline coordinate DOFs, which enter the point-particle action in 
equation \eqref{oneparteft}, and account for the simplified case of a 
non-rotating point mass coupled to gravity. 

Next, we proceed in section \ref{spinparticle} to take into account the 
actual case of a rotating object coupled to gravity by adding rotational 
worldline DOFs to the point-particle action. As it turns out, the extension 
of the point-particle formulation to describe a spinning particle is quite 
challenging. This is due to the fundamental conflict between an actual 
rotating gravitating object, which must have an extended finite size for its 
rotational velocity to not surpass the speed of light, and its view in the 
EFT as a point particle. 

Finally, dissipative effects, which occur at the small scale of the single 
object, such as the absorption of gravitational radiation on the horizon of 
BHs, or the dynamical tidal deformations of NSs, should also be taken into 
account. Therefore, in section \ref{dissipat} we discuss the addition of new 
worldline DOFs to the point-particle action, which are dissipative and hence 
dynamical.

\subsection{Point mass}
\label{pointmass}

We consider first the basic case of a non-spinning massive object 
\cite{Goldberger:2004jt,Goldberger:2007hy}.
In this case the only DOFs to add to the gravitational field modes are the 
worldline coordinates, $y^{\mu}$, and the relevant symmetries of the theory 
are the following:
\begin{enumerate}
\item \emph{General coordinate invariance}, including \emph{parity 
invariance}, where the latter plays a key role beyond minimal coupling;
\item \emph{Worldline reparametrization invariance};
\item \emph{Internal Lorentz invariance} of the local frame field.
\end{enumerate} 
The latter is an additional gauge freedom of the tetrad field that covers 
the spacetime manifold, and it is further considered in the gauge choices 
discussed in section \ref{gauges}. 

Let us first consider the minimal coupling in the point-particle action. Then 
we have the following term:
\begin{align} \label{eq:mmc}
-m \int d\tau&=-m \int \sqrt{\bar{g}_{\mu\nu}dy^{\mu}dy^{\nu}}\nn\\
&=-m \int d\sigma \, 
\sqrt{\bar{g}_{\mu\nu}\frac{dy^{\mu}}{d\sigma}\frac{dy^{\nu}}{d\sigma}}
=-m \int d\sigma \, \sqrt{u^2},
\end{align}
where $\tau$ is the proper time along the worldline, and 
$u^{\mu}\equiv \frac{dy^{\mu}}{d\sigma}$ is the coordinate velocity. 
This leading term includes the Newtonian interaction at the 0PN order, and 
provides higher-order PN corrections. Though usually the term `point-mass' 
refers to the approximation, that consists only of this leading term, in the 
EFT context it refers to all terms in the action, which are induced by the 
presence of the mass, considered as a point particle. 

Hence, we proceed to the non-minimal coupling terms in the point-particle  
action, namely to mass-induced higher multipoles, which account for finite 
size effects, where the internal structure of the objects starts to play a 
role. Considering operators with some dependence on the Riemann tensor and 
covariant derivatives, the first option that arises involves the Ricci 
tensor, $R_{\mu\nu}$. However, the leading EOMs of the gravitational field are 
sourceless, i.e.~of a vacuum spacetime, so $R_{\mu\nu}=0$. Therefore, 
operators constructed with the Ricci tensor are what is referred to in 
EFT terminology as \emph{redundant} operators, and can be practically omitted 
from the effective action. Formally, one can redefine the field in the 
operator, such that its Wilson coefficient is made to vanish. Suppose, for example, 
that we consider the operator $c_R\int d\tau \,R$. Then we can redefine the 
metric, which also appears in the Einstein-Hilbert term, according to 
$g_{\mu\nu} \to g_{\mu\nu}+\delta g_{\mu\nu}$, as follows:
\be
\delta  g_{\mu\nu}=-16\pi G \,\cdot \epsilon_R \int d\tau 
\,\frac{\delta^d(y-y(\tau))}{\sqrt{g}} \,\,g_{\mu\nu},
\ee
where $\epsilon_R$ is some unfixed coefficient. Plugging this redefinition 
of the field into the gravitational action results in a shift of the Wilson 
coefficient, $c_R\to c_R+\epsilon_R$, so by adjusting $\epsilon_R$ we can 
make $c_R$ vanish. This well-known property in EFTs is analogous to the 
well-known application of leading EOMs of coordinates in higher-order 
perturbative Lagrangians, e.g.~in PN ones, which is also found to be formally 
equivalent to the redefinition of coordinates, as is further discussed in detail 
in section \ref{redacteomh}. The property, that operators, which vanish 
`on-shell', can be omitted from the effective action, is very general, and it
is repeatedly used throughout the construction of the point-particle action. 

Subsequently, we consider operators that contain the Riemann tensor. The 
curvature tensor is thus decomposed into its electric and magnetic 
components of definite parity, which are defined as follows:
\bea
E_{\mu\nu}&\equiv& R_{\mu\alpha\nu\beta}u^{\alpha}u^{\beta}, \label{eq:E}\\
B_{\mu\nu}&\equiv& \frac{1}{2} \epsilon_{\alpha\beta\gamma\mu} 
R^{\alpha\beta}_{\,\,\,\,\,\,\,\delta\nu}u^{\gamma}u^{\delta}\label{eq:B},
\eea
that is of even and odd parity, respectively. These tensors are usually defined 
with the Weyl tensor, which due to the Ricci flatness is equivalent in our 
case to the Riemann tensor. The leading mass-induced higher-order operators are 
thus quadratic in the Riemann tensor, such that the point-particle action 
takes the following form: 
\be \label{sppm}
S_{\text{pp}}= -m\int d\sigma \, \sqrt{u^2}\,
+\, c_E \int d\sigma 
\frac{E_{\mu\nu}^{\,\,2}(y^{\alpha}(\sigma))}{[\sqrt{u^2}]^3} \,
+\, c_B \int d\sigma  
\frac{B_{\mu\nu}^{\,\,2}(y^{\alpha}(\sigma))}{[\sqrt{u^2}]^3}.
\ee
These new operators stand for the mass-induced quadrupolar tidal deformation 
of the extended object by the gravitational field. Each added derivative in 
an operator scales as $1/r$, and is preceded by a Wilson coefficient with the 
proper power of $r_s$, so that overall each term scales as powers of 
$r_s/r = v^2$. Thus, from dimensional analysis one easily finds that the 
leading Wilson coefficients $c_{E,B}$ in equation \eqref{sppm}, which are 
equivalent to the `Love numbers' originally defined in Newtonian gravity, 
scale as $r_s^5$ at LO, and that these finite size operators enter at the 5PN 
order. For NSs however, $r_s^5\gg (Gm)^5$, so the PN scaling breaks down, and the 
numerical value of the Wilson coefficients of tidal couplings can be large. 
This is also discussed in section \ref{dissipat}, where dissipative effects 
at the scale of the single object are considered.

More generally, an effective action for mass-induced tidal deformations has 
been constructed in \cite{Bini:2012gu}. Furthermore, the electric Love 
numbers of static BHs, including the $c_E$ coefficient in equation \eqref{sppm}, 
have been studied in the EFT framework in an arbitrary dimension in 
\cite{Kol:2011vg}. There, it was found that these Wilson coefficients, 
including $c_E$, vanish for BHs at $d=4$, where $d$ is the number of 
spacetime dimensions. Note that these coefficients are gauge invariant, 
though their matching is always done in some specific gauge, which implies 
that when they vanish, they vanish at all scales, and there is no RG 
running of this coupling. Interestingly, it was also found in 
\cite{Kol:2011vg}, that at higher dimensions, i.e.~at $d>4$, these coupling 
constants do not vanish, and that for a half-integral value of 
$\frac{l}{d-3}$, where $l$ is the multipole exponent, e.g.~for the case of 
$c_{E(d=7)}$, the couplings exhibit a (classical) RG flow, consistent with 
the divergences of the one-particle EFT. 

All in all, we conclude that for the non-spinning case of a massive particle the 
effective action is given by equation \eqref{sppm} up to the 5PN order.

\subsection{Spinning particle}
\label{spinparticle}

Let us now proceed to consider the actual, real case of a 
\emph{spinning} gravitating object, which is much more challenging to handle. 
Unlike the mass of the object, which is a feature readily compatible with the 
EFT perspective of a point particle, incorporating the spin of the object 
into the EFT framework is far from evident. This is essentially since a 
spinning object in relativity must be an extended one, i.e.~it has a 
non-vanishing finite size, with the lower bound for its size being the radius 
of the ring singularity of a Kerr BH, below which the rotational velocity of the 
spinning object would have exceeded the speed of light. This conflict with the 
point-particle perspective is actually related to the fundamental problem of how 
to define a sensible `center' of an object in relativistic physics, since the 
unique notion of the `center of mass' becomes ambiguous once we leave Newtonian 
physics. 

Thus, there is in fact in this case an important additional gauge freedom, 
that should be carefully taken into account: the choice of an internal point 
within the extended object, whose trajectory along the linear motion is to be 
followed, and which is to designate the evolving position of the spinning object. 
Worldline tetrads, as well as tetrad fields equivalent to the metric field, 
need to be invoked to describe the rotation of the object in curved 
spacetime. We define all multipoles of the single object beyond the mass 
monopole, namely its spin dipole and higher spin-induced multipoles, coupled 
to the gravitational field, through the rotating DOFs of the object. As was 
explicitly shown in \cite{Levi:2015msa}, the choice of the timelike vector 
of the worldline tetrad of the object is equivalent to the abovementioned 
choice of the internal point within the object, and in turn equivalent to the 
choice of the spin variable of the object. 

Therefore, from an EFT viewpoint, there are additional DOFs and related 
symmetries, that need to be properly identified and correctly considered in 
order to construct the effective action, beyond those mentioned in 
section \ref{pointmass} for the point-mass case. 
First, it is assumed that the isolated object has no intrinsic permanent 
multipole moments beyond the spin dipole.
Then, for the DOFs related with the rotation of the object, which are discussed 
in detail in what follows, the following symmetries are formulated 
\cite{Levi:2015msa}:
\begin{enumerate}
\item \emph{$SO(3)$ rotational invariance} of the worldline spatial triad;
\item \emph{Spin gauge invariance}, which is an invariance under the choice of
a completion to a tetrad of the worldline spatial triad through a timelike 
vector.
\end{enumerate} 

The action for a spinning particle in relativity was first considered 
decades ago in both flat and curved spacetimes, e.g.~in \cite{Hanson:1974qy} 
and \cite{Bailey:1975fe}, respectively, though not within an EFT framework. It 
was first tackled with an EFT approach in \cite{Porto:2005ac} in order to 
reproduce PN effects at LO, and to approach effects at NLO, linear in the spins 
of the objects \cite{Porto:2006bt}. To that end, a Routhian approach from 
\cite{Yee:1993ya}, devised for up to the quadratic order in the spin, was later 
adopted in \cite{Porto:2008tb,Porto:2008jj} in order to produce the EOMs of 
the objects at NLO. 

An independent application within the EFT approach was made in 
\cite{Levi:2008nh,Levi:2010zu} for spin effects up to NLO, where it was 
explicitly noted that the spin gauge can be fixed at the level of the 
one-particle EFT, and then one would be left only with the physical $SO(3)$ 
rotational DOFs. Following these works, and the first treatment of spin 
effects at NNLO within the EFT approach in \cite{Levi:2011eq}, further formal 
clarifications for a proper treatment of spin within an EFT framework were 
made in \cite{Levi:2014sba}, and an EFT for a spinning particle was presented 
in \cite{Levi:2015msa}. The latter enabled a proper rederivation of all the spin 
effects in the conservative sector up to NLO, the obtainment of further effects up 
to the quartic order in the spin \cite{Levi:2014gsa}, and the completion of the 
effects up to the quadratic order in the spins at NNLO 
\cite{Levi:2015uxa,Levi:2015ixa,Levi:2016ofk}. 
These constitute the current state of the art in the conservative sector with 
spins, namely at the level of the EFT of the composite particle, where the 
additional ingredients that make up the advances at this subsequent level are 
the topic of section \ref{compeft}.

It should be noted that yet another independent formulation to handle 
arbitrary gravitating spinning objects, based on a coset construction for 
spacetime symmetries, was put forward in \cite{Delacretaz:2014oxa}. The 
effective action developed in the latter contains only physical rotational 
DOFs. However, this action is an expansion in $v_{\text{rot}}/c_s$, where 
$v_{\text{rot}}$ is the rotational velocity, and $c_s$, the speed of sound, 
which scales as $c_s\sim c$ for relativistic matter. For this reason this 
formulation is ideal for slowly rotating objects, i.e.~whenever the angular 
frequency is smaller than the characteristic frequencies of the object, such 
as for NSs. This formulation was applied to tackle dissipative effects, 
which are discussed in section \ref{dissipat}, for spinning objects in an EFT 
approach in \cite{Endlich:2015mke,Endlich:2016jgc}, though these effects were 
not specified for the PN context.  

Let us then proceed to outline the construction of the effective action of a 
spinning particle starting from the minimal coupling part, as already 
formulated in \cite{Hanson:1974qy,Bailey:1975fe}, and pressing on to the 
formal EFT developments provided in \cite{Levi:2015msa,Levi:2017oqx} for both 
the minimal and non-minimal coupling parts of the effective action.

\subsubsection{Minimal coupling}
\label{spinmc}

We start by adding to the worldline DOFs of the single object the tetrad, 
$e_{A}^{\mu}(\sigma)$, an orthonormal set consisting of a timelike 
future-oriented vector and three spacelike vectors, which satisfies 
$\eta^{AB}e_A{}^\mu(\sigma)e_B{}^\nu(\sigma)=g^{\mu\nu}$. From the 
coordinate of the object, $y^{\mu}(\sigma)$, we already have the 
four-velocity, $u^{\mu}\equiv \frac{dy^{\mu}}{d\sigma}$, and from the tetrad 
we further define the angular velocity tensor, 
$\Omega^{\mu\nu}\equiv e^{\mu}_A\frac{De^{A\nu}}{D\sigma}$, which generalizes 
the flat spacetime definition, 
$\Omega^{ab}\equiv \Lambda^{a}_A\frac{d\Lambda^{Ab}}{d\sigma}$, with Lorentz 
transformations, $\Lambda^{a}_A$.

Since the action is reparametrization invariant, the Lagrangian is required 
to be a homogeneous function of degree 1 in the velocities $u^{\mu}$ and 
$\Omega^{\mu\nu}$. Therefore if we define the linear momentum, $p_{\mu}$, and 
the antisymmetric spin tensor, $S_{\mu\nu}$, as follows:
\bea
p_{\mu} &\equiv& -\frac{\partial L_{\text{pp}}}{\partial u^{\mu}},\\
S_{\mu\nu} &\equiv& 
-2 \frac{\partial L_{\text{pp}}}{\partial \Omega^{\mu\nu}},
\eea
where the sign is fixed according to the NR limit, then from Euler's theorem 
we get the following form for the minimal coupling part of the action 
\cite{Hanson:1974qy,Bailey:1975fe}:
\be
L_{\text{pp}}=-p_{\mu} u^{\mu}-\frac{1}{2}S_{\mu\nu}\Omega^{\mu\nu},
\ee
where the linear momentum satisfies 
$p^\mu=m \frac{u^\mu}{\sqrt{u^2}} + {\cal{O}}(S^2)$, independent of the gauge 
chosen for the rotational DOFs, $e_{A}^{\mu}(\sigma)$. In fact, a unique 
covariant gauge, which eliminates the unphysical DOFs from the rotational 
DOFs, was provided by Tulczyjew in \cite{Tulczyjew:1959b}. This gauge reads 
as follows:
\be
e_0^{\mu}=\frac{p^{\mu}}{\sqrt{p^2}},
\ee
which corresponds to what is referred to as a `spin supplementary 
condition' (SSC), given by 
\be
S_{\mu\nu}p^{\nu}=0,
\ee
so we see that this unique gauge fixes the spin to be a spatial $SO(3)$ 
tensor in the body-fixed tetrad frame. 

Therefore, we highlight that the spin is considered as a further dynamical 
worldline DOF, which serves as a classical source from a QFT perspective.
Hence, the starting point for the whole action of a spinning particle reads 
as follows \cite{Levi:2015msa}:
\be \label{totspinaction}
S_{\text{pp}}
\left[\bar{g}_{\mu\nu}(y^{\mu}),u^{\mu},e_{A}^{\mu},S_{\mu\nu}\right](\sigma)=
\int d\sigma \left[-m\sqrt{u^2}-\frac{1}{2} S_{\mu\nu}\Omega^{\mu\nu} 
+L_{\text{NMC}}
\left[\bar{g}_{\mu\nu}(y^{\mu}),u^{\mu},S_{\mu\nu}\right]\right],
\ee
where the label `NMC' here refers to the non-minimal coupling part of the 
action induced by the presence of spin, and the covariant gauge is implied.

First, we would like to restore to the action the gauge freedom of the 
rotational variables \cite{Levi:2015msa}, in the spirit of the Stueckelberg 
action; see, e.g.~\cite{Ruegg:2003ps}. Rather than resorting to changing the 
gauge of the spin variable at a later stage, we apply a generic 
transformation to the worldline tetrad in order to make the action manifestly 
gauge invariant to the choice of gauge of rotational variables. We do this by 
applying an effectively covariant boost to the worldline tetrad, and then 
considering how the rotational minimal coupling term, 
$\frac{1}{2} S_{\mu\nu}\Omega^{\mu\nu}$, and the non-minimal coupling part of 
the action, are affected. 

Using a boost in its four-dimensional covariant form, $L^\mu_\nu(w,q)$, we 
transform the tetrad, $e_A^{\mu}$, from some gauge $e_{0}^{\mu}=q^\mu$ to a 
generic gauge for the tetrad as follows:
\be \label{tetgengauge}
\hat{e}_{0}^{\mu}=w^\mu, 
\ee 
with the transformation $\hat{e}_{A}^{\mu}=L^\mu_\nu(w,q)e_{A}^{\nu}$. Then, 
starting from the covariant gauge, i.e.~taking $q^\mu=\tfrac{p^\mu}{\sqrt{p^2}}$, the 
generic gauge for the tetrad in equation \eqref{tetgengauge} satisfies the following 
generic SSC:
\be \label{genssc}
\hat{S}^{\mu\nu}(p_\nu+\sqrt{p^2}\hat{e}_{0\nu})=0.
\ee 
The generic spin variable, $\hat{S}^{\mu\nu}$, is then related to 
$S^{\mu\nu}$ by
\be
\hat{S}^{\mu\nu}=S^{\mu\nu}-\delta z^\mu p^\nu + \delta z^\nu p^\mu, 
\ee
where $\delta z^\mu\equiv\hat{z}^\mu-y^\mu$, so that we see that indeed this 
gauge freedom also corresponds to a choice of an object's `center', 
$\hat{z}$. Assuming then, that we started from the covariant gauge, we get 
the following for the minimal coupling term:
\be \label{spinmctrans}
\frac{1}{2} S_{\mu\nu}\Omega^{\mu\nu}
=\frac{1}{2} \hat{S}_{\mu\nu}\hat{\Omega}^{\mu\nu}+
\frac{\hat{S}_{\mu\nu} p^\nu}{p^2}\frac{Dp^\mu}{D\sigma}.
\ee
The new extra term on the RHS of equation \eqref{spinmctrans} contributes to 
finite size effects with spin, yet it carries no Wilson coefficient. This is 
since this term does not encapsulate any UV physics of the structure of the 
object, rather it just accounts for the fact that a spinning object is of a 
finite size. An elaborate discussion of this extra term is found in 
\cite{Levi:2015msa}.

We note that for a maximally rotating object, i.e.~which rotates at the speed 
of light, the spin of the compact object scales as $S\sim mr_s$. With the 
spin being derivatively coupled, the leading PN spin contribution from 
equation \eqref{spinmctrans} then enters at order $r_s/r$ relative to the 
Newtonian term in equation \eqref{eq:mmc}, i.e.~at the 1PN order for a rapidly 
rotating compact object. We will see in section \ref{feynman} that this term 
actually starts contributing at the leading spin-orbit interaction of the 
composite particle, which enters only at the 1.5PN order. 

As for the effect of restoring gauge invariance to the action in 
equation \eqref{totspinaction} in its non-minimal coupling part (that we analyze 
in detail below), which depends only on the spin, $S_{\mu\nu}$, we only need to use 
the following transformation to the generic spin variable:
\be
S_{\mu\nu} = \hat{S}_{\mu\nu}-\frac{\hat{S}_{\mu\rho}p^{\rho} p_{\nu}}{p^2} 
+ \frac{\hat{S}_{\nu\rho}p^{\rho} p_{\mu}}{p^2}.
\ee

\subsubsection{Non-minimal coupling}
\label{spinnmc}

Let us then proceed to consider how to construct the non-minimal coupling 
part of the action of a spinning particle \cite{Levi:2015msa}.
We begin by noting that parity invariance plays a key role in fixing the 
non-minimal couplings. Hence, in order to construct the spin-induced ones in 
terms of ingredients of definite parity, let us first define the spin vector, 
$S^\mu$, using the dual to the spin tensor, 
$*S_{\alpha\beta} \equiv \frac{1}{2}\epsilon_{\alpha\beta\mu\nu} S^{\mu\nu}$:
\be
S^\mu \equiv *S^{\mu}_{\,\,\nu}\frac{p^\nu}{\sqrt{p^2}}
\simeq *S^{\mu}_{\,\,\nu}\frac{u^\nu}{\sqrt{u^2}}.
\ee
From this spacelike vector the spin length, $S^2$, is defined as 
$S^2\equiv -S_\mu S^\mu=\frac{1}{2}S_{\mu\nu} S^{\mu\nu}$. 

From the Cayley-Hamilton theorem higher powers of the spin tensor are
expected to be dependent. Indeed, considering higher powers of the spin 
tensor in the sense of matrix multiplication, one finds that the minimal 
polynomial of the spin tensor is of order $3$: $x(x+iS)(x-iS)=0$ (as both the 
spin vector, $S^\mu$, and the four-velocity, $u^\mu$, are eigenvectors of the 
spin tensor, corresponding to the degenerate eigenvalue $0$). Since we also 
find that a contraction of two spin tensors is equivalent to the direct 
product of two spin vectors $S^{\alpha}S^{\beta}$ (when contracted with 
traceless tensors, which are also orthogonal to $u^{\mu}$), we conclude that 
in order to construct the higher order spin-induced operators, we should use 
direct products of the spin vector, $S^\alpha$, which will have alternating 
parity. The scalar of spin length, $S^2$, is absorbed in the mass 
parameter, and renormalizes the Wilson coefficients. Considered in the 
body-fixed frame, the spin-induced multipoles are $SO(3)$ irrep tensors, 
since we recall that we start from the covariant gauge, 
$e_0^\mu\simeq\frac{u^\mu}{\sqrt{u^2}}$, so that $e_i^\mu u_\mu=0$. It is 
thus inferred that the spin-induced multipoles are symmetric, traceless, and 
spatial constant tensors in the body-fixed frame. 

The even and odd spin-induced higher multipoles then are coupled to the even 
and odd parity electric and magnetic curvature components, $E_{\mu\nu}$ and 
$B_{\mu\nu}$, from equations \eqref{eq:E} and \eqref{eq:B}, respectively, and their 
covariant derivatives. Here, we consider only operators linear in the Riemann 
tensor, namely not taking into account dissipative tidal effects (these 
are discussed in section \ref{dissipat}, and are found in general to 
contribute at higher PN orders). From the symmetries of the Riemann tensor, 
the first Bianchi identity, and the leading vacuum field solution, it is easy 
to find, that $E_{\mu\nu}$ and $B_{\mu\nu}$ are both symmetric, traceless, 
and orthogonal to $u^\mu$. Hence, like the spin-induced higher multipoles, 
these are $SO(3)$ tensors, that we consider in the body-fixed frame, where 
they are spatial. 

The covariant derivatives are also projected to the body-fixed frame, 
$D_i=e_i^\mu D_\mu$, and the time derivatives, 
$D_0\simeq u^\mu D_\mu \equiv D/D\sigma$, can be ignored to linear order in 
the Riemann tensor \cite{Levi:2015msa}. The indices of the covariant 
derivatives are also symmetric among themselves, and with respect to the 
indices of the electric and magnetic curvature tensors. The latter symmetry 
can be deduced from the differential Bianchi identity, where one obtains the 
following, in analogy to Maxwell's equations: 
\bea
\epsilon_{ikl}D_k E_{lj}&=&\dot{B}_{ij},\\
\epsilon_{ikl}D_k B_{lj}&=&-\dot{E}_{ij},
\eea
and as we noted above regarding the time derivatives, the RHS can then be 
taken as vanishing. From further contracting the last equations we also find 
that traces involving the covariant derivatives vanish, i.e.
\be 
D_i E_{ij} =D_i B_{ij}=0,
\ee
and similarly 
\be
\Box E_{ij}=\Box B_{ij}=0.
\ee
Therefore, all in all, the electric and magnetic curvature tensors and their 
covariant derivatives form $SO(3)$ tensors like the spin-induced higher 
multipoles to which they couple. 

Based on the above analysis we can write down the LO spin-induced non-minimal 
couplings to all orders in spin as follows:
\bea \label{eq:spinnmc}
L_{\text{NMC}}&
=&\sum_{n=1}^\infty \frac{(-1)^n}{(2n)!}\frac{C_{ES^{2n}}}{m^{2n-1}}
D_{\mu_{2n}}\cdots D_{\mu_{3}}\frac{E_{\mu_{1}\mu_{2}}}{\sqrt{u^2}}
S^{\mu_1}S^{\mu_2}\cdots S^{\mu_{2n-1}}S^{\mu_{2n}}\nn\\
&+&\sum_{n=1}^\infty \frac{(-1)^n}{(2n+1)!}\frac{C_{BS^{2n+1}}}{m^{2n}}
D_{\mu_{2n+1}}\cdots D_{\mu_{3}}\frac{B_{\mu_{1}\mu_{2}}}{\sqrt{u^2}}
S^{\mu_1}S^{\mu_2}\cdots S^{\mu_{2n}}S^{\mu_{2n+1}},
\eea
with new spin-induced Wilson coefficients, which are identified as unity for 
Kerr BHs, but should in fact be properly matched using the full UV theory. 

In particular, let us extract from the general sum in equation \eqref{eq:spinnmc} 
the three leading terms, which contribute to finite size effects, due to the 
spin-induced quadrupole, octupole, and hexadecapole, given by the following, 
respectively
\bea
L_{ES^2}&=&-\frac{C_{ES^2}}{2m}\frac{E_{\mu\nu}}{\sqrt{u^2}} S^\mu S^\nu, 
\label{es2}\\
L_{BS^3}&=&-\frac{C_{BS^3}}{6m^2}D_{\lambda}\frac{B_{\mu\nu}}{\sqrt{u^2}} 
S^\mu S^\nu S^\lambda, \label{bs3}\\
L_{ES^4}&=&
\frac{C_{ES^4}}{24m^3}D_{\lambda}D_{\kappa}\frac{E_{\mu\nu}}{\sqrt{u^2}} 
S^\mu S^\nu S^\lambda S^\kappa. \label{es4}
\eea
Again, it is easy to see that for each spin-induced higher multipole a spin 
vector (divided by a mass factor) is added along with an additional 
derivative. Hence, for a maximally rotating compact object the additional 
relative factor scales as $r_s/r$, and thus the quadrupole, octupole, and 
hexadecapole, enter at the 2PN, 3PN, and 4PN orders, respectively. However, 
similar to what we noted above for the odd parity spin dipole, and as we will 
see in section \ref{feynman}, the odd parity spin-induced octupole starts 
contributing at the leading spin-induced octupole-orbit interaction of the 
composite particle, which actually enters only at the 3.5PN order. 

To conclude, we see that spin-induced finite size effects significantly 
dominate over finite size effects, which are induced by the mass.

\subsection{Dissipative DOFs}
\label{dissipat}

In order to capture dissipative effects, which take place at the scales of 
the single object, such as the absorption of gravitational energy by the 
horizons of BHs, or the dissipative tidal deformations of NSs, further 
worldline DOFs should be added to the EFT beyond the coordinate and 
rotational DOFs \cite{Goldberger:2005cd}. In principle the complete theory 
would also contain a part that specifies the theory of the microscopic DOFs 
associated with the dissipation. However, even if the underlying detailed nature 
of these DOFs is unknown, it is possible to use the symmetries of spacetime 
and of the object to construct the related infinite set of operators of 
non-minimal coupling in the point-particle action.

For BHs these DOFs would be localized on the horizon to account for horizon 
absorption effects, which is related to fundamental questions on the nature 
of the BH horizon and the microphysics of BH entropy. To begin with, using 
the $SO(3)$ symmetry of a static spacetime, the couplings of the horizon 
modes to gravity read as follows \cite{Goldberger:2005cd}:
\be \label{sppdiss}
S_{\text{pp(diss)}} = -\int d\tau \, Q_{ab}^E E^{ab} 
-\int d\tau \, Q_{ab}^B B^{ab} + \cdots,
\ee
where $E^{ab}$, $B^{ab}$, are the electric and magnetic curvature components 
from equations \eqref{eq:E} and \eqref{eq:B}, respectively, projected onto the 
locally flat frames, and $Q_{ab}^E$, $Q_{ab}^B$, are quadrupolar composite 
operators formed in some manner from the horizon DOFs. We note that it was 
suggested in \cite{Kol:2008hc} to decompose the metric field at this stage 
into three (rather than two) components of Fourier modes, 
$g_{\mu\nu}=g^s_{\mu\nu}+g^{\text{hor}}_{\mu\nu}+\bar{g}_{\mu\nu}$, 
where the additional component, $g^{\text{hor}}_{\mu\nu}$, represents the low 
frequency modes of the field at the near horizon zone and is identified as 
the dissipative DOFs on the horizon. 

According to the classical optical theorem (see, e.g.~\cite{Jackson:1998nia}, 
or \cite{Peskin:1995ev}) for the quantum counterpart, it turns out that the 
power loss of the composite binary system is related to the imaginary part of 
its effective action, as follows \cite{Goldberger:2004jt}:
\be \label{optthe} 
2\,\text{Im}[S_{\text{eff(comp)}}] = 
\int dt \, \int \frac{d\omega}{E(\omega)} \,\frac{dP}{d\omega},
\ee
where $S_{\text{eff(comp)}}$ is the effective action of the composite particle
from equations \eqref{seff2part} and \eqref{seffcomp}, on which we elaborate in 
sections \ref{compeft} and \ref{togwobs}, and $dP/d\omega$ is the power spectrum 
observable. The power spectrum, $dP_{\text{abs}}/d\omega$, which gets 
absorbed by the single objects through their dissipative DOFs in 
equation \eqref{sppdiss}, can be directly related to the absorption cross section, 
$\sigma_{\text{abs}}(\omega)$. 

Let us see then how this relation transpires. Since $Q_{ab}^{E,B}$ are 
dissipative DOFs, they dynamically propagate along the worldline. Due 
to rotational invariance $\langle Q^{E,B}_{ab}\rangle$ vanishes; then the LO 
contribution of the worldline operators in equation \eqref{sppdiss} to the 
absorption cross section and to equation \eqref{optthe} involves the 
correlators of these operators. One should then consider the 
two-point function along the worldline, which can be written as follows:
\be \label{horizoncorr}
\int d\tau e^{-i\omega\tau} \langle T \,Q_{ab}^{E,B}(\tau) 
Q_{cd}^{E,B}(0)\rangle
= -\frac{i}{2}\left[\delta_{ac}\delta_{bd}+\delta_{ad}\delta_{bc}
-\frac{2}{3}\delta_{ab}\delta_{cd}\right]F(\omega).
\ee
Here, $F(\omega)$ is identified as the response function that encodes the 
propagator of the dissipative DOFs (namely it encodes aspects of strong 
gravity), and it arises in the graviton absorption cross section as follows 
\cite{Goldberger:2005cd}:
\be
\sigma_{\text{abs}}(\omega)=16\pi G\,\omega^3\,\text{Im}[F(\omega)],
\ee
which is also known from purely classical gravitational physics, such that 
the response function is found to be given by
\be 
\text{Im}[F(\omega)]=\frac{16}{45}G^5m^6\left|\omega\right|.
\ee 

Hence, the propagator in equation \eqref{horizoncorr} also enters the leading 
dissipative interaction potential of the binary composite, and then from 
equation \eqref{optthe} the power loss due to the absorption on the horizon is 
found. From power counting the response function, recalling that $\omega\sim 
v/r$, and the operators in equation \eqref{sppdiss}, it is easily found that this 
leading dissipative interaction potential enters only at the 6.5PN order. 
Having found the power loss, we can now infer a generic relation between the 
power spectrum and the cross section of the absorption, where the cross 
section can be explicitly computed given a model for the internal structure 
of the single compact objects. 

The generalization of \cite{Goldberger:2005cd} for absorption by rotating BHs 
was approached in \cite{Porto:2007qi}. For rotating BHs, superradiance (the 
radiation enhancement due to interaction with a rotating object) yields 
the leading PN dissipative effect, which was reproduced in 
\cite{Porto:2007qi}, and is enhanced by $v^{-3}$ with respect to the 
non-rotating case, namely entering at the 5PN order. Dissipative effects for 
arbitrary gravitating rotating objects were treated more generally in 
\cite{Endlich:2015mke} and \cite{Endlich:2016jgc}, following the effective 
action developed in \cite{Delacretaz:2014oxa}, while also adding dissipative 
DOFs. In this approach, which followed \cite{Delacretaz:2014oxa}, the tidal 
distortions and superradiance of rotating objects are both governed by the 
same dissipative couplings, and furthermore the matching procedure in the 
non-rotating case also holds with spin.

To treat dynamical tidal deformations of NSs, in particular resonances in 
non-rotating extended objects, a similar approach to \cite{Goldberger:2005cd} 
was invoked in \cite{Chakrabarti:2013lua,Chakrabarti:2013xza,
Steinhoff:2016rfi}, which captures effects such as tidal heating, tidal 
disruption, tidal locking, and resonances in bound binaries. The worldline 
action is supplemented with gravitational multipoles as dissipative DOFs, for 
which the dynamics is encoded in the propagator, i.e~some response function. 
However, for generic stars the tower of effective theories, presented in section 
\ref{towerefts}, actually starts from also having a fluid description of 
matter, that is, a theory for the matter should be added to 
equation \eqref{sgfull}, e.g.~an ideal fluid action. For idealized Newtonian stars 
\cite{Chakrabarti:2013xza}, and approximately also for stars in GR 
\cite{Chakrabarti:2013lua}, the scale of the object can then be explicitly 
integrated out. This leads to a point-mass action augmented by harmonic 
oscillator DOFs, corresponding to oscillation modes of the star, which gives 
rise to resonances. Other dynamical tidal deformations are described by terms 
in the worldline action such as $(\dot{E}_{ab})^2\equiv (u^{\mu} 
D_{\mu}E_{ab})^2$.

\section{EFT of a composite particle} 
\label{compeft}

In this section we discuss in detail the obtainment of the EFT of the  
composite particle, as defined in equation \eqref{seff2part}, namely via the 
top-down procedure, using standard perturbative QFT methods, involving a 
diagrammatic expansion and Feynman calculus. This EFT removes the scale of 
orbital separation between the components of the binary, $r$. It is then 
matched onto the effective action constructed bottom-up, as presented in 
equation \eqref{seffcomp}, with the point-particle action now being that of the 
composite particle, in terms of multipole moments. More precisely, in this 
section we focus on the conservative sector, where one is concerned with the 
computation of two-particle interactions, which make up the mass/energy 
monopole of the composite particle. We also discuss how to work out standard 
results and observables from these two-body interactions. The matching 
from the full two-particle theory at the orbital scale onto the effective 
theory of the composite  particle with multipole moments, involving the 
radiative sector, is discussed in the beginning of section \ref{radrg}.

We open in section \ref{gauges} by describing all the choices, and gauges, 
that are made in order to facilitate computation. Then, in section 
\ref{feynman} we detail and review the Feynman rules (which are extracted 
from the two-particle theory in equation \eqref{twoparteft}), the Feynman diagrams 
corresponding to various interaction sectors, and the main issues in the 
Feynman calculus required for their evaluation. We review the buildup of 
these computations tackled in the works \cite{Goldberger:2004jt, Kol:2007bc,
Cardoso:2008gn, Gilmore:2008gq,Chu:2008xm,Foffa:2011ub,Foffa:2012rn,
Foffa:2016rgu,Damour:2017ced} for the two-particle point-mass sector, which 
involves only the non-minimal coupling from section \ref{pointmass}, from the 
1PN order to the recently attained state of the art at the 4PN order. We note 
here related works, which have discussed 
further aspects of tackling the computations \cite{Kol:2009mj,Kol:2013ega}, 
or their interpretation \cite{Porto:2017dgs}. Moreover, we note the progress 
made in the computations in the two-particle sector with spins from section 
\ref{spinparticle}, which involve both minimal and non-minimal couplings, in 
the works \cite{Porto:2005ac,Porto:2006bt, Porto:2008tb,Levi:2008nh, 
Porto:2008jj,Perrodin:2010dy,Porto:2010tr,Levi:2010zu,Levi:2011eq,
Levi:2014sba,Levi:2014gsa,Levi:2015msa,Levi:2015uxa,Levi:2015ixa,Levi:2016ofk,
Levi:2017oqx}, where the state of the art in PN theory was actually pushed forward. 

Finally, in section \ref{redacteomh} we discuss how to process the outcome of 
integrating out the orbital modes for the EFT of the composite particle in 
the conservative sector. In order to bring the effective action into a 
standard form, higher-order time derivatives should be handled. EOMs should 
be derived for the positions and spins of the constituents of the binary. 
Moreover, the Hamiltonian, which is the crucial input for the EOB framework, 
as well as other conserved integrals of motion, should be derived. The 
developments in the EFT formulation, computation, and their outcome, as well 
as further gauge invariant relations, e.g.~among the binding energy of the 
binary and its total angular momentum, have been all put into a public code, 
dubbed `EFTofPNG', in \cite{Levi:2017kzq}. We elaborate on this public PN 
code later in section \ref{eftofpngcode}.

It should be stressed in particular for the sector including spins, that the 
resulting effective action should contain \emph{no remaining potential field 
DOFs} \cite{Levi:2008nh,Levi:2010zu,Levi:2014sba,Levi:2015msa}. The 
observation that this key aspect of an EFT should be upheld was crucial in 
order to formally and computationally advance the state of the art in this 
sector, and it pertains to having a clear identification and separation 
between field and particle DOFs. Hence, to prepare for this EFT at the next stage, 
where we get the two-particle interactions by explicitly integrating over the 
orbital field modes, we need to disentangle the field DOFs from the particle 
DOFs, and to fix all relevant rotational gauges. 

First, we have for the worldline tetrad the following factorization:
\be 
\hat{e}^\mu_A=\hat{\Lambda}^{b}_{A}\tilde{e}^{\mu}_b,
\ee
so that the worldline tetrad is decomposed into rotational particle DOFs and 
field DOFs, namely the Lorentz matrices in the locally flat frames, defined by
$\eta^{AB}\hat{\Lambda}^{a}_{A}\hat{\Lambda}^{b}_{B}=\eta^{ab}$, and the 
tetrad field, which covers the manifold, defined by 
$\eta_{ab}\tilde{e}^{a}_{\mu}\tilde{e}^{b}_{\nu}=g_{\mu\nu}$, respectively. 
For the latter an additional internal Lorentz invariance should be noted. 
Then, we rewrite the minimal coupling rotational term in terms of the new 
DOFs as follows \cite{Levi:2010zu}:
\be \label{scoupterm}
\frac{1}{2}\hat{S}_{\mu\nu}\hat{\Omega}^{\mu\nu}=
\frac{1}{2}\hat{S}_{ab}\hat{\Omega}^{ab}_{\text{LF}}+
\frac{1}{2}\hat{S}_{ab}\omega_\mu^{ab}u^{\mu},
\ee
where $\hat{\Omega}^{ab}_{\text{LF}}\equiv
\hat{\Lambda}^{aA}\frac{d\hat{\Lambda}_A^b}{d\sigma}$ 
is the locally flat angular velocity, and 
$\omega_{\mu}^{ab}\equiv \tilde{e}^{b}_{\nu} D_\mu \tilde{e}^{a\nu}$ are the 
Ricci rotation coefficients, so that the eventual particle rotational 
variables are the Lorentz matrices, $\hat{\Lambda}_{A}^{a}$, and the spin, 
$\hat{S}_{ab}$, in the locally flat frame. 

Our generic gauge from equations \eqref{tetgengauge} and \eqref{genssc}, for the 
rotational variables then reads as follows \cite{Levi:2015msa}:
\begin{align}
\hat{\Lambda}_{0}^{a} & = w^a,\\
\hat{S}^{ab}\left(p_b+\sqrt{p^2}\hat{\Lambda}_{0b}\right) & = 0.
\end{align}
Therefore, the separation of field and particle DOFs is not yet complete, 
since $w^a = \tilde{e}^{a}_{\mu}w^{\mu}$, and the temporal spin components, 
$\hat{S}^{0i}$, contain further field dependence. It is, in fact, not 
surprising from the EFT viewpoint that specifying the point-particle 
action necessitates fixing the gauge of rotational variables at the level of 
the one-particle action. This is due to the fact that only once this choice is 
made is the view of the extended object as a point particle actualized. 
Hence, only once the gauge of the rotational variables is fixed, can the 
field be completely disentangled from the particle DOFs, and all orbital 
field modes can be explicitly integrated out, as required at this stage 
\cite{Levi:2008nh,Levi:2014sba,Levi:2015msa}.

\subsection{KK metric and gauge fixing}
\label{gauges}

First, let us recall from equation \eqref{eq:kmodes} that for the orbital field 
modes we have $k_0\simeq v/r$, whereas $|\vec{k}|\simeq 1/r$, so that for the 
propagator in momentum space, given by
\be
\int \frac{d^4k}{\left(2\pi\right)^4} \,e^{-ikx}\frac{1}{k^2}=
\int \frac{d^4k}{\left(2\pi\right)^4}e^{-ik_0t
+i\vec{k}\cdot\vec{x}}\frac{1}{k_0^2-{\vec{k}}^2},
\ee
the denominator can be expanded in the PN approximation as follows 
\cite{Goldberger:2007hy}:
\be \label{pnpotprops}
\frac{1}{k_0^2-{\vec{k}}^2}=
-\frac{1}{{\vec{k}}^2}\left(1+\frac{k_0^2}{{\vec{k}}^2}+\cdots\right)=
-\frac{1}{{\vec{k}}^2}\left(1+{\cal{O}}(v^2)\right),
\ee
and we get the following for the propagator of the orbital component of the 
field: 
\be \label{potgravprops}
\int \frac{dk_0}{2\pi} \, e^{-ik_0t} 
\int \frac{d^3\vec{k}}{\left(2\pi\right)^3}
\frac{e^{i\vec{k}\cdot\vec{x}}}{{\vec{k}}^2}=
\delta(t)\int \frac{d^3\vec{k}}{\left(2\pi\right)^3}
\frac{e^{i\vec{k}\cdot\vec{x}}}{{\vec{k}}^2},
\ee
so that the propagator is instantaneous, and the relativistic time 
corrections to the propagator are considered as quadratic perturbations 
\cite{Goldberger:2007hy}.

Thus, in the context of the NR regime of gravity it is useful to switch to a 
unique space+time parametrization of the metric, since in the NR limit the 
time dimension can be regarded as compact in comparison to the spatial 
dimensions \cite{Kol:2007bc,Kol:2010ze}. Therefore it makes sense to reduce 
over the time coordinate in the metric in a Kaluza-Klein (KK) fashion, so 
that the metric is rewritten as follows:
\begin{align}\label{kk}
ds^2=g_{\mu\nu}dx^{\mu}dx^{\nu}\equiv 
e^{2\phi}\left(dt-A_idx^i\right)^2-e^{-2\phi}\gamma_{ij}dx^idx^j,
\end{align}
which defines the KK fields: $\phi$, $A_i$, and 
$\gamma_{ij}\equiv\delta_{ij}+\sigma_{ij}$, identified as the Newtonian 
scalar, the gravito-magnetic vector, and the tensor fields, respectively, 
where $\gamma^{ij}\gamma_{jk}\equiv \delta^i_k$, and 
$A^i\equiv \gamma^{ij}A_j$\footnote{We note that an exponential 
parametrization of the metric coefficients was also introduced in 
\cite{Damour:1990pi}.}. 

Let us then proceed to fix the gauge of the various DOFs. For the purely 
gravitational action, $S_g$, the (fully) harmonic gauge is chosen, so that 
the action reads as follows:
\bea \label{sgravpure}
S_g &=& S_{EH}+S_{GF} \nn \\
&=& -\frac{1}{16\pi G} \int d^4x \sqrt{g} \, R 
+ \frac{1}{32\pi G} \int d^4x \sqrt{g} \, g_{\mu\nu}\Gamma^\mu\Gamma^\nu,
\eea
where $\Gamma^{\mu}\equiv \Gamma^{\mu}_{\rho\sigma}g^{\rho\sigma}$. Working 
within the background field method, on which we elaborate in section \ref{radrg}, 
covariant derivatives are taken with respect to the background field 
\cite{Goldberger:2004jt,Peskin:1995ev}. However, in the conservative sector that 
we are now considering, no background radiation modes are present, and so the 
covariant derivative is just the standard one, i.e.~taken with respect to the 
asymptotic flat spacetime. Using Cartan's method of two-forms, the full 
gravitational action can be computed in terms of the KK parametrization 
\cite{Kol:2010si}, and the propagators and self-interaction vertices, which 
we present in section \ref{feynman}, are readily obtained from the action in 
this parametrization. 

We still have to fix the internal Lorentz gauge of the tetrad field, which 
couples to the worldline multipoles. Again, in the NR regime, where the time 
direction is singled out, it is convenient to choose Schwinger's time gauge 
\cite{Schwinger:1963re}, which reads as follows:
\be
\tilde{e}^0_i(x)=0,
\ee
so that in terms of the KK fields from equation \eqref{kk}, the tetrad field reads 
as follows \cite{Levi:2015msa}: 
\be
\tilde{e}^a{}_{\mu} = \left( \begin{array}{cc}
e^{\phi} & - e^{\phi} A_i \\
0 & e^{-\phi} \sqrt{\gamma}_{ij}
\end{array} \right),
\ee
where $\sqrt{\gamma_{ij}}$ is the symmetric square root of $\gamma_{ij}$.

For the worldline parameter we choose the time coordinate, $t=y^0$, such that 
$\sigma=t$, $u^0=1$, and $u^i\equiv\frac{dy^i}{dt}=v^i$. Hence in terms of 
the KK metric in equation \eqref{kk} the mass coupling in equation \eqref{eq:mmc} is 
rewritten as follows:
\be \label{mcoup}
-m \int dt \, \sqrt{g_{\mu\nu}\frac{dx^{\mu}}{dt}\frac{dx^{\nu}}{dt}}
=-m \int dt \left[ 
e^{\phi}\sqrt{\left(1-A_iv^i\right)^2-e^{-4\phi}\gamma_{ij}v^iv^j}\right],
\ee
which is then expanded in the velocity, $v$, and the KK fields.

Finally, for the rotational variables we recall that one starts by assuming 
the covariant gauge, namely $\hat{\Lambda}_{0a}=\tfrac{p_a}{\sqrt{p^2}}$ for 
the Lorentz matrix; however, it is found to be beneficial to switch to the 
`canonical gauge' \cite{Levi:2015msa}. The latter is the general 
relativistic generalization of what is known as the Pryce-Newton-Wigner SSC 
in special relativity \cite{Pryce:1948pf,Newton:1949cq}, and it reads as follows:
\begin{align}
\hat{\Lambda}_{0}^{a} &= \delta_0^a, \\
\implies \hat{S}^{ab}\left(p_b+\sqrt{p^2}\delta_{0b}\right) & = 0. 
\end{align}
This gauge enables one to completely integrate out the orbital field modes as 
required at this stage, and leads to physical spin variables, $\hat{S}^{ij}$, 
which eventually satisfy canonical Poisson brackets, as we further discuss in 
section \ref{redacteomh}.

\subsection{Feynman graphs and calculus}
\label{feynman}

After we applied a beneficial parametrization for the metric and fixed all 
gauges in the previous section, we can proceed to extract the Feynman rules 
from the gauge-fixed two-particle effective action. These rules constitute 
the building blocks of the Feynman graphs, which make up the perturbative 
expansion of the exponent in the functional integral in 
equation \eqref{seff2part}. The Feynman rules can also be generated automatically 
in a comprehensive manner using the `EFTofPNG' public code, on which we 
elaborate in section \ref{eftofpngcode}. All Feynman rules are given in 
position space in what follows.

Following from equation \eqref{potgravprops} we start by extracting the 
propagators from the quadratic time-independent part of the gravitational 
action in terms of the KK metric. The propagators of the KK fields then 
read as follows:
\begin{align}
\label{eq:prphi} \langle{~\phi(x_1)}~~{\phi(x_2)~}\rangle
& = \parbox{18mm}{\includegraphics[scale=0.5]{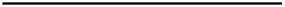}} = 
~~~~4\pi G  \times\,\delta(t_1-t_2) \int_{\vec{k}} 
\frac{e^{i \vec{k}\cdot\left(\vec{x}_1 - \vec{x}_2\right)}}{{\vec{k}}^2},\\ 
\label{eq:prA} \langle{A_i(x_1)}~{A_j(x_2)}\rangle 
& = \parbox{18mm}{\includegraphics[scale=0.5]{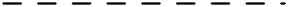}} = 
-16\pi G \times \,\delta(t_1-t_2)\int_{\vec{k}} 
\frac{e^{i\vec{k}\cdot\left(\vec{x}_1 - \vec{x}_2\right)}}{{\vec{k}}^2} 
~\delta_{ij},\\ 
\label{eq:prsigma}  \langle{\sigma_{ij}(x_1)}{\sigma_{kl}(x_2)}\rangle 
& = \parbox{18mm}{\includegraphics[scale=0.5]{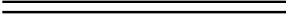}} = 
~~32\pi G \times\,\delta(t_1-t_2)\int_{\vec{k}}
\frac{e^{i\vec{k}\cdot\left(\vec{x}_1 - \vec{x}_2\right)}}{{\vec{k}}^2} 
~P_{ij;kl},
\end{align}
where we abbreviate $\int \frac{d^3\vec{k}}{\left(2\pi\right)^3}$ as 
$\int_{\vec{k}}$, and $P_{ij;kl}\equiv\frac{1}{2}\left(\delta_{ik}\delta_{jl}
+\delta_{il}\delta_{jk}-2\delta_{ij}\delta_{kl}\right)$. Hence, the KK 
decomposition allows one to use simplified propagators, which reduce the overload 
of tensor indices. For the relativistic time correction vertices on the 
propagators from equation \eqref{pnpotprops} we then have the  following:
\begin{align}
\label{eq:prtphi}  \parbox{18mm}{\includegraphics[scale=0.5]{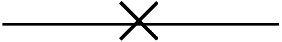}}
= & ~~~~\frac{1}{8\pi G}~~\int d^4x~\left(\partial_t\phi\right)^2, \\ 
\label{eq:prtA}   \parbox{18mm}{\includegraphics[scale=0.5]{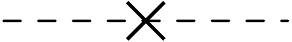}}
= & -\frac{1}{32\pi G} \int d^4x~\left(\partial_tA_i\right)^2, \\ 
\label{eq:prtsigma} \parbox{18mm}{\includegraphics[scale=0.5]{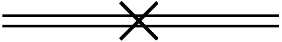}}
= & ~~\frac{1}{128\pi G} \int d^4x~\left[2(\partial_t\sigma_{ij})^2
-(\partial_t\sigma_{ii})^2\right], 
\end{align}
where the crosses represent these quadratic self-gravitational vertices,  
containing two time derivatives.
However, in GR there are also non-linear self-interaction vertices, which give 
rise to loop integrals, where $n$-graviton vertices correspond to 
($n-$2) loops. For example, the following cubic vertices can be easily read 
from the static part of the gravitational action in equation \eqref{sgravpure} in 
terms of the KK parametrization:
\begin{align}
\label{eq:sigmaphi2} \parbox{18mm}{\includegraphics[scale=0.35]{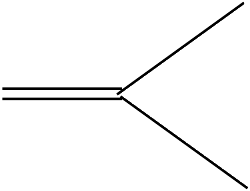}}
& = ~ \frac{1}{16\pi G}\int  d^4x~\left[
2\sigma_{ij}\partial_i\phi\partial_j\phi
-\sigma_{ii}\partial_j\phi\partial_j\phi\right],\\
\label{eq:phiA2} \parbox{18mm}{\includegraphics[scale=0.35]{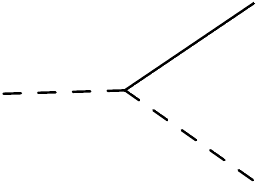}}
& = \frac{1}{8\pi G}\int d^4x~\phi\left[
\partial_iA_j\left(\partial_iA_j-\partial_jA_i\right)
+\left(\partial_iA_i\right)^2\right].
\end{align}

Next, we proceed to consider the worldline graviton couplings, which arise 
from the point-particle action. From equation \eqref{mcoup} we get the following 
one-graviton couplings for the mass couplings of the objects:
\begin{align}
\label{eq:mphi} \parbox{12mm}{\includegraphics[scale=0.35]{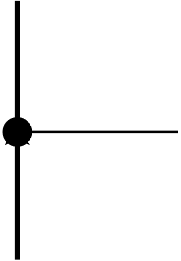}}
& = - m \int dt~\phi~\left[1+\frac{3}{2}v^2+ {\cal{O}}(v^4)\right], \\ 
\label{eq:mA} \parbox{12mm}{\includegraphics[scale=0.35]{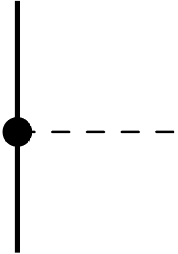}}
& = ~m \int dt~A_iv^i~\left[1+ {\cal{O}}(v^2) \right],\\ 
\label{eq:msigma} \parbox{12mm}{\includegraphics[scale=0.35]{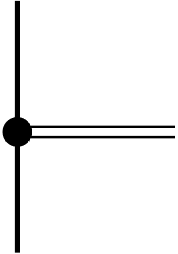}}
& = \frac{m}{2} \int dt~\sigma_{ij}v^iv^j~\left[1+{\cal{O}}(v^2)\right],
\end{align}
where a bold vertical line represents a worldline, a mass coupling is 
represented by a black circle, and there is an infinite expansion in $v^2$. 
We can already see from these couplings one of the benefits of the KK 
decomposition, for example, in organizing the PN hierarchy of the coupling to 
the mass of the graviton field components. From further expanding 
equation \eqref{mcoup} in the fields the $n$-graviton mass couplings are obtained, 
e.g. the following leading two-graviton mass coupling:
\begin{align}
\label{eq:mphi2}  \parbox{12mm}{\includegraphics[scale=0.35]{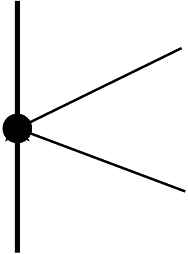}}
& = -\frac{1}{2}m \int dt~\phi^2~\left[1 + {\cal{O}}(v^2)\right].
\end{align}

For the spin couplings from equations \eqref{spinmctrans} and \eqref{scoupterm}, 
which originate from the minimal coupling, it should be stressed first, that 
just as the mass sector contains kinetic terms with no field couplings (see 
equation \eqref{mcoup}), then there are also kinematic contributions involving 
spin, given by
\be \label{frskin}
L_{\text{pp[S(kin)]}}=-\vec{S}\cdot \vec{\Omega} + \frac{1}{2} 
\vec{S}\cdot\vec{v}\times\vec{a}\left(1+\frac{3}{4}v^2+{\cal{O}}(v^4)\right),
\ee
where $a^i\equiv\dot{v}^i$, and all indices are Euclidean. We also recall 
that at this stage there are only physical spin and angular velocity 
components, namely $S_i\equiv\frac{1}{2}\epsilon_{ijk}S_{jk}$, 
$\Omega_i\equiv\frac{1}{2}\epsilon_{ijk}\Omega_{jk}$, where $\epsilon_{ijk}$ 
is the three-dimensional Levi-Civita symbol, and
$\Omega_{ij}\equiv -\Lambda_{ik} \dot{\Lambda}_{kj}$, where the Lorentz 
matrices are now the $SO(3)$ rotation matrices. Then, there are the following 
one-graviton spin couplings, e.g.:
\begin{align}
\label{eq:sA}  \parbox{12mm}{\includegraphics[scale=0.35]{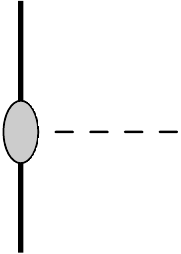}}
& = \int dt \,\,\epsilon_{ijk}S_{k}\left(\frac{1}{2}\partial_iA_j + 
\frac{3}{4} v^iv^l 
\left(\partial_lA_j-\partial_jA_l\right)+v^i\partial_tA_j+\cdots\right),\\ 
\label{eq:sphi}   \parbox{12mm}{\includegraphics[scale=0.35]{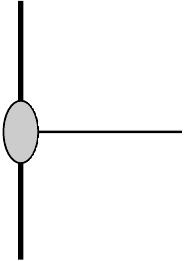}}
& = \int dt \,\,2\epsilon_{ijk}S_{k}\left(v^i\partial_j\phi + \cdots\right),
\end{align} 
where the gray oval represents a spin coupling, and the ellipsis stands for 
higher PN order terms. 

Proceeding to the non-minimal coupling, we get from equation \eqref{es2}, 
for example, the following one-graviton spin-squared coupling:
\begin{align}
\label{eq:sqphi}   \parbox{12mm}{\includegraphics[scale=0.35]{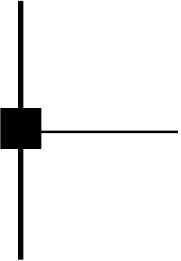}}
& = \int dt \,\,\left[\frac{C_{ES^2}}{2m} S^{i}S^{j} \partial_i\partial_j\phi 
+ \cdots\right],
\end{align}
where the black square box represents the spin-squared quadrupole. From 
equation \eqref{bs3} we get, for example, the following couplings to the 
spin-induced octupole:
\begin{align}
\label{eq:s3A}  \parbox{12mm}{\includegraphics[scale=0.35]{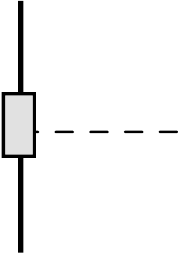}}
=& \int dt \,\, \left[-\frac{C_{\text{BS}^3}}{12m^2}
S^{i}S^{j}\epsilon_{klm}S^{m}\partial_i\partial_j\partial_k A_l 
+\cdots\right],\\ 
\label{eq:s3phi}   \parbox{12mm}{\includegraphics[scale=0.35]{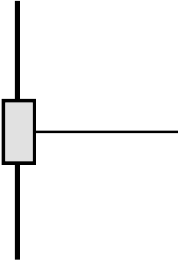}}
=& \int dt \,\,\left[\frac{C_{\text{BS}^3}}{3m^2}
S^{i}S^{j}\epsilon_{klm}S^{m}\partial_i\partial_j\partial_k\phi\,v^l
+\cdots\right],
\end{align}
where the gray rectangular boxes represent the octupole. Finally, from equation 
\eqref{es4} the coupling to the quartic spin hexadecapole reads as follows:
\begin{align}
\label{eq:s4phi}   \parbox{12mm}{\includegraphics[scale=0.35]{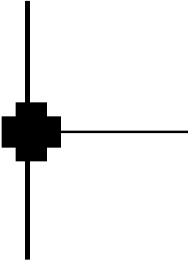}}
 =& \int dt \,\,\left[-\frac{C_{\text{ES}^4}}{24m^3}S^{i}S^{j}S^{k}S^{l}
 \partial_i\partial_j\partial_k\partial_l\phi+\cdots\right],
\end{align} 
where the black crossed box represents the hexadecapole. 

\begin{figure}[t]
\centering
\includegraphics[width=0.5\linewidth]{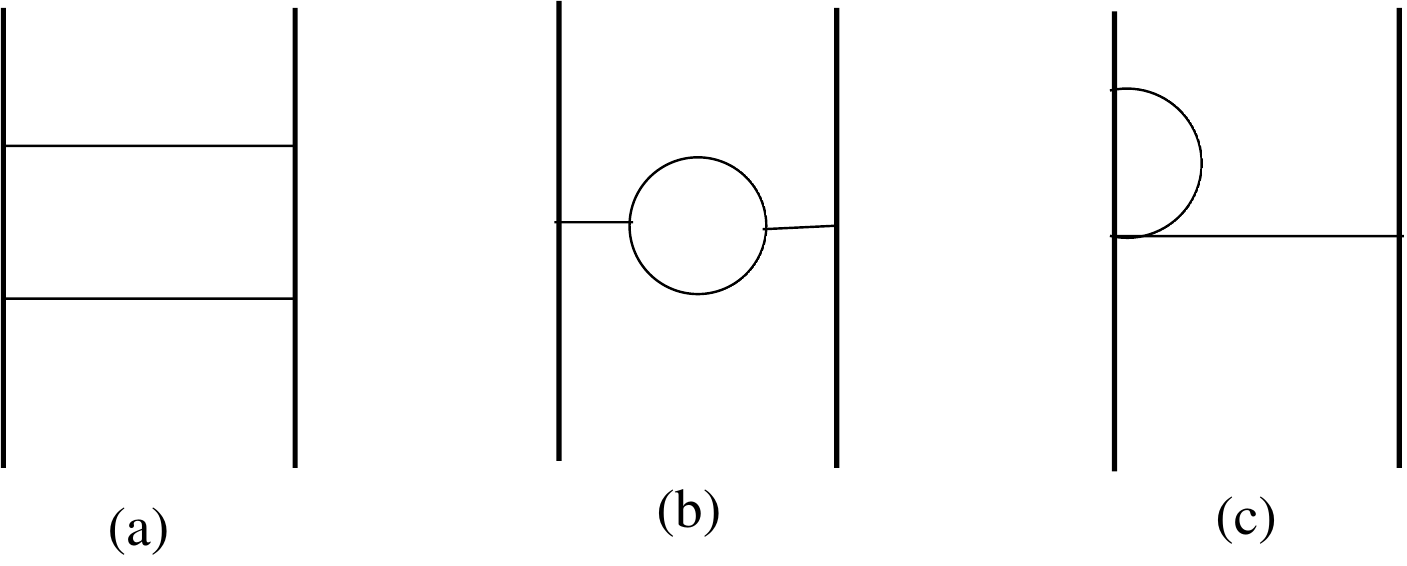}
\caption{Graph topologies that are excluded from the diagrammatic expansion 
of the two-particle interaction for the EFT of the composite particle 
\cite{Goldberger:2004jt,Goldberger:2007hy}: 
(a) Graph with more than a single connectivity component (where worldlines 
are stripped off); 
(b) Graph containing a graviton loop; 
(c) Graph renormalizing the UV divergence related with the Wilson 
coefficients of the one-particle EFT, e.g.~the mass. 
We note that the bold vertical lines represent the worldlines, where time 
flows up, rather than from left to right as is customary in particle physics.}
\label{exclude} 
\end{figure}

We are now in a position to consider the construction of the Feynman graphs, 
which contribute to the various two-particle interactions 
\cite{Goldberger:2004jt,Goldberger:2007hy}. All such graphs would contain the 
two worldlines of the components of the binary, and gravitons, which 
propagate and mediate the interaction between them. In addition, the
worldlines are to be interchanged, as the interaction is symmetric under the 
exchange of the two particles. Next, we note that there are some graph 
topologies that will be excluded from the diagrammatic expansion, as in 
the examples displayed in figure \ref{exclude}: 1.~Graphs that contain more 
than a single connectivity component (where worldlines are stripped off the 
graph), since by definition the effective action is the exponent in 
equation \eqref{seff2part}; 2.~Graphs that contain graviton loops (i.e.~again, 
worldlines stripped off), as we are concerned here with classical gravity, 
and hence only tree graphs are relevant; 3.~Graphs that UV renormalize the Wilson 
coefficients of the one-particle EFT, i.e.~the mass, spin, or the 
coefficients of the higher induced multipoles\footnote{These 
power-divergent diagrams are set to zero in dimensional regularization.}. 
We note that in figure \ref{exclude} and henceforth, graphs 
show the time flowing up in accordance with the way spacetime is drawn 
in relativity (unlike in figure \ref{phi4eftgr}, or in many works in the 
field, where time flows from left to right, according to the convention in 
particle physics).

\begin{figure}[t]
\centering
\includegraphics[width=0.1\linewidth]{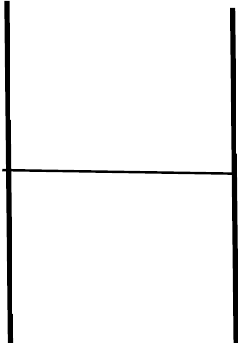}
\caption{The single topology at ${\cal{O}}(G^1)$: One-graviton exchange with 
no gravitational self-interaction.}
\label{g1topo} 
\end{figure}

\begin{figure}[t]
\centering
\includegraphics[width=0.28\linewidth]{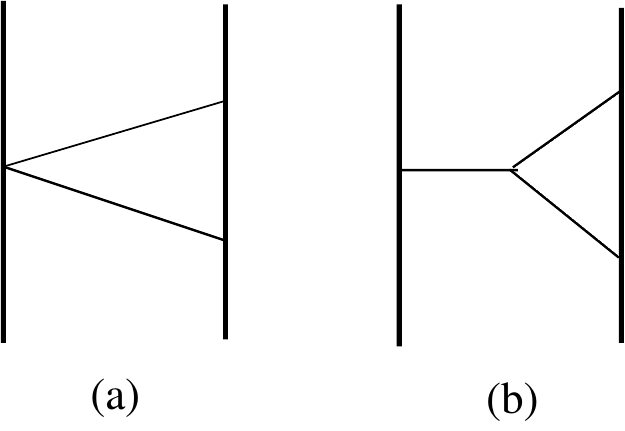}
\caption{The topologies at ${\cal{O}}(G^2)$: (a) Two-graviton exchange; 
(b) Single cubic self-interaction. This is a one-loop topology. When using 
the KK metric only topology (a) enters at the 1PN order \cite{Kol:2007bc}; 
see figure \ref{01pn}. Both topologies enter at the 2PN order of the 
point-mass sector \cite{Gilmore:2008gq}.}
\label{g2topo} 
\end{figure}

Next, let us consider the bare graph topologies at each order of the 
gravitational constant, $G$. To this end we consider the generic power 
counting of the fields as inferred from the Feynman rules for the 
propagators, where at first each $n$-graviton self-interaction vertex scales 
as $G^{\frac{n}{2}-1}$, and each $n$-graviton worldline coupling scales as 
$G^{\frac{n}{2}}$ \cite{Gilmore:2008gq}. Therefore at the order of $G^1$ 
there is only a single possible topology of a one-graviton exchange, as can 
be seen in figure \ref{g1topo}. At the order of $G^2$ there are two possible 
topologies, where there is already one topology with a cubic 
self-interaction, namely a one-loop graph, as can be seen in figure 
\ref{g2topo}. Generally, at the order of $G^n$, which corresponds to the 
($n-$1)PN order, we first encounter ($n-$1)-loop topologies. 

\begin{figure}[t]
\centering
\includegraphics[width=0.4\linewidth]{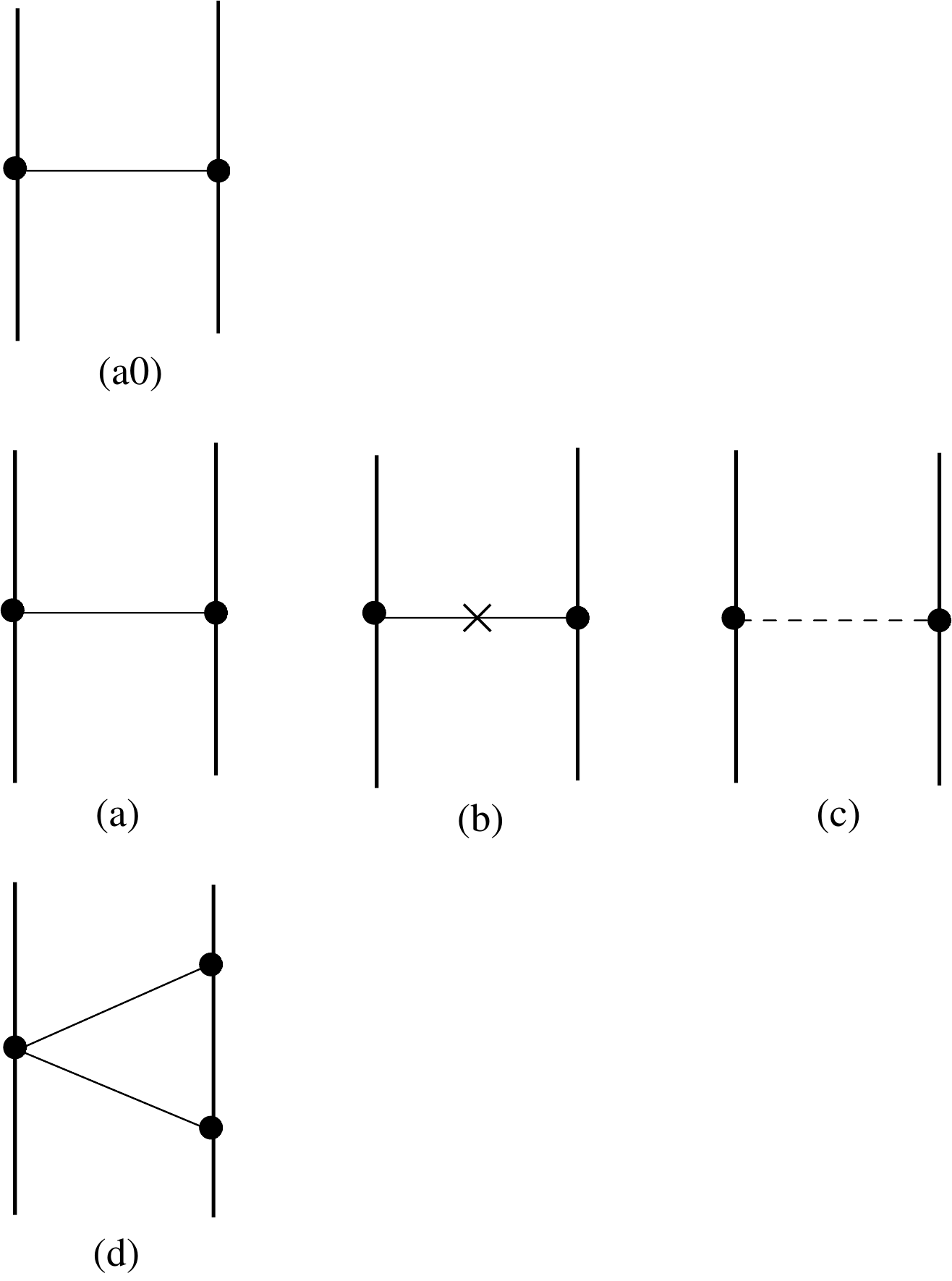}
\caption{The Newtonian (0PN) and first post-Newtonian (1PN) order 
interactions \cite{Goldberger:2004jt,Kol:2007bc}: 
(a0) The Newtonian interaction is mediated by the scalar field, $\phi$, of 
the KK metric; 
(a)-(c) At the 1PN order we start to take into account higher-order mass 
couplings, as in (a) and (c), and the relativistic correction to the 
instantaneous propagator in (b);
(d) At the non-linear level we expect in general to also have a one-loop 
graph \cite{Goldberger:2004jt}, but with the KK metric it is deferred to the 
next PN order, and there is only a single diagram of two-graviton exchange 
\cite{Kol:2007bc, Gilmore:2008gq}.}
\label{01pn} 
\end{figure}

We then proceed to dress these bare topologies with actual vertices from the 
Feynman rules, that are PN-weighed according to their power in $v$. The 
evaluation of the resulting Feynman diagrams involves the usual Wick 
contractions, symmetry factors, and Feynman integrals, familiar from QFT. For 
example, the Newtonian interaction, which appears in figure \ref{01pn}(a0), 
is evaluated as follows:
\bea
\text{Figure \ref{01pn}(a0)} & = &-m_1 \int dt_1 
\,\contraction[2.5ex]{}{\phi(y_1)} {\times (-m_2) \int dt_2\,} {\phi(y_2)} 
\phi(y_1) \times (-m_2) \int dt_2 \, \phi(y_2)\nn \\
& = & 4\pi G \times m_1m_2 \int dt_1 dt_2 \,\delta(t_1-t_2)\int_{\vec{k}} 
\frac{e^{i \vec{k}\cdot\left(\vec{y}_1(t_1) - \vec{y}_2(t_2)\right)}}
{{\vec{k}}^2} \nn \\
& = & \int dt \,\, \frac{G m_1 m_2}{r},
\eea
where $r\equiv|\vec{r}|\equiv|\vec{y}_1-\vec{y}_2|$, and the action of the 
familiar Newtonian interaction is obtained. After Wick contracting the scalar 
fields using the propagator in equation \eqref{eq:prphi}, we integrate over time 
with the delta function, and for the Fourier integral the following scalar 
master integral is needed: 
\be \label{eq:0loop}
I\equiv
\int\frac{d^d\vec{k}}{(2\pi)^d}
\frac{e^{i\vec{k}\cdot\vec{r}}}{(\vec{k}^2)^{\alpha}}=
\frac{1}{(4\pi)^{\frac{d}{2}}}
\frac{\Gamma(\frac{d}{2}-\alpha)}{\Gamma(\alpha)} 
\left(\frac{r^2}{4}\right)^{\alpha-\frac{d}{2}}, 
\ee
where dimensional regularization is always used throughout, and eventually 
the limit $d\to3$ is taken. This master integral can be easily derived using 
Schwinger parameters \cite{Smirnov:2006ry}. In general, the master integral 
in equation \eqref{eq:0loop} is required for all computations, where the related 
tensorial integrals are simply obtained by differentiating the scalar 
integral with respect to $\vec{r}$.

The diagrams that make up the 1PN order interaction are shown in figure 
\ref{01pn}(a)-(d), where the KK metric is used \cite{Kol:2007bc}. These 
contain higher-order mass couplings and the leading relativistic correction 
to the Newtonian instantaneous propagators. In general, a one-loop graph 
enters at the 1PN order \cite{Goldberger:2004jt,Cardoso:2008gn}. However, if 
the KK metric is used, a one-loop graph implies that all mass couplings must be 
of the scalar field, $\phi$, since only these couplings scale as $\sim v^0$, 
and the cubic self-interaction should be a static one, with only $\phi$ 
fields. However, in the KK metric the cubic vertex with $\phi$ fields is of the 
form $\phi(\partial_t\phi)^2$, i.e.~it is time dependent, and since each time 
derivative scales as $v^1$, one-loop graphs are postponed to the 2PN order 
\cite{Kol:2007bc,Gilmore:2008gq}. In general, the $n$PN order in the 
point-mass sector requires computation at the $n$-loop level, but using the 
KK metric only the $2\lfloor n/2\rfloor$-loop level is required. For the sector 
with spins the $n$-loop computation is shifted to the ($n+$1.5)PN order or 
higher for rapidly rotating objects. 

A crucial component in the evaluation of diagrams is the treatment of the 
time derivatives, which appear already at the 1PN order; see figure 
\ref{01pn}(b). To this end the following generic identity is central:
\be 
\int dt_1 dt_2 \partial_{t_1}\delta(t_1-t_2)f(t_1)g(t_2)= 
-\int dt_1 dt_2 \partial_{t_2}\delta(t_1-t_2)f(t_1)g(t_2),
\ee
where repeated integration by parts (IBP) is used to drop the time 
derivatives on the worldline couplings. Each integral in position space 
coming from the bulk vertices transforms as usual into a delta function 
conserving momentum in the vertex, and for a one-loop integral the following 
scalar master integral is used:
\bea
J&\equiv&
\int \frac{d^d\vec{k}}{(2\pi)^d}\frac{1}{[\vec{k}^2]^{\alpha}
[(\vec{k}-\vec{q})^2]^\beta}\nn\\
&=& \frac{1}{(4\pi)^{d/2}}
\frac{\Gamma(\alpha+\beta-d/2)}{\Gamma(\alpha)\Gamma(\beta)}
\frac{\Gamma(d/2-\alpha)\Gamma(d/2-\beta)}{\Gamma(d-\alpha-\beta)}
\left(q^2\right)^{d/2-\alpha-\beta}.\label{eq:1loop}
\eea
This formula can easily be derived using Feynman and Schwinger parameters 
\cite{Smirnov:2006ry}, and the related tensor integrals are similarly derived.

\begin{figure}[t]
\centering
\includegraphics[width=0.3\linewidth]{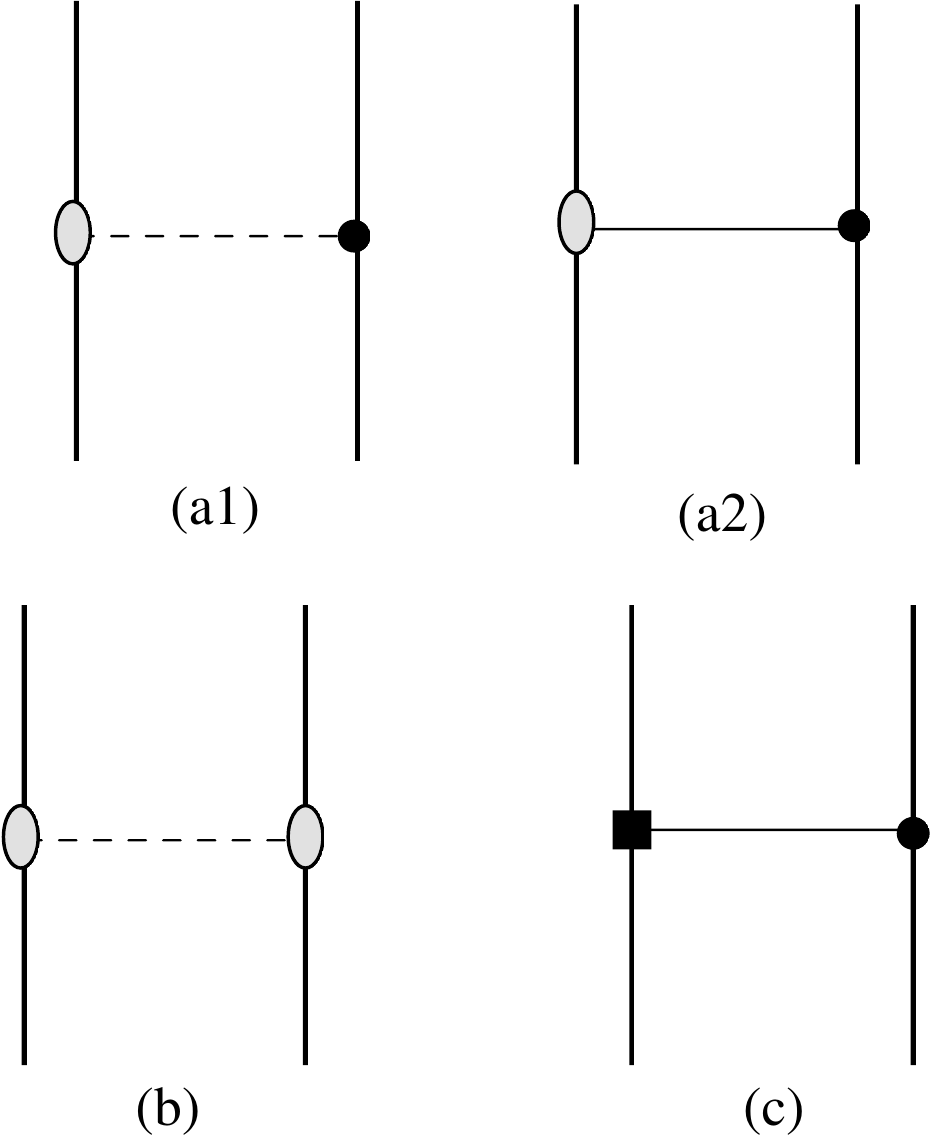}
\caption{The LO PN interactions up to quadratic order in the spins: 
(a1)-(a2) The LO spin-orbit interaction. This is the leading PN correction 
with spins, which contributes at the 1.5PN order for rapidly rotating compact 
objects \cite{Porto:2005ac,Levi:2010zu,Levi:2015msa}; 
(b) The LO spin1-spin2 interaction between the spins of the two objects 
\cite{Porto:2005ac,Levi:2008nh}; 
(c) The LO spin-squared interaction, containing the spin-induced quadrupole 
of the object. This is the leading PN interaction from finite size effects, 
which contains a non-trivial Wilson coefficient that should be matched 
\cite{Porto:2008jj,Levi:2014gsa}. 
The interactions, which are quadratic in the spins, contribute at the 2PN 
order for rapidly rotating compact objects.}
\label{lospin} 
\end{figure}

Let us continue and consider interactions involving the spins of the objects 
in the binary. These interactions enter as of the 1.5PN order for rapidly 
rotating objects, i.e.~the next PN correction, after the 1PN order from 
the point-mass sector of figure \ref{01pn}, with the spin-orbit interaction 
being the leading one in the sector with spins. The LO PN interactions up to 
quadratic order in the spins can be seen in figure \ref{lospin} 
\cite{Porto:2005ac,Levi:2010zu,Levi:2015msa,Levi:2008nh,Porto:2008jj,
Levi:2014gsa}. It is important to note that already at the LO spin-orbit 
interaction, shown in graphs (a1)-(a2) of figure \ref{lospin}, the gauge of 
the rotational variables needs to be addressed, unlike the subleading LO 
spin1-spin2 and spin-squared interactions, shown in graphs (b) and (c) of 
figure \ref{lospin}, respectively, which contribute at the 2PN order for the 
case of rapidly rotating compact objects. Also noteworthy is that from the LO 
spin-orbit interaction (and from the 2PN order correction in the point-mass 
sector), one starts to encounter accelerations, and higher-order time derivatives, 
which should be properly removed via variable redefinitions, as we elaborate 
in section \ref{redacteomh}. 

\begin{figure}[t]
\centering
\includegraphics[width=0.28\linewidth]{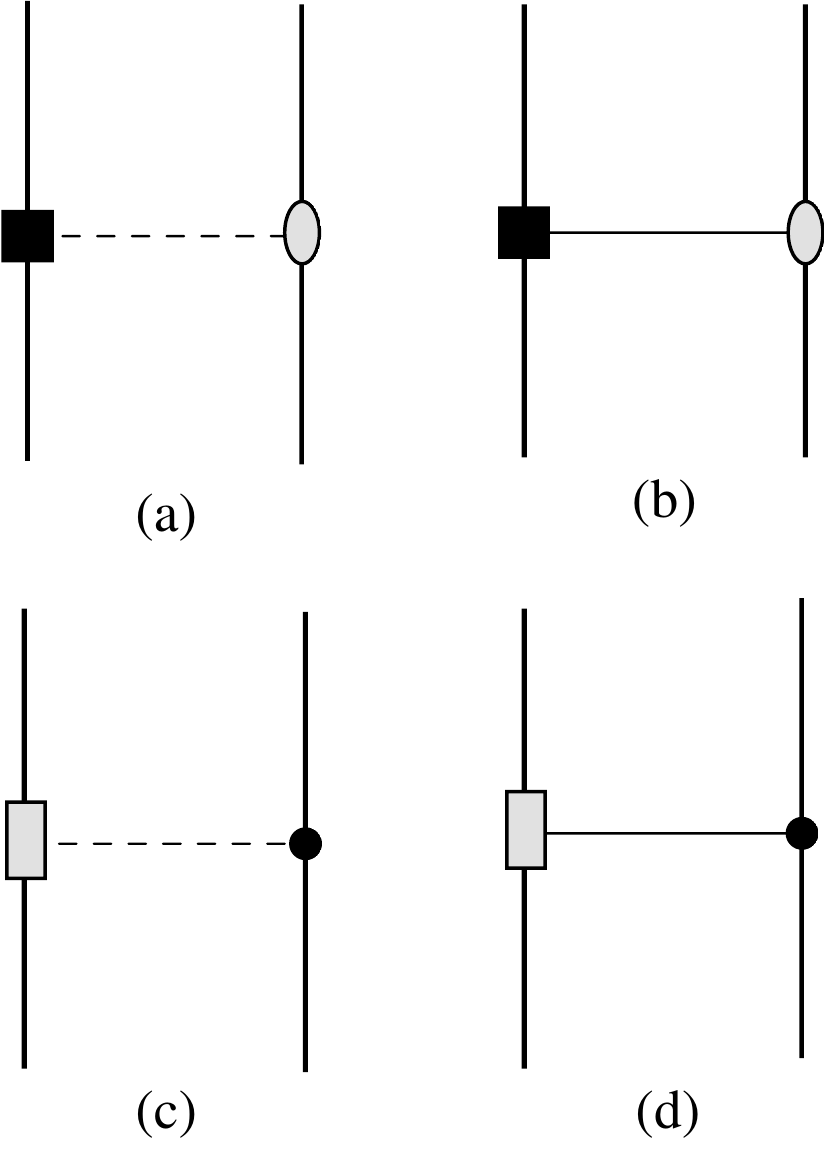}
\caption{The LO interaction at cubic order in the spins, which contributes at 
the 3.5PN order for rapidly rotating compact objects \cite{Levi:2014gsa,
Levi:2015msa}. This interaction is in analogy with the LO spin-orbit 
interaction in figure \ref{lospin}(a1)-(a2) according to the parity of the 
spin-induced multipoles: 
(a)-(b) The spin-induced quadrupole interacting with the spin dipole; 
(c)-(d) The spin-induced octupole interacting with the mass monopole.}
\label{los3} 
\end{figure}

\begin{figure}[t]
\centering
\includegraphics[width=0.45\linewidth]{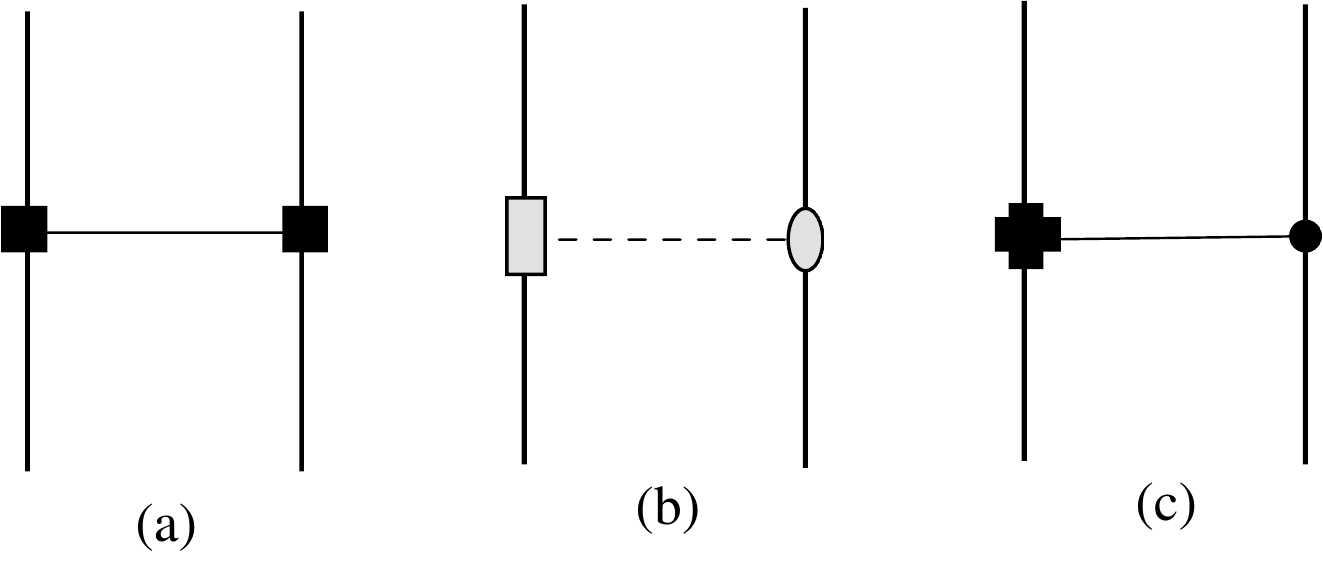}
\caption{The LO interaction at quartic order in the spins, which contributes 
at the 4PN order for rapidly rotating compact objects \cite{Levi:2014gsa, 
Levi:2015msa}: 
(a) and (c) Two spin-induced quadrupoles interacting and a spin-induced 
hexadecapole interacting with a mass monopole, respectively, both in analogy 
with the LO spin-squared interaction in figure \ref{lospin}(c); 
(b) Spin-induced octupole interacting with a spin dipole in analogy with the 
LO spin1-spin2 interaction in figure \ref{lospin}(b).}
\label{los4} 
\end{figure}

The leading non-trivial finite size effect is thus spin-induced, namely the LO 
spin-squared interaction with the spin-induced quadrupole in figure 
\ref{lospin}(c), and it is preceded by a Wilson coefficient, which 
encapsulates the internal physics of the compact object, and should be 
matched. Further spin-induced finite size effects at LO are shown in figures 
\ref{los3} and \ref{los4}, corresponding to interactions at cubic and quartic 
order in the spins, respectively, also including the spin-induced octupoles 
and hexadecapoles \cite{Levi:2014gsa,Levi:2015msa}, and further Wilson 
coefficients to be matched. The interactions at cubic and quartic order in 
the spins contribute at the 3.5PN and 4PN orders, respectively, for rapidly 
rotating compact objects. It is easy to see the analogy of these interactions 
with those in figure \ref{lospin}, which are lower in the order of the spins, 
corresponding to the parity of the spin-induced multipoles.

\begin{figure}[t]
\centering
\includegraphics[width=0.4\linewidth]{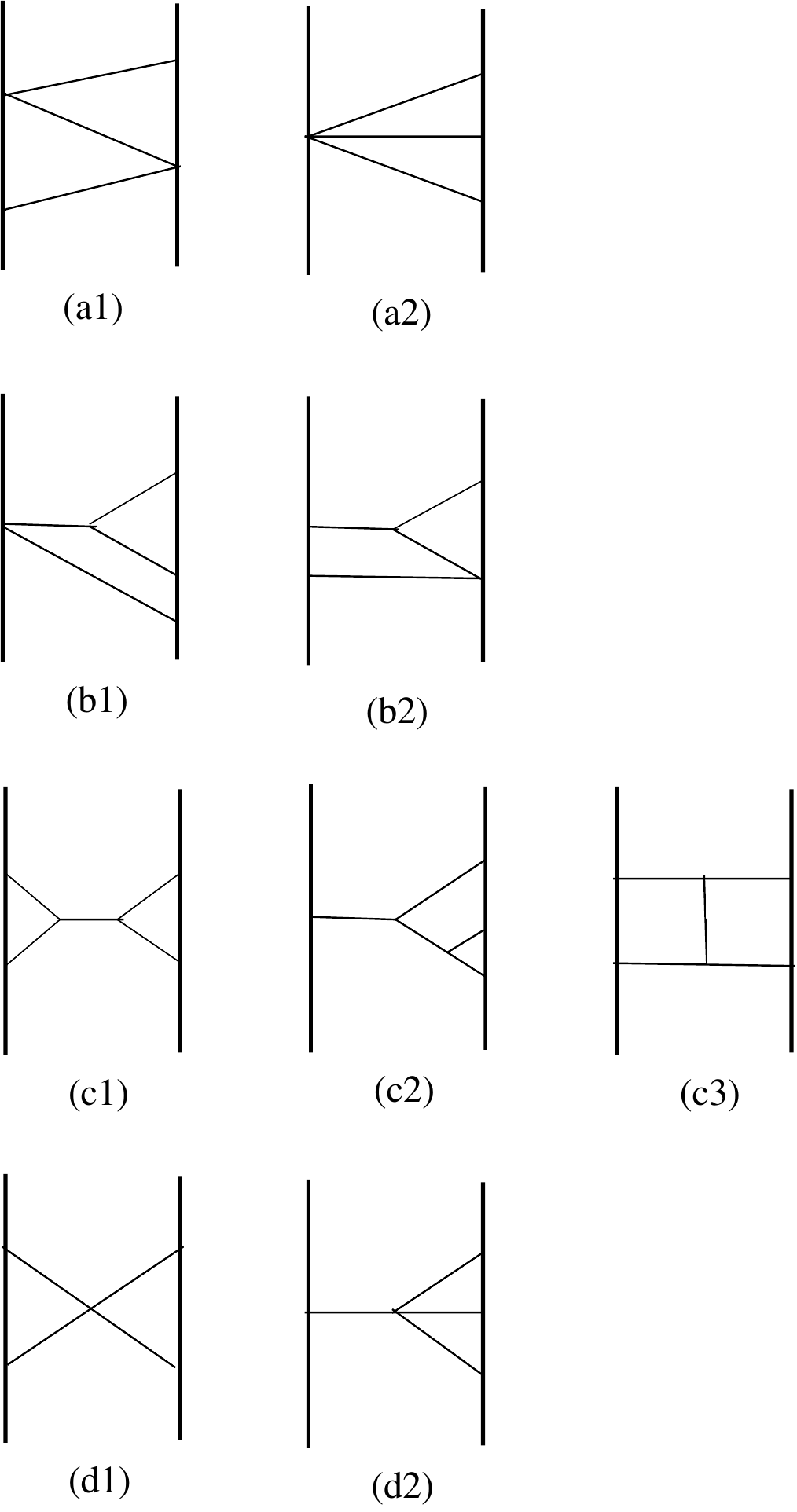}
\caption{The topologies at ${\cal{O}}(G^3)$ \cite{Gilmore:2008gq,Chu:2008xm}: 
(a) Up to three-graviton exchanges; 
(b) One cubic self-interaction with a two-graviton exchange; 
(c) Two cubic self-interactions; 
(d) One quartic self-interaction. 
The graphs in (c) and (d) are the two-loop topologies. 
The analogy in the evaluation of graphs (c1) and (c2), with graphs (d1) and 
(d2), respectively, can be easily seen from the topologies. 
It can also be easily seen that graph (c1) is factorizable, i.e.~it 
factorizes into 2 one-loop topologies, and that graph (c2) is nested, i.e.~it 
consists of a one-loop topology nested within a one-loop topology. 
On the other hand, graph (c3) is not trivially reducible to one-loop topologies. 
When using the KK metric only the five topologies in graphs (a) and (c) enter 
at the 2PN order of the point-mass sector \cite{Gilmore:2008gq}. All of the 
topologies enter at the 3PN order \cite{Foffa:2011ub}.}
\label{g3topo} 
\end{figure}

Let us proceed to the order of $G^3$ to gain more insight into the topologies 
and computations that arise at higher PN orders. The topologies at the order 
of $G^3$ are shown in figure \ref{g3topo} \cite{Gilmore:2008gq,Chu:2006ce}. 
We encounter new topologies up to the one-loop level, which can be clearly 
factorized into topologies that appeared in lower orders, as in figures 
\ref{g3topo}(a1)-(b2), whereas figures \ref{g3topo}(c1)-(d2) display two-loop 
topologies. Nevertheless, it can be easily seen that graphs (c1) and (d1) can 
be factorized into 2 one-loop topologies, and that in graphs (c2) and (d2) there 
is a one-loop topology nested within a one-loop topology. The topology, which 
appears in figure \ref{g3topo}(c3) thus remains the only topology at this order 
that represents a new computational feature. Hence there is only one kind of 
two-loop integral, corresponding to graph (c3), which cannot be readily broken 
down into a one-loop computation \cite{Gilmore:2008gq,Kol:2009mj,Levi:2011eq}. In 
such topologies we encounter two-loop integrals, which contain products of five 
propagators that cannot be trivially disentangled. 

However, these two-loop integrals can be reduced to a sum of factorizable and 
nested two-loop integrals \cite{Gilmore:2008gq,Levi:2011eq,Foffa:2013qca}, 
using the IBP method \cite{Smirnov:2006ry}, which yields the following useful 
reduction relation:
\bea
F(a_1,a_2,a_3,a_4,a_5)&\equiv&
\int_{\vec{k}_1,\vec{k}_2} \frac{1}{{[\vec{k}_1^2]}^{a_1}
{[(\vec{k}-\vec{k}_1)^2]}^{a_2}{[\vec{k}_2^2]}^{a_3}
{[(\vec{k}-\vec{k}_2)^2]}^{a_4}{[(\vec{k}_1-\vec{k}_2)^2]}^{a_5}}\nn\\
&=&
\frac{a_1\left[F(a_1+,a_3-)-F(a_1+,a_5-)\right]
+[1\leftrightarrow2,3\leftrightarrow4]}
{a_1+a_2+2a_5-d}, 
\label{eq:2loopibp}
\eea 
where we use abbreviated notation, 
e.g.~$F(a_1+,a_3-)\equiv F(a_1+1,a_2,a_3-1,a_4,a_5)$, and $1\leftrightarrow2$ 
is an exchange of the exponent labels. This reduction relation yields 
intermediate expressions with explicit poles in $d=3$, but these cancel out 
in the dimensional regularization. When using the KK metric only the five 
topologies in graphs (a) and (c) of figure \ref{g3topo} enter at the 2PN 
order of the point-mass sector \cite{Gilmore:2008gq}. All of the topologies 
at ${\cal{O}}(G^3)$ enter at the 3PN order \cite{Foffa:2011ub}.

\begin{figure}[t]
\centering
\includegraphics[width=0.5\linewidth]{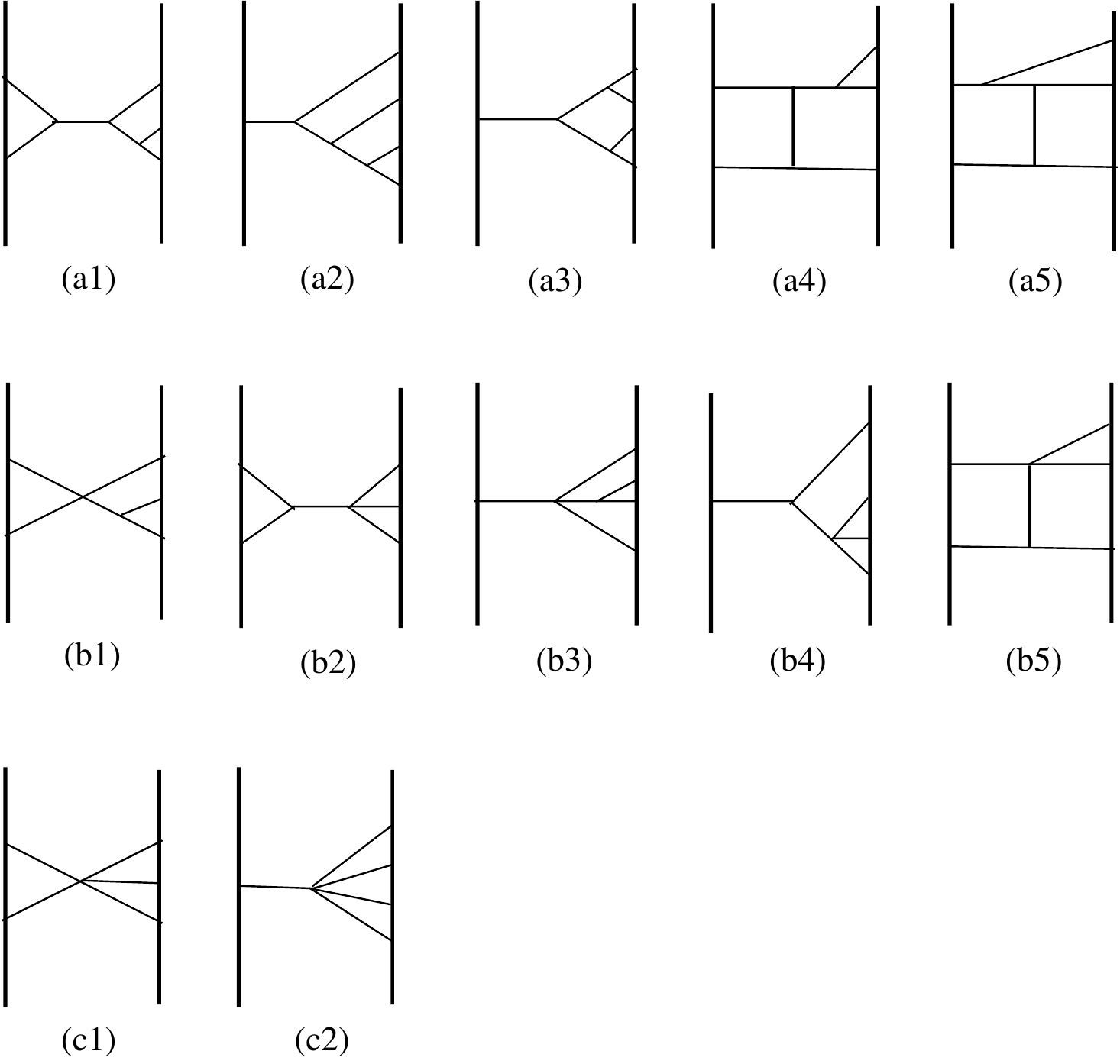}
\caption{The three-loop topologies at ${\cal{O}}(G^4)$ (out of the total 32 
topologies at this order): 
(a) Three cubic vertices; 
(b) One cubic vertex and one quartic vertex; 
(c) One quintic vertex. 
As in figure \ref{g3topo}, it can be readily identified which topologies are 
factorizable and/or nested two-loop topologies. The few graphs, which are not trivially 
reduced to one-loop topologies, are reduced in a similar manner as in graph (c3) 
of figure \ref{g3topo}, i.e.~by using the IBP relation in equation \eqref{eq:2loopibp} 
\cite{Foffa:2013qca}. Regardless, when using the KK metric, none of these 
three-loop topologies enter at the 3PN order of the point-mass sector 
\cite{Foffa:2011ub}, but they all enter at the 4PN order of the point-mass 
sector \cite{Foffa:2013qca}.}
\label{g4topo} 
\end{figure}

As it turns out there are no new computational features that appear at 
${\cal{O}}(G^4)$, and hence at the 3PN order. The most complex graphs at 
${\cal{O}}(G^4)$ are the three-loop topologies, which are all shown in figure 
\ref{g4topo}. As in ${\cal{O}}(G^3)$ in figure \ref{g3topo}, the topologies 
that are factorizable and/or nested two-loop topologies are readily recognized. The 
few, which are not trivially reduced to one-loop topologies, are reduced 
similarly to graph (c3) of figure \ref{g3topo}, i.e.~by using the IBP relation 
in equation \eqref{eq:2loopibp} \cite{Foffa:2013qca}. Regardless, when using the KK 
metric, none of these three-loop topologies enter at the 3PN order of the 
point-mass sector \cite{Foffa:2011ub}, and they all enter only at the 4PN 
order of the point-mass sector \cite{Foffa:2013qca}. Let us underline that 
notably thus far all of the topologies up to ${\cal{O}}(G^4)$, i.e.~up to the 
three-loop level, reduce to one-loop integrals.

\begin{figure}[t]
\centering
\includegraphics[width=0.5\linewidth]{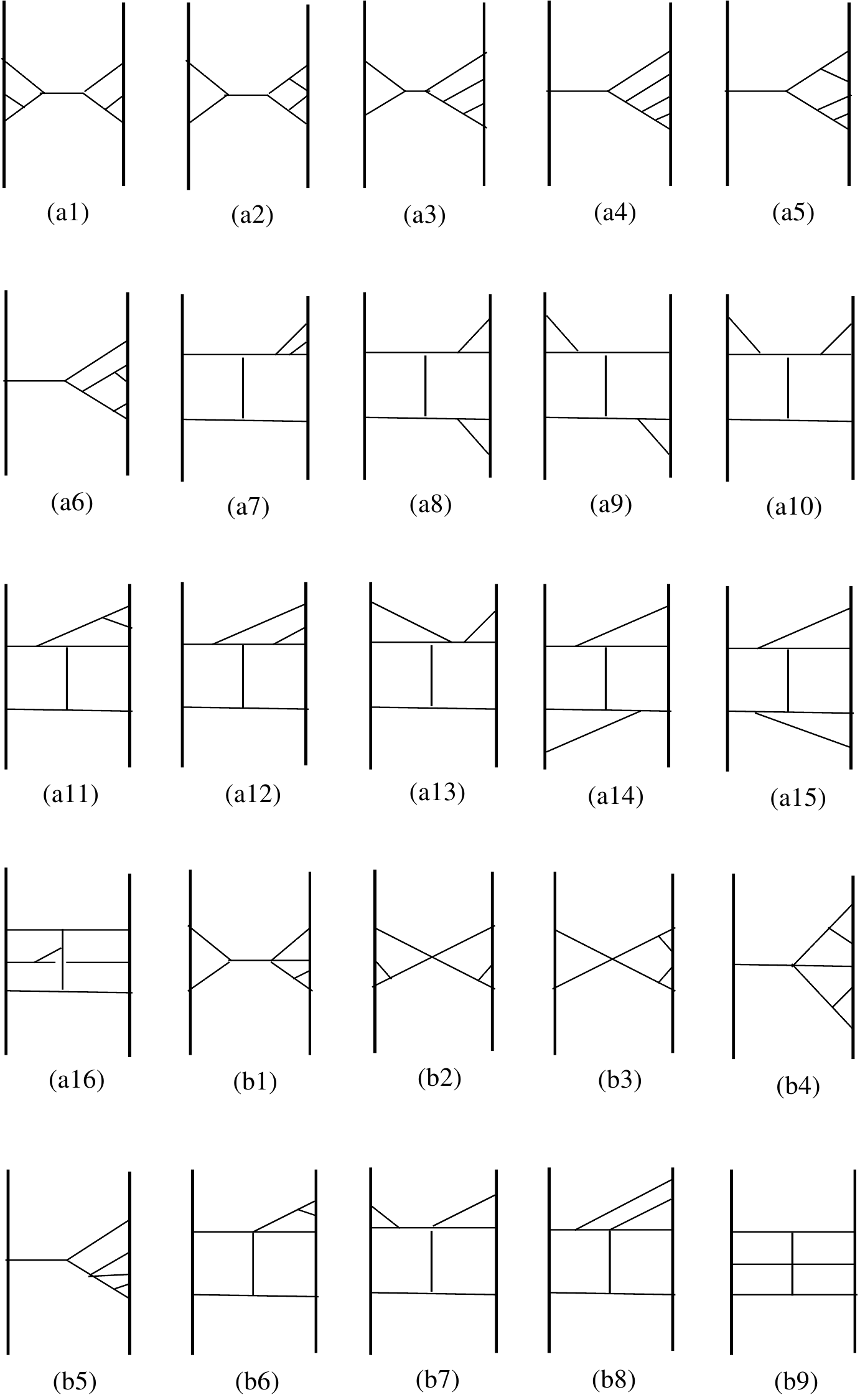}
\caption{The four-loop topologies at ${\cal{O}}(G^5)$, which enter at the 4PN 
order using the KK metric: 
(a) Four cubic vertices; (b) Two cubic vertices and one quartic vertex. 
As in figures \ref{g3topo} and \ref{g4topo}, the topologies, which are 
factorizable and/or nested, are easily identified. It is also visibly 
recognizable that graphs (a9)-(a16) and (b7)-(b9) have a more complex 
structure to compute, whereas the few other graphs, which are not trivially 
reducible, are reduced to one-loop integrals using IBP \cite{Foffa:2016rgu}. 
The complex graphs require knowledge of certain four-loop master integrals 
in the limit $d\to3$. Out of these, graphs (a16) and (b9) in particular, which 
are also represented as the two-point graphs in figure \ref{4loop2pt}, 
require a specific four-loop master integral, which was evaluated in analytic 
form in the GR context in \cite{Foffa:2016rgu} and \cite{Damour:2017ced} 
independently.}
\label{g5topo} 
\end{figure} 

However, at the 4PN order, which involves four-loop topologies at 
${\cal{O}}(G^5)$, four-loop master integrals are required 
\cite{Foffa:2013qca,Foffa:2016rgu}. The four-loop topologies at 
${\cal{O}}(G^5)$, which enter at the 4PN order using the KK metric, are shown 
in figure \ref{g5topo}. Since the computational complexity at this order goes 
considerably beyond the one-loop level, it is useful to bring up at this 
point an analogy between the graphs in our gravitational two-particle EFT and 
two-point functions of massless gauge theory in QFT \cite{Kol:2013ega}. 
Since our classical sources on the worldlines do not 
propagate, any graph at ${\cal{O}}(G^n)$ in our EFT can be mapped onto a 
($n-$1)-loop two-point function with massless internal lines, or pictorially
\begin{align}
\label{eq:2body2pt}  
\parbox{18mm}{\includegraphics[scale=0.4]{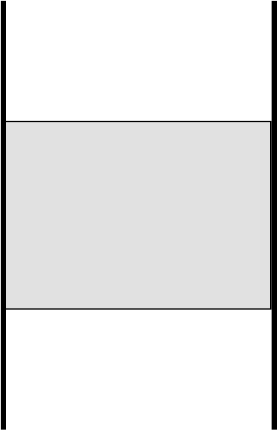}}
& \equiv \hspace{0.5cm} 
\parbox{18mm}{\includegraphics[scale=0.4]{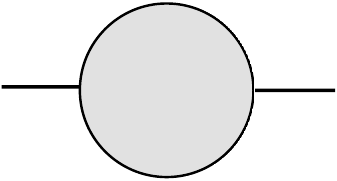}},
\end{align}
with external momentum $p$, $p^2\neq0$, which is the momentum transfer of the 
source, i.e.~just the Fourier transform momenta in the EFT. This analogy 
enables one to evaluate the graphs in figure \ref{g5topo}, using standard QFT 
multiloop techniques based on IBPs, and known four-loop master integrals.

\begin{figure}[t]
\centering
\includegraphics[width=0.3\linewidth]{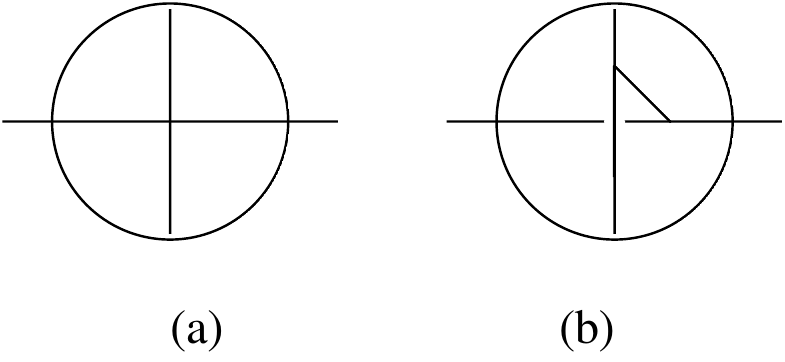}
\caption{The four-loop two-point topologies at the 4PN order using the KK 
metric, corresponding to the topologies in figure \ref{g5topo}, that 
contain the four-loop master integral represented by graph (a), which is 
evaluated in the GR context in analytic form in \cite{Foffa:2016rgu} and 
\cite{Damour:2017ced} independently: 
(a) Topology with a quartic vertex, corresponding to graph (b9) in figure 
\ref{g5topo}; 
(b) Topology with only cubic vertices, corresponding to graph (a16) in figure 
\ref{g5topo}.}
\label{4loop2pt} 
\end{figure}

As in figures \ref{g3topo} and \ref{g4topo}, the topologies in figure 
\ref{g5topo}, which are factorizable and/or nested, are easily identified. 
It is also readily recognized, that graphs (a9)-(a16) and (b7)-(b9) in figure 
\ref{g5topo} have a more complex structure, whereas the few other graphs, 
which are not trivially reducible, are reduced to one-loop integrals using IBP 
\cite{Foffa:2016rgu}. As for the complex graphs (a9)-(a16) and (b7)-(b9) in 
figure \ref{g5topo}, these are reduced, using IBP technology, to linear 
combinations of seven specific four-loop master integrals 
\cite{Foffa:2016rgu}. Of these four-loop master integrals, one, which 
actually contributes in the $d\to3$ limit, and is represented by the 
two-point graph (a) in figure \ref{4loop2pt}, was reproduced in analytic form 
as an expansion in $d-3$, and evaluated in the GR context independently in 
\cite{Foffa:2016rgu} and \cite{Damour:2017ced}. Specifically, graphs (a16) 
and (b9) in figure \ref{g5topo}, which are also represented as the two-point 
graphs in figure \ref{4loop2pt}, require this particular four-loop master 
integral. 

To conclude, the current state of the art in the conservative sector of the 
EFT of the composite particle is at the 4PN order. It has been recently 
tackled and completed in the point-mass sector in multiple works via various 
methods \cite{Blanchet:2013haa}, including the ADM canonical formalism 
\cite{Jaranowski:2012eb, Jaranowski:2013lca, Damour:2014jta,Damour:2016abl}, 
the Fokker action \cite{Bernard:2015njp,Bernard:2016wrg}, and the EFT 
approach \cite{Foffa:2012rn,Foffa:2016rgu,Damour:2017ced}, where the works 
using EFT have presented partial results up to ${\cal{O}}(G^2)$, and at 
${\cal{O}}(G^5)$. 
Ambiguities, which appeared between the results obtained via the ADM 
Hamiltonian approach, and intermediate results of the Fokker action approach, 
related to IR divergences, were further discussed in  \cite{Porto:2017dgs, 
Bernard:2017bvn}. In addition to the graph topologies shown in this section 
in figures \ref{g1topo}, \ref{g2topo}, and \ref{g3topo}-\ref{g5topo}, a 
contribution, that is non-local in time, which also enters at the 
conservative 4PN order, and originates from the leading non-linear 
radiation-reaction effect, is shown in graph (c) of figure \ref{radreaction}. 
This effect involves interaction with radiation, and hence its discussion is 
deferred to section \ref{radrg}. 

On par with the accuracy attained in the non-spinning case, the 4PN order has 
also been completed in the conservative sector with spins, 
particularly with the completion of the full LO interaction, which is quartic 
in the spins, in \cite{Levi:2014gsa}, and of the NNLO spin-squared 
interaction in \cite{Levi:2015ixa}. These interactions were exclusively 
derived via the EFT approach, and are based on \cite{Levi:2015msa}, where the 
formal developments, which led up to this current state of the art with 
spins, were described in section \ref{spinparticle}. Rather than the strictly 
computational challenge in the point-mass sector, with spins the main hurdle 
lies at the formal obtainment of an effective theory for a spinning particle, 
as discussed in section \ref{spinparticle}, on which the computations with 
the two-particle theory rely. At the 4PN order the bump in the point-mass 
sector is in the high loop level required in the computations within the 
two-particle theory, essentially, in the leap from the one-loop level, 
practically sufficient for lower PN orders, to the four-loop level. Notably, 
contrary to the point-mass sector, where the minimal coupling of the 
one-particle theory is sufficient for such high PN order, with spins the 
finite size effects from spin-induced higher multipoles, originating from 
non-minimal coupling, have to also be taken into account.

\subsection{Standard action, EOMs, and potentials}
\label{redacteomh}

After the progression presented in section \ref{oneeft} and the current 
section is carried out, in particular with the spins of the objects, we 
obtain interaction potentials, which contain \textit{only} physical DOFs, and 
higher-order time derivatives of the variables, similar to the outcome in the 
non-spinning sector. From this point on the action is handled in the standard 
manner in the PN context. Higher-order time derivatives can be eliminated via 
redefinitions of the position and spin variables already at the level of the 
action \cite{Levi:2015msa}. The EOMs of the position and spin variables are 
then directly obtained via a proper variation of the action 
\cite{Levi:2014sba,Levi:2015msa}. Furthermore, Hamiltonian potentials are 
derived in a straightforward manner with a standard Legendre transform of the 
position variables \cite{Levi:2015msa}. Alternatively, if higher-order time 
derivatives are not treated at the level of the action, one can also derive 
the EOMs first via a proper variation of a generalized action. In this case 
the higher-order time derivatives can be equivalently removed at the level of 
the EOMs, so that the EOMs are eventually well-defined \cite{Levi:2015msa}. 

Indeed, in the PN perturbative scheme the resulting action contains higher-order 
time derivatives, such as $\vec{a}_1\equiv\dot{\vec{v}}_1$. A common 
procedure to reduce the action to a standard one is to eliminate the 
higher-order time derivatives via their substitution, using EOMs of lower PN 
orders, at the level of the action. Such a substitution entails a redefinition of 
variables \cite{Barker:1980,Schafer:1984mr,Damour:1985mt,Damour:1990jh}, 
which can easily be seen, considering that a small variable shift results in 
a variation of the action, which exactly yields the EOM of the variable. In 
fact, the same idea applies in any perturbative context, and we note the 
analogy to the concept of redundant operators in EFTs \cite{Levi:2014sba}, 
which vanish on-shell, as was also noted in section \ref{pointmass}. Let us 
review the procedure for eliminating higher-order time derivatives of position 
variables, and its extension for a spin variable, which was provided in 
\cite{Levi:2014sba}. 

A shift of the position, $\Delta x$, adds to the Lagrangian the EOM 
multiplied by $\Delta x$, where higher orders in the shift are neglected. 
Thus, it is clear that we can set the shift to cancel terms with 
acceleration in the Lagrangian. This is precisely what we would get from 
substituting the EOM directly, with the notable subtlety, that there are also 
contributions, which are non-linear in the shift, 
$\Delta L({\cal{O}}(\Delta x)^2)$. These contributions will enter at higher 
perturbative orders, and thus one should keep track of them. For further 
higher-order time derivatives the procedure should be iterated until all 
higher-order time derivatives are eliminated. As for the rotational 
variables, the Lorentz rotation matrix, $\Lambda^{ij}$, is transformed with a 
generator of rotations, $\omega^{ij}$, and the redefinitions of the spin 
variables are specified so as to rotate similarly to the rotation matrices. 
Then, similarly as with the position variables, we can fix the rotation generator 
to cancel the terms with a time derivative of the spin, $\dot{S}^{ij}$, so that 
the shift in the Lagrangian is just what we would get from substituting in the 
spin EOMs, up to higher-order contributions in the rotation generator. Again, 
one should keep a record of these non-linear contributions, which will affect 
higher PN orders.

For the EOMs of the spin variables, we should consider the spin-dependent 
part of the full action, recalling the rotational kinetic term 
\cite{Levi:2010zu}, so that the action has the following form:
\be
S_{\text{eff(comp)}}=
\int dt \left[ 
\sum_{a=1}^2\frac{m_a}{2} v_a^2
-\sum_{a=1}^2\frac{1}{2}[S_a]_{ij}[\Omega_a]_{ij}
-V\left(\vec{x}_a,\dot{\vec{x}}_a, \ddot{\vec{x}}_a, \dots, 
[S_a]_{ij}, [\dot{S}_a]_{ij}, \dots \right)\right]. 
\ee
Such a form for the action is associated with a canonical so$(3)$ algebra for 
the physical spin variable. Varying this action independently with respect to 
the spin, and to its conjugate, the angular velocity, the following EOMs for 
the three-dimensional physical spin variable are obtained:
\begin{equation} 
\dot{S}_a^{ij} 
= -4 S_a^{k[i} \delta^{j]l} \frac{\delta\int{dt\,V}}{\delta S_a^{kl}} 
= -4 S_a^{k[i} \delta^{j]l} \left[ \frac{\partial V}{\partial S_a^{kl}} 
- \frac{d}{dt} \frac{\partial V}{\partial \dot{S}_a^{kl}} + \dots \right]. 
\label{S3EOM}
\end{equation}
Therefore, similar to the EOMs for the position, the EOMs for the spin should 
be obtained from a variation of the action, in terms of which the EFT is 
naturally formulated \cite{Levi:2014sba,Levi:2015msa}.

After time derivatives of the spin (and velocity) have been removed, the EOMs 
of the spin take on the following simple form:
\begin{equation}
\dot{S}_a^{ij}
= - 4 S_a^{k[i} \delta^{j]l} \frac{\partial V_s}{\partial S_a^{kl}},
\label{canSeom}
\end{equation}
where $V_s$ denotes the potential reduced to standard form, and this is a 
restricted case of equation \eqref{S3EOM}. If we impose on the spin variables the 
equal-time Poisson brackets, namely 
\begin{equation}
\{ S_a^{ij}, S_a^{kl} \} = S_a^{ik} \delta^{jl} - S_a^{il} \delta^{jk}
+ S_a^{jl} \delta^{ik} - S_a^{jk} \delta^{il},
\end{equation}
then we can rewrite the EOMs in equation \eqref{canSeom} in the following form:
\begin{equation}
\dot{S}_a^{ij} = \{ S_a^{ij}, -V_s \}.
\end{equation}
Therefore, after the elimination of time derivatives of the spin, one arrives 
at canonical spin variables.

Finally, the conservative action of the composite particle is invariant under 
global Poincar\'e transformations. This invariance gives rise to conserved 
integrals of motion (see, e.g.~in \cite{Levi:2016ofk}), including the 
Hamiltonian, which is crucially important as an input for the EOB framework. 
Since the higher-order time derivatives have been eliminated, we can perform 
a standard Legendre transform with respect to the position variables to 
obtain the Hamiltonian directly from the effective interaction potential. 
Related to that we note that physically equivalent Hamiltonians are obtained 
using canonical transformations \cite{Goldstein}, similar to the addition of 
total time derivatives to obtain equivalent Lagrangians.

\section{Effective theory of dynamical multipoles} 
\label{togwobs}

In this section we are concerned with the final stage of our EFT scheme, 
dealing with effects that take place at the radiation level, and completing  
the connection of the whole EFT framework to GW emission observables. 

In section \ref{radrg} we start by considering the EFT of the composite 
particle in the radiative sector, i.e.~where radiation modes are present, 
which was first treated in detail in \cite{Goldberger:2009qd}. As the 
bottom-up approach was taken to set up the theory in equation \eqref{seffcomp}, we 
first discuss in detail the matching of the multipole moments, i.e.~the 
Wilson coefficients, of the EFT of the composite particle at the radiation 
scale from the full two-particle theory at the orbital scale. We then present 
hereditary tail effects, which were reproduced up to subleading corrections 
in \cite{Goldberger:2009qd}. The various singularities which arise in these 
tail effects are noted, of which a classical RG flow of the effective theory 
emerges for the mass quadrupole. 

We take time to introduce further advanced QFT concepts and tools, required 
at this stage, as the radiating system is dissipative, and time reversal no 
longer holds for the EFT in the radiative sector. Thus we present the closed 
time path (CTP) formalism \cite{Schwinger:1960qe, Keldysh:1964ud, 
Calzetta:2008iqa}, and then we can proceed, via the top-down approach, to 
integrate out the radiation modes from the CTP effective action. We go on to 
present the analysis of the RG flow of the total mass of the system as it
was recovered in \cite{Goldberger:2012kf}, which is related with the leading 
tail effect and is also manifested in the leading non-linear radiation 
reaction. Thus, the progress made within the EFT approach in the treatment of 
the radiation reaction, which was initiated in \cite{Galley:2009px, 
Galley:2012qs}, is then presented. 

We end in section \ref{eftofpngcode} with an overview of the `EFTofPNG' 
code, which constitutes the first public comprehensive automatization of PN 
theory, and also covers the whole of the EFT framework, as outlined in this 
review, from section \ref{oneeft} to the current section. The code provides 
high precision computation for PN inspirals, so as to ultimately serve as an 
accessible pipeline to GW observables for the diverse gravity community, and 
in particular the GW community.

\subsection{Radiation, radiation reaction, and RG flows} 
\label{radrg}

The EFT of the composite particle in the presence of radiation modes was 
first discussed in \cite{Goldberger:2004jt,Goldberger:2007hy}, and formulated 
in \cite{Goldberger:2009qd}, using the bottom-up approach. The latter work 
studied tail effects, which constitute radiative PN corrections, up to the 
subleading ones, and the related divergences that arise. This analysis led to 
the first demonstration of a classical RG flow in the theory, of the mass 
quadrupole. Moreover, the matching of the theory at the radiation scale, 
i.e.~of the multipole moments, to the full two-particle theory at the orbital 
scale was formulated and displayed up to NLO, and the effect of the tails on 
the emitted power was also derived. Furthermore, in \cite{Ross:2012fc} a 
general multipole expansion at the level of the action provided general 
expressions for the mass and current multipole moments, the radiated power, 
and the gravitational waveform amplitude. 

Let us proceed to outline the primary developments made in 
\cite{Goldberger:2009qd}. We begin by once again building on the bottom-up 
approach presented in equation \eqref{seffcomp}. We assume that the orbital 
separation has been integrated out, and can no longer be resolved, and we 
write down an effective action, which contains the appropriate DOFs and 
satisfies the proper symmetries at the radiation scale. This theory is 
actually similar to that which we encountered in the one-particle EFT in 
section \ref{oneeft}, only that the remaining field modes at this stage are the 
radiation modes, and we now assume higher-order multipole moments, $l\ge2$, 
which are sourcing the radiation. In fact, as our system is radiating, time 
reversal is lost, and later on we address this crucial point for the 
obtainment of the proper effective action. For now we start by further 
specifying the theory from equation \eqref{seffcomp}.

We recall that the effective action of the single composite object coupled to 
the gravitational radiation field, 
$\widetilde{g}_{\mu\nu}\equiv\eta_{\mu\nu}+\widetilde{h}_{\mu\nu}$, reads 
\cite{Goldberger:2009qd,Ross:2012fc}:
\be \label{seffcomp1}
S_{\text{eff(comp)}}[\widetilde{g}_{\mu\nu},y_c^\mu,e_{c\,A}^{\,\,\mu}]
=-\frac{1}{16\pi G}\int d^4x
\sqrt{\widetilde{g}}\,R\left[\widetilde{g}_{\mu\nu}\right] 
\,+\, S_{\text{pp(comp)}}
[\widetilde{g}_{\mu\nu}(y_c),y_c^\mu,e_{c\,A}^{\,\,\mu}](\sigma_c),
\ee
where $S_{\text{pp(comp)}}$ is the effective worldline action at the 
radiation scale, describing the composite object with some `center of the 
object' coordinate, $y_c^\mu(\sigma_c)$, and tetrad, 
$e_{c\,A}^{\,\,\mu}(\sigma_c)$. The worldline action of the composite 
particle is then of the following form \cite{Goldberger:2009qd,Levi:2010zu, 
Ross:2012fc}: 
\begin{align}\label{sppcomp}
S_{\text{pp(comp)}}[\widetilde{h}_{\mu\nu},y_c^\mu,e_{c\,A}^{\,\,\mu}](t)=
-\int dt \sqrt{\widetilde{g}_{00}}\,\biggr[ & M(t)
+\frac{1}{2}\epsilon_{ijk}L^{k}(t)\left(\Omega_{\text{LF}}^{ij}
+\omega_\mu^{ij}u^\mu\right) \nn\\
&-\sum_{l=2}^{\infty}
\left(\frac{1}{l!}I^{L}(t)\nabla_{L-2}E_{i_{l-1}i_{l}}\right.\nn\\
&-\left.\frac{2l}{(l+1)!}J^{L}(t)\nabla_{L-2}B_{i_{l-1}i_{l}}\right)\biggr],
\end{align}
where also here the worldline is parametrized in terms of the time 
coordinate, $t$. $M$ here is the total mass of the composite object, and the 
spin connection, $\omega_\mu^{ij}$, couples here to the \textit{total} 
angular momentum, $L^{ij}\equiv\epsilon_{ijk}L^{k}$, of the composite object. 
The $SO(3)$ tensors $I^L$ and $J^{L}$, with the superscript $L$ as an 
abbreviated notation for $i_1\cdots i_l$ ($l\geq2$), which are symmetric and 
trace free (STF) with respect to the spatial Euclidean metric, are commonly 
referred to as the mass and current multipoles, and are of even and odd 
parity, respectively.

These moments constitute the full set of higher multipole moments of the 
composite object, similar to the induced higher multipoles in the non-minimal 
coupling in the effective action for a spinning particle, constructed in 
sections \ref{pointmass} and \ref{spinparticle}. However, since 
these DOFs, localized on the worldline, encode the internal physics of the 
composite particle, which yields the dissipative radiation, similar to 
the dissipative DOFs in section \ref{dissipat}, they are dynamical. In 
particular, the $l=0,2$, multipole moments of the composite system, i.e.~the 
total mass, $M(t)$, and the mass quadrupole, $I_{ij}(t)$, are the first two 
`Wilson coefficients' of the theory which exhibit classical RG flows, 
related with well-known tail effects, as discussed below. As the sources 
of gravitational radiation, the multipole moments are coupled to the radiation 
modes, $\widetilde{h}_{\mu\nu}$, where $E_{i_{l-1}i_{l}}$, 
$B_{i_{l}i_{l+1}}$, are as in equations \eqref{eq:E} and \eqref{eq:B}, the electric 
and magnetic projections of the Riemann tensor onto its even and odd parity 
components, respectively. 

At this stage we are concerned with expectation values, and hence we proceed to 
consider an effective action for the composite particle, which is a 
functional of the radiation mode, regarded as a mean field, serving as a 
fixed background, using the background field method \cite{Goldberger:2004jt, 
Peskin:1995ev,Calzetta:2008iqa}. In this case the potential modes play the 
role of the fluctuating `quantum' field, as the integration variable of the 
functional integral in the computation of the effective action, and the 
radiation modes serve as the `classical' fixed background. The background 
field method reflects the spirit of Wilson's idea of integrating out the high 
momentum DOFs, while taking proper care to preserve the gauge invariance of 
the remaining low momentum modes, when we compute the effective action for a 
slowly varying background field. Therefore, the gauge-fixing in the 
background field method is modified in the radiative sector with respect to 
the conservative sector.  Thus, in equation \eqref{sgravpure} covariant 
derivatives with respect to the background radiation field are used; this 
preserves the coordinate invariance with respect to the radiation field. The 
gauge-fixing condition is covariant with respect to the background field, so 
that when we functionally integrate over the fluctuating field to compute the 
effective action, the result still has general covariance of the background 
field.

Let us briefly illustrate what the reserved term `effective action' used here 
actually designates in order to avoid ambiguity with the various 
effective actions, synonymous with effective theories, which are implied 
throughout \cite{Peskin:1995ev}. Let us recall, for example, that for a QFT 
of a scalar field, $\phi$, in the presence of an external source, $J$, the 
generating functional, $Z[J]$, is defined as follows:
\begin{align}
Z[J] &\equiv \text{exp}^{iW[J]}\nn\\
 & \equiv \int {\cal{D}}\phi \,\,\text{exp}\left[i\int d^4x 
\left[{\cal{L}}[\phi]+J(x)\phi(x)\right]\right],
\label{genfunc}
\end{align}
where $Z[J]$ is the basic object of the functional formalism, and is the 
generating functional of correlation functions, and 
$W\equiv-i\,\text{log}\,Z$ is the generating functional of connected 
correlation functions. Thus, the functional derivative of the exponent with 
respect to the external source yields the one-point function of the field, as 
follows:
\be
\frac{\delta}{\delta J(x)} W[J] = 
\langle\phi(x)\rangle\equiv \phi_{\text{mean}}(x),
\ee
where we define the one-point function as a mean field. Now, we can proceed 
to obtain the exact one-point function, $\langle \phi \rangle$, in the full 
quantum theory as the minimum of some function, as if our field is still 
only classical. We just define the Legendre transform of $W[J]$ as follows:
\be
\Gamma[\phi_{\text{mean}}]\equiv W[J]
- \int d^4y \,J(y) \phi_{\text{mean}}(y),
\label{EA}
\ee
where $\Gamma[\phi_{\text{mean}}]$ is known as the `effective action', 
which satisfies
\be
\frac{\delta}{\delta \phi_{\text{mean}}(x)} \Gamma[\phi_{\text{mean}}]=-J(x),
\ee
and then if the external source is set to vanish, the effective action 
satisfies 
\be
\frac{\delta}{\delta \phi_{\text{mean}}(x)} \Gamma[\phi_{\text{mean}}]=0.
\ee
Then, the solutions to this equation are the values of the exact one-point 
function, $\langle \phi(x)\rangle$, in the full theory.

Now, in order to match for the EFT of the composite particle with multipole 
moments of equations \eqref{seffcomp1} and \eqref{sppcomp}, the specific form of 
the multipole moments should be computed, given the details of the short distance 
gravitational physics of the binary system, which makes up the internal 
structure of the composite particle \cite{Goldberger:2009qd}. This is a 
standard procedure in dealing with EFTs, and is carried out explicitly in 
this case. In order to do the matching, it is sufficient to consider the 
single graviton emission amplitudes in the full theory at the orbital scale. 
These amplitudes arise in terms of the following term in the effective 
action: 
\be \label{oneptrad}
\Gamma[\widetilde{h}_{\mu\nu}]=-\frac{1}{2}
\int d^4x \,T^{\mu\nu}\widetilde{h}_{\mu\nu},
\ee
where $T^{\mu\nu}$ is the flat energy-momentum pseudotensor of matter plus 
gravity, which interacts with the background radiation field. It satisfies 
the conservation law, $\partial_\mu T^{\mu\nu}(x)=0$, due to the Ward 
identities for graviton amplitudes. 

The multipole expansion is then generated by taking the long wavelength 
limit, $|\vec{k}|\to 0$, or in position space by expanding the radiation 
field as follows:
\be \label{fieldmult}
\widetilde{h}_{\mu\nu}(x)=\sum_{n=0}^{\infty} 
\frac{1}{n!}\vec{x}^{i_1}\cdots\vec{x}^{i_n}
\partial_{i_1}\cdots\partial_{i_n}\widetilde{h}_{\mu\nu}(x^0,\vec{0}),
\ee 
where the point $\vec{x}=\vec{0}$ is taken to be the center of the composite 
object. This expansion is truncated at any finite PN order, as we have that 
$\vec{x}\cdot\vec{\nabla}\sim r/\lambda\sim v$. Substituting this expansion 
into equation \eqref{oneptrad}, and decomposing the moments,
$\int d^3x \,T^{\mu\nu}\vec{x}^{i_1}\cdots\vec{x}^{i_n}$, into irreducible 
representations of the rotation group, yields the classical multipole 
moments, which appear in equation \eqref{sppcomp}. For example, at LO we have the 
total energy of the isolated binary system, $M=\int d^3x\,T^{00}$, which is 
the mass monopole term in equation \eqref{sppcomp}; at the next order in the 
expansion we find the total angular momentum, 
$L^{ij}=-\int d^3x\,(T^{0i}x^j-T^{0j}x^i)$. At the next order in the 
velocity, $v$, we then find for the mass (or `electric') quadrupole
\be
I^{ij}=\left[\int d^3x \,(T^{00}+T^{kk}) x^ix^j\right]_{TF},
\ee
where TF denotes the traceless part of the tensor. 

\begin{figure}[t]
\centering
\includegraphics[width=0.45\linewidth]{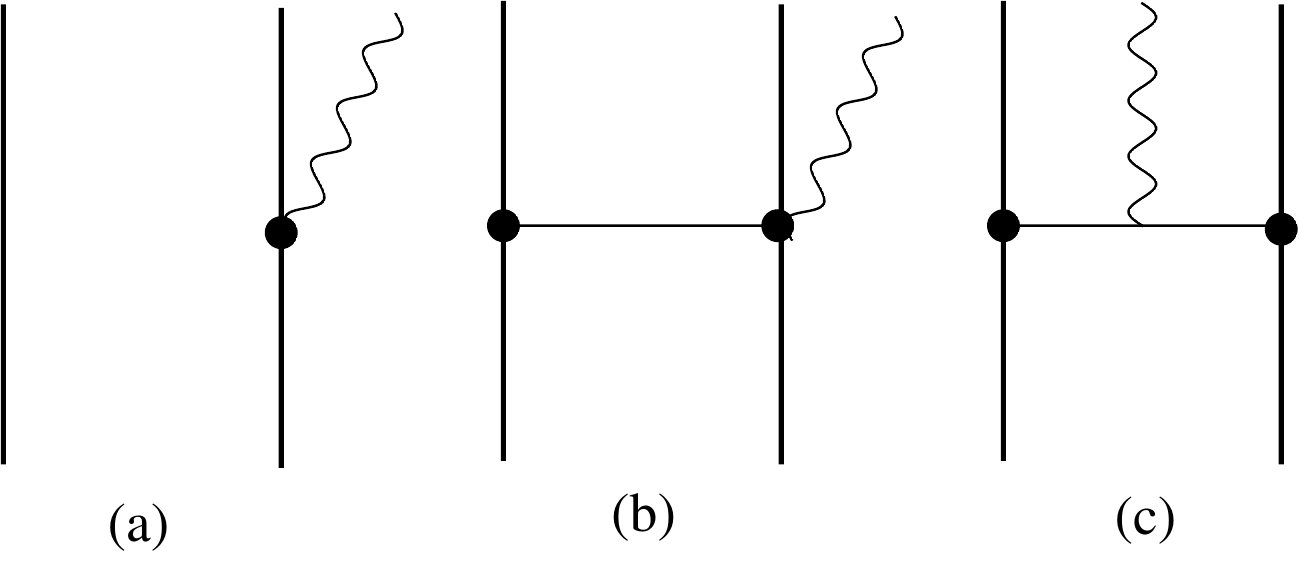}
\caption{The Feynman diagrams that contribute up to the 1PN order of the mass 
quadrupole, $I^{ij}$ \cite{Goldberger:2004jt,Goldberger:2009qd}. 
Unlike the conservative sector of the two-particle EFT, which contains graphs 
with internal potential gravitons only, and yields interaction potentials, as 
in figures \ref{01pn}-\ref{los4} (and in the topologies of figures 
\ref{g1topo}, \ref{g2topo}, and \ref{g3topo}-\ref{g5topo}), in the radiative 
sector we consider diagrams with external radiation modes. The latter are 
represented here by wiggly lines, whereas hereafter, all orbital modes are 
represented by solid lines. Contributing to the leading orders of the 
energy-momentum pseudotensor amplitude are the following diagrams: 
(a) Contribution to the leading $T^{\mu\nu}$ and subleading $T^{00}$ 
component; 
(b) Contribution to the subleading $T^{00}$ component;
(c) Contribution to the subleading $T^{00}$ and leading $T^{ij}$ components. 
Diagram (c), which contains cubic self-interaction, is omitted when using the 
KK parametrization \cite{Birnholtz:2013nta}, similar to the 1PN order 
correction in the conservative sector; see figure \ref{01pn}.}
\label{matchrad} 
\end{figure}

The energy-momentum pseudotensor, $T^{\mu\nu}$, which is essentially the 
off-shell emission amplitude, should thus be computed in terms of the orbital 
DOFs. It is computed as a sum of Feynman graphs with internal potential 
gravitons and a single external (off-shell) radiation graviton, so that the 
potential modes are integrated out as in equation \eqref{seff2part} and in section 
\ref{feynman}. Figure \ref{matchrad} shows the graphs that contribute up to 
the 1PN order of the mass quadrupole, $I_{ij}$. To compute these diagrams one 
needs to have the Feynman rules of the propagators of the potential modes as 
in section \ref{feynman}, but also of the worldline couplings and 
self-interaction vertices, which contain modes of the two types, potential 
and radiation, such as $mH\widetilde{h}$ on the worldline, or the cubic bulk 
vertex $\widetilde{h}H^2$. The latter would be extracted from the purely 
gravitational action with the harmonic gauge fixing, using the background 
field method \cite{Goldberger:2004jt}.

After this matching of the multipole moments is done, one can proceed to 
compute the PN corrections to the radiation observables, such as the GW 
energy flux (or emitted power) or GW amplitude. In \cite{Goldberger:2009qd} 
it was demonstrated how to undertake this matching of the multipole moments at the 
1PN order as shown in figure \ref{matchrad}, and subsequently the 1PN order 
correction for the emitted power was reproduced. In \cite{Ross:2012fc} a 
general multipole expansion of the action linear in the radiation field 
specified the mass and current multipole moments to all orders in terms of 
the energy-momentum pseudotensor, and consequently the radiated power and 
the radiation amplitude. In \cite{Birnholtz:2013nta,Birnholtz:2014fwa} the 
1PN order correction to all mass multipoles was explicitly derived. In 
addition, multipole moments at NLO, dependent on the spins of the compact 
constituents, were tackled in \cite{Porto:2010zg,Porto:2012as}.

\begin{figure}[t]
\centering
\includegraphics[width=0.45\linewidth]{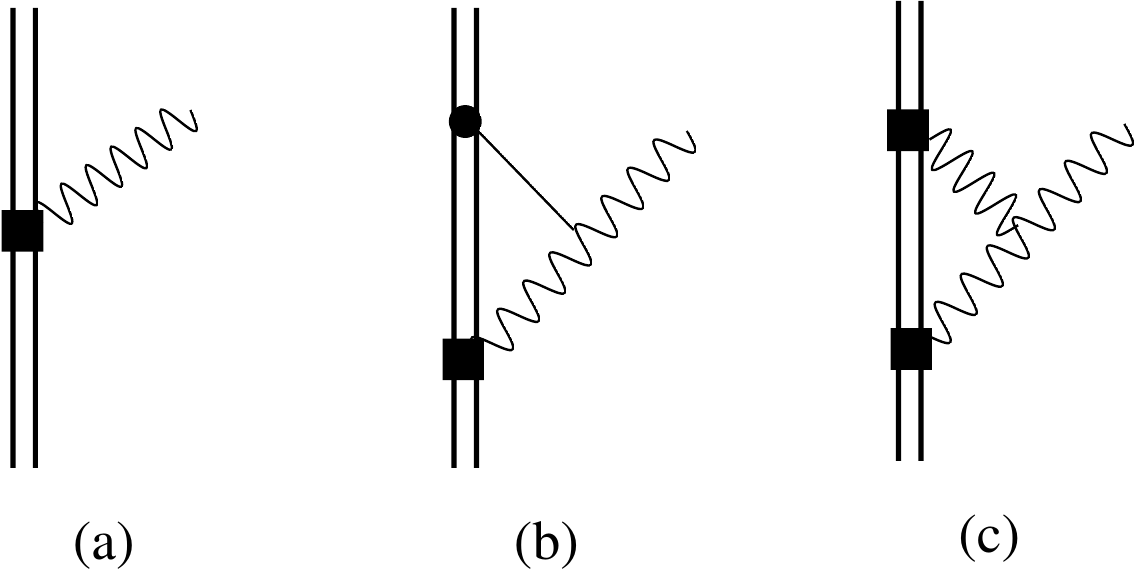}
\caption{Radiation and hereditary effects in the EFT of the composite 
particle. Here the black square boxes and the sphere represent the 
time-dependent mass quadrupole, $I_{ij}$, and total mass monopole, $M$, 
moments of the binary, respectively: 
(a) The diagrams in figure \ref{matchrad} match onto this diagram in the 
effective theory of the composite particle, which depicts the quadrupole 
radiation \cite{Goldberger:2004jt};
(b) The leading tail effect, which enters at the relative 1.5PN order. The 
subleading corrections at the relative 3PN order, as will be shown in figure 
\ref{nlotail}, were also reproduced in \cite{Goldberger:2004jt, 
Goldberger:2009qd};
(c) The leading memory effect, which enters at the relative 2.5PN order, 
related with the LO radiation reaction; see figures \ref{quadrad}(b2), 
\ref{radreaction}(a) in the following \cite{Galley:2009px,Goldberger:2012kf}.}
\label{tailmemory} 
\end{figure}

\begin{figure}[t]
\centering
\includegraphics[width=0.45\linewidth]{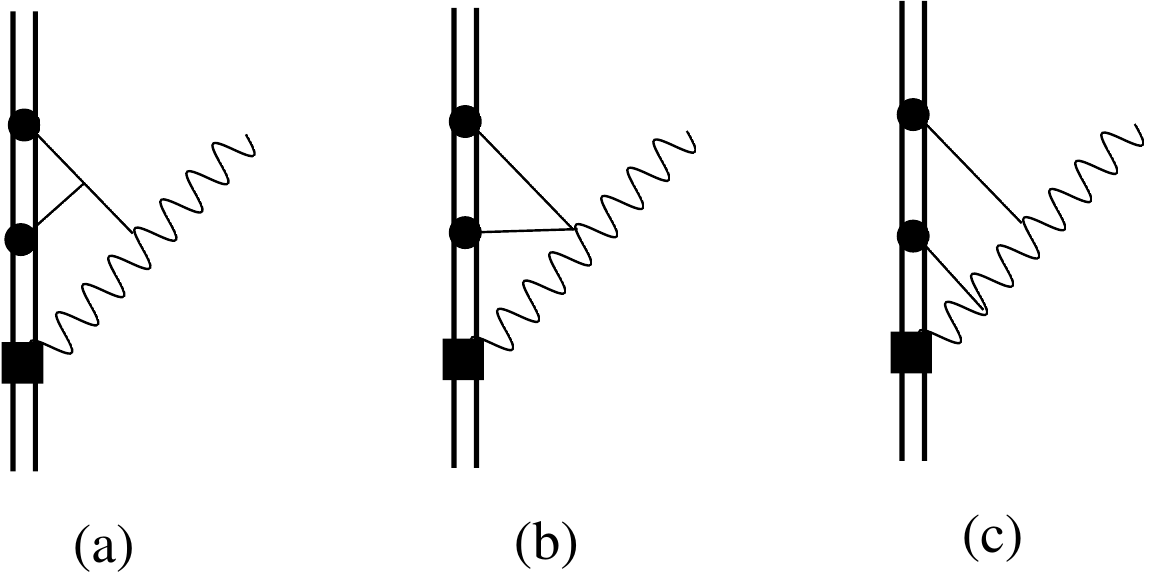}
\caption{Radiation and subleading tail effects in the EFT of the composite 
particle. The NLO corrections to the tail effect in figure 
\ref{tailmemory}(b), which enter at the relative 3PN order, are reproduced in 
\cite{Goldberger:2004jt,Goldberger:2009qd}. All of the graphs are 
logarithmically UV divergent, related with the classical RG flow of the mass 
quadrupole moment, $I_{ij}$, from the orbital to the radiation scales 
\cite{Goldberger:2009qd}.} 
\label{nlotail} 
\end{figure}

The linear coupling of radiation gravitons to the multipole moments, e.g.~to 
the mass quadrupole, as shown in figure \ref{tailmemory}(a), has non-linear 
corrections due to gravitational self-interactions, as shown in figures 
\ref{tailmemory}, \ref{nlotail}, commonly referred to as hereditary effects. 
Not only are these relevant PN corrections, but they also involve logarithmic 
singularities, related with classical RG flows of the multipole moments from 
the orbital to the radiation scales \cite{Goldberger:2009qd}. We recall that 
the relevant expansion parameter of the EFT of a composite particle is 
$r/\lambda\sim v$, which just amounts to a multipole expansion, where the 
radiation sourced by the $l$-pole moment is suppressed by a relative factor 
of $v^l$. However, tail effects at the radiation scale, which arise from 
non-linear interaction with the gravitational potential binding the composite 
particle, scale as $GM/\lambda\sim v^3$. Hence, the leading non-linear 
correction in figure \ref{tailmemory}(b), which shows the scattering of 
outgoing radiation gravitons from potential gravitons sourced by the total 
mass of the system, enters at the relative 1.5PN order. 

Long distance infrared (IR) logarithmic singularities already appear at the 
leading non-linear order, but they are eventually seen to explicitly cancel out 
from physical observables \cite{Goldberger:2009qd}. However, the 
subleading non-linear corrections to the tail effect in figure 
\ref{tailmemory}(b), which are shown in figure \ref{nlotail}, also involve 
logarithmic short distance UV singularities. These corrections enter at the 
relative 3PN order and are dubbed the `tail of tail'  \cite{Goldberger:2009qd}. 
The UV singularities concerned represent a real UV divergence of the EFT of the 
composite particle, and involve renormalization of the theory. All these UV 
divergences can be rendered finite by a renormalization of the multipole moments. 
The renormalized moments exhibit scale dependence described by a renormalization 
group (RG) equation. This RG equation generates the entire series in powers 
of $v^3$ of leading UV logarithms. This set of terms is generic, that is, 
independent of the short distance physics of the composite particle. It is a 
prediction of the RG equation that this series of terms appears in long 
wavelength gravitational radiation from any compact binary system.

These subleading tail corrections and logarithms were reproduced and further 
understood via the EFT approach in \cite{Goldberger:2009qd}, and amount to a 
two-loop computation, where graphs (a) and (b) of figure \ref{nlotail} in 
fact easily reduce to nested one-loop computations. A scale dependence for a 
renormalized moment was then introduced, in order to compensate for the 
explicit scale dependence of the logarithm, through the following RG equation 
for the mass quadrupole moment:
\be \label{rgq}
\mu\frac{d}{d\mu}I_{ij}(\omega,\mu)=
-\frac{214}{105}\left(GM\omega\right)^2I_{ij}(\omega,\mu).
\ee
This equation then has the following solution:
\be \label{losol}
I_{ij}(\omega,\mu)=\left(\mu/\mu_0\right)^{-\frac{214}{105}
\left(GM_0\omega\right)^2} I_{ij}(\omega,\mu_0),
\ee
where the scale $\mu_0$ is taken as the characteristic UV scale of the 
composite object size, that is, $\mu_0\sim1/r$, where in the PN regime it 
holds that $\mu/\mu_0\sim v$. Thus we have a running of the theory of the 
composite particle between the scales $r$ and $r/v$, and a classical RG flow 
of physical observables, where the RG analysis is universal, and applies to 
gravitational radiation from any localized composite particle. The RG flow is 
shown here for the quadrupole moment but would apply similarly to further 
multipole moments. However, let us suspend the analysis of the subleading RG flow 
of the total mass of the composite particle 
\cite{Goldberger:2012kf,Galley:2015kus}, 
in order to introduce an essential formal progression for this stage. 

As we noted the radiating system is dissipative, as there is no incoming 
radiation at the asymptotic past, and one should proceed to impose boundary 
conditions, which are not symmetric in time, in order to obtain real time 
observables such as the gravitational waveform or radiation reaction 
effects. Using the effective action of the composite particle presented in 
equations \eqref{seffcomp1} and \eqref{sppcomp}, and the consequent standard Feynman 
propagators, which account for a pure in-(out-)going wave at past (future) 
infinity, would lead to a non-causal evolution. The related in-out generating 
functional, $Z$, as in equation \eqref{genfunc}, which is useful for the 
calculation of scattering amplitudes, yields matrix elements rather than the 
real time expectation values. Further, the in-out effective action, $\Gamma$, as 
in equation \eqref{EA}, generates EOMs for the worldline DOFs, which are in 
general neither real nor causal as they should be \cite{Galley:2008ih, 
Galley:2009px,Galley:2012hx}. In order to consistently implement retarded 
boundary conditions, and proceed to use retarded propagators when integrating 
out the radiation field, we need to construct an action in which time 
reversal is broken. In fact, there exists a consistent framework in QFT to 
define such an action to describe non-conservative evolution, known as the 
\textit{closed time path} (CTP; or the `in-in', or Schwinger-Keldysh) 
formalism \cite{Schwinger:1960qe,Keldysh:1964ud}, commonly used in 
non-equilibrium QFT \cite{Calzetta:2008iqa}. 

In the CTP formalism the time contour in the path integral formulation goes 
from the asymptotic past hypersurface to a constant hypersurface at some 
possibly finite time, and back to the asymptotic past on another time branch. 
Therefore in the CTP action the actions of two different histories of the 
DOFs are compared by computing their difference, where in the first history 
there is forward time evolution, and the second history evolves backward 
through time from its final to initial configuration. It is important 
to stress that the values of the DOFs at the final time are set to be equal, which 
is referred to as the `equality condition' (or the `CTP boundary condition'), 
but their value is not fixed. Hence, the CTP formulation is an initial 
value formulation, as opposed to the in-out formulation, which is a boundary 
value formulation, where the 
paths go from the asymptotic past to the asymptotic future. However, in both 
formulations there is only one relevant physical history, so the two 
histories in the CTP formulation are both set equal to the physical one, 
after taking the variations of the CTP action, which is referred to as the 
`physical limit'. 

Let us review the CTP formalism to gain a more concrete understanding of how it 
works, where here we consider its quantum formulation, and assume that it is 
eventually applied in the classical limit. As one would expect, the latter 
limit is found to be equivalent to various independent classical 
formulations, which were made for non-conservative evolution 
\cite{Galley:2012hx,Galley:2014wla,Birnholtz:2013nta}, yet here we want to 
avoid restricting the discussion to the specific classical context, but 
rather to keep it on a more generic fundamental level. Given a general action 
for a scalar field theory, $S[\phi]$, the DOFs are doubled, and the following 
action is defined as follows:
\be
S_{\text{CTP}}[\phi_1,\phi_2]\equiv S[\phi_1]-S[\phi_2]^*.
\ee
The CTP generating functional of correlation functions of such a scalar field 
theory, coupled to a source $J_1$ during the forward evolution in time and 
a source $J_2$ for the evolution backward in time, is defined by the 
following path integral: 
\begin{align}
Z_{\text{CTP}}[J_1,J_2]
&\equiv \text{exp}^{iW_{\text{CTP}}[J_1,J_2]}\nn\\
&\equiv \int {\cal{D}}\phi_1{\cal{D}}\phi_2\, 
\text{exp} \left[i S_{\text{CTP}}[\phi_1,\phi_2]
+i \int d^{4}x\,\left(J_1\phi_1-J_2\phi_2\right)\right],
\end{align}
with the equality condition, $\lim \limits_{t\to t_f}\phi_1(x)-\phi_2(x)=0$, 
at some final time, $t_f$.

Two mean fields are considered to be derived from the following 
variations: 
\begin{align}
+\frac{\delta}{\delta J_1(x)}W_{\text{CTP}}=\langle\phi_1(x)\rangle,\quad
-\frac{\delta}{\delta J_2(x)}W_{\text{CTP}}=\langle\phi_2(x)\rangle,
\end{align}
after which one sets $J_1=J_2=0$, and then 
$\langle \phi_1 \rangle = \langle \phi_2 \rangle\equiv\phi_{\text{mean}}$, 
which amounts to the physical constraint, that $\phi_1=\phi_2$ on-shell.
We can also introduce the CTP effective action as the Legendre transform of 
$W_{\text{CTP}}$:
\be
\Gamma_{\text{CTP}}[\langle\phi_{1}\rangle,\langle\phi_{2}\rangle]
=W_{\text{CTP}}[J_1,J_2]
-\int d^4x\left[J_1\langle\phi_1\rangle-J_2\langle\phi_2\rangle\right],
\ee 
and for the mean fields we have the equations
\begin{align}
\frac{\delta\Gamma_{\text{CTP}}}{\delta\langle\phi_1\rangle}=-J_1,\quad
\frac{\delta\Gamma_{\text{CTP}}}{\delta\langle\phi_2\rangle}=+J_2, 
\end{align}
where we set $J_1=J_2=0$ to obtain the EOMs for the physical mean field, 
$\phi_{\text{mean}}$, and the two equations turn out to be equivalent.

The propagators between $\phi_A$ and $\phi_B$, where $A, B = 1, 2$ 
denote the time branch within the time contour, are ordered in the following 
matrix: 
\begin{align}\label{gcpt}
G_{AB}=
\left(
\begin{array}{cc}
 G_F & -G_-\\
-G_+ & [G_F]^*
\end{array} \right),
\end{align}
where $G_F$ is the Feynman propagator given by
\be 
G_{F}(x, x')
=\theta(t-t')\Delta_+(x, x')+\theta(t'-t)\Delta_-(x, x'),
\ee
and the propagators $G_{\pm}\equiv\Delta_{\pm}$ read as follows:
\be
\Delta_{\pm}(x, x')=
\int\frac{d^3\vec{k}}{(2\pi)^3} \,e^{\pm ik_0(t-t')} 
\frac{e^{-i\vec{k}\cdot(\vec{x}-\vec{x}')}}{2k_0}.
\ee
So $G_{AB}$ in equation \eqref{gcpt} stands for the `path-ordered' propagators: 
Path-ordering is equivalent to time-ordering for points on the first time branch, 
anti-time-ordering on the second time branch, and placing points on the second 
branch to the left of points on the first branch.

It is useful to change the basis to the Keldysh representation by rewriting the 
sources and fields in terms of their average and the relative difference between 
the two time branches, namely $J_+\equiv(J_1+J_2)/2$, $J_-\equiv J_1-J_2$, 
and $\phi_+\equiv(\phi_1+\phi_2)/2$, $\phi_-\equiv\phi_1-\phi_2$. The `$+$' 
field variable evolves forward in time, and satisfies the initial conditions, 
whereas the `$-$' field variable evolves backward in time with the equality 
condition at the final time. When taking the physical limit, $\phi_+\to\phi$, 
i.e.~the average of the two histories is identified with the physical one, 
while their difference vanishes, $\phi_-\to0$. The propagator matrix in the 
Keldysh basis, where now $A, B$ take the $+, -$ labels, is then given by
\begin{align}\label{propsctp}
G_{AB}=
\left(
\begin{array}{cc}
0 & -iG_{\text{adv}}\\
-iG_{\text{ret}} & \tfrac{1}{2}G_H
\end{array} \right),
\end{align}
where the retarded and advanced propagators are given by 
\begin{align}
-iG_{\text{ret}}(x, x')&
=\theta(t-t')\left(\Delta_+(x, x')-\Delta_-(x, x')\right),\\
G_{\text{adv}}(x, x')& 
=G_{\text{ret}}(x', x),
\end{align}
and the diagonal entry in the matrix in equation \eqref{propsctp}
can in fact be disregarded, i.e.~taken as $0$, in the classical limit. 

The generalization of the above formulation from a scalar field theory to our 
gravitational effective theory of a composite particle is rather straightforward 
\cite{Galley:2009px}. In our theory the generating functional, 
$Z_{\text{CTP}}$, depends on current densities, $J^{\mu\nu}$ and $j_\mu$, 
which couple linearly to the radiation field, $\widetilde{h}_{\mu\nu}$, and 
the particle worldline coordinate, $x^\mu$, respectively, where both field 
and worldline DOFs are doubled. Variation of the corresponding generating 
functional, $W_\text{{CTP}}$, with respect to the current density, 
$J^{\mu\nu}$, gives the one-point function of the radiation field. A partial 
Legendre transform of $W_\text{{CTP}}$ with respect to $j_\mu$ gives the 
corresponding effective action functional, $\Gamma_\text{{CTP}}$, whose 
variation yields the EOMs for the orbital motion of the compact objects. For 
the computation of Feynman diagrams within the CTP formalism, as in figures 
\ref{quadrad}-\ref{radreaction} below: CTP labels should also be included at 
each worldline-graviton coupling, with $n$ labels for a worldline coupling 
with $n$ gravitons, in order to keep track of the time branches in the CTP 
path integral, where a two-point function matrix similar to 
equation \eqref{propsctp} should be used, and all CTP indices should be summed 
over.

As we noted classical formulations to handle dissipative systems were 
put forward, motivated by the radiating binary inspiral problem. Notably, a 
formulation, which extends Hamilton's principle to classical causal actions 
for generic non-conservative systems, was introduced in \cite{Galley:2012hx}, 
and later extended for generic classical field theories in 
\cite{Galley:2014wla}. This provided an independent classical formulation, 
which was tested successfully against the CTP formalism of nonequilibrium QFT 
in the classical limit. Furthermore, a study of the classical origin of the 
quantum CTP formalism at the level of the action, which can also be generally 
applied at the level of the EOMs, was carried out independently in 
\cite{Birnholtz:2013nta,Birnholtz:2014fwa}. This study was also extended for 
a general spacetime dimension in \cite{Birnholtz:2013ffa,Birnholtz:2015hua}.

Let us resume the discussion on the renormalization of the total mass of the 
composite particle. The classical RG flow induces time dependence, and thus 
the time evolution of the renormalized mass can be studied by computing the 
expectation value of the energy-momentum pseudotensor, using its conservation 
from the Ward identities and the CTP formalism adapted to the classical 
system \cite{Goldberger:2012kf}. The RG flow of the total mass can also be 
inferred from an analysis of the radiation reaction as in 
\cite{Galley:2015kus}, using the purely classical dissipative formulation 
devised in \cite{Galley:2012hx,Galley:2014wla}; see figure \ref{radreaction} 
and the related discussion below. 

Let us outline here the analysis in terms of the stress-energy pseudotensor, 
following the work in \cite{Goldberger:2012kf}. First, the metric is expanded 
into a weak background, 
$\tilde{g}_{\mu\nu}=\eta_{\mu\nu}+\tilde{h}_{\mu\nu}$, and a fluctuating 
field, $\hat{h}_{\mu\nu}=\widetilde{g}_{\mu\nu}-\tilde{g}_{\mu\nu}$ (note 
that the weak background, $\tilde{g}$, is denoted with a tilde, whereas the 
total radiation metric, $\widetilde{g}$, is denoted with a \emph{wide} 
tilde). Then, the fluctuations are integrated out according to the CTP 
formalism to obtain the effective action as follows:
\be
\text{exp}^{i\Gamma[\tilde{h}_+,\tilde{h}_-,M,I]}=
\int {\cal{D}}\hat{h}_+ {\cal{D}}\hat{h}_- 
\,\,\text{exp}^{i\left[S[\widetilde{g}_+,M,I]-S[\widetilde{g}_-,M,I]\right]},
\ee
where the field DOFs have been doubled, and transformed into the Keldysh 
representation, 
and the CTP boundary conditions are imposed on the fluctuating modes. Then, 
the real time expectation value of the effective energy-momentum pseudotensor 
is given by
\be
\langle T_{\mu\nu}(x)\rangle
=-\frac{2}{\sqrt{\widetilde{g}_+}}
\frac{\delta\Gamma}{\delta\tilde{g}_+^{\mu\nu}(x)}
\left|_{\tilde{h}_+=\tilde{h}_-=0},\right.
\ee 
where after the variation we set the external graviton, e.g.~$\tilde{h}_+$, 
to $0$.

Hence, we find the expectation value $\langle T_{\mu\nu}\rangle$ by 
considering Feynman graphs with one (off-shell) external graviton, as can be 
seen in figures \ref{quadrad} and \ref{rgflow2}. The first graphs to 
consider, shown in figure \ref{quadrad}, give the leading contribution to the 
expectation value, and yield the time evolution of the total mass, from which 
one recovers the following well-known quadrupole radiation formula:
\be
\langle\dot{M}\rangle=-\frac{G}{5}\langle I_{ij}^{(3)} I_{ij}^{(3)} \rangle,
\ee
which accounts for the energy loss of the system to gravitational radiation.
This corresponds to the LO radiation reaction depicted in graph (a) of figure 
\ref{radreaction} \cite{Galley:2009px}. 

\begin{figure}[t]
\centering
\includegraphics[width=0.4\linewidth]{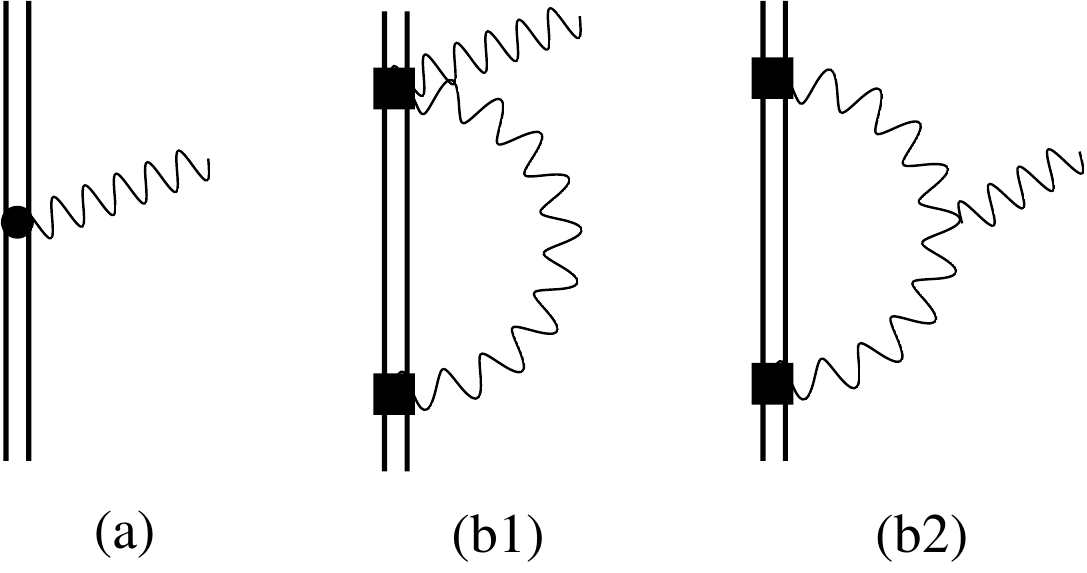}
\caption{The leading time dependence of the mass quadrupole and total mass 
can be found by considering the expectation value $\langle T^{\mu\nu}\rangle$ 
\cite{Goldberger:2012kf}. These diagrams are not logarithmically divergent, 
and they yield the well-known LO quadrupole radiation formula, which also 
corresponds to the LO radiation reaction, depicted in figure 
\ref{radreaction}(a) \cite{Galley:2009px}.}
\label{quadrad} 
\end{figure}

\begin{figure}[t]
\centering
\includegraphics[width=0.45\linewidth]{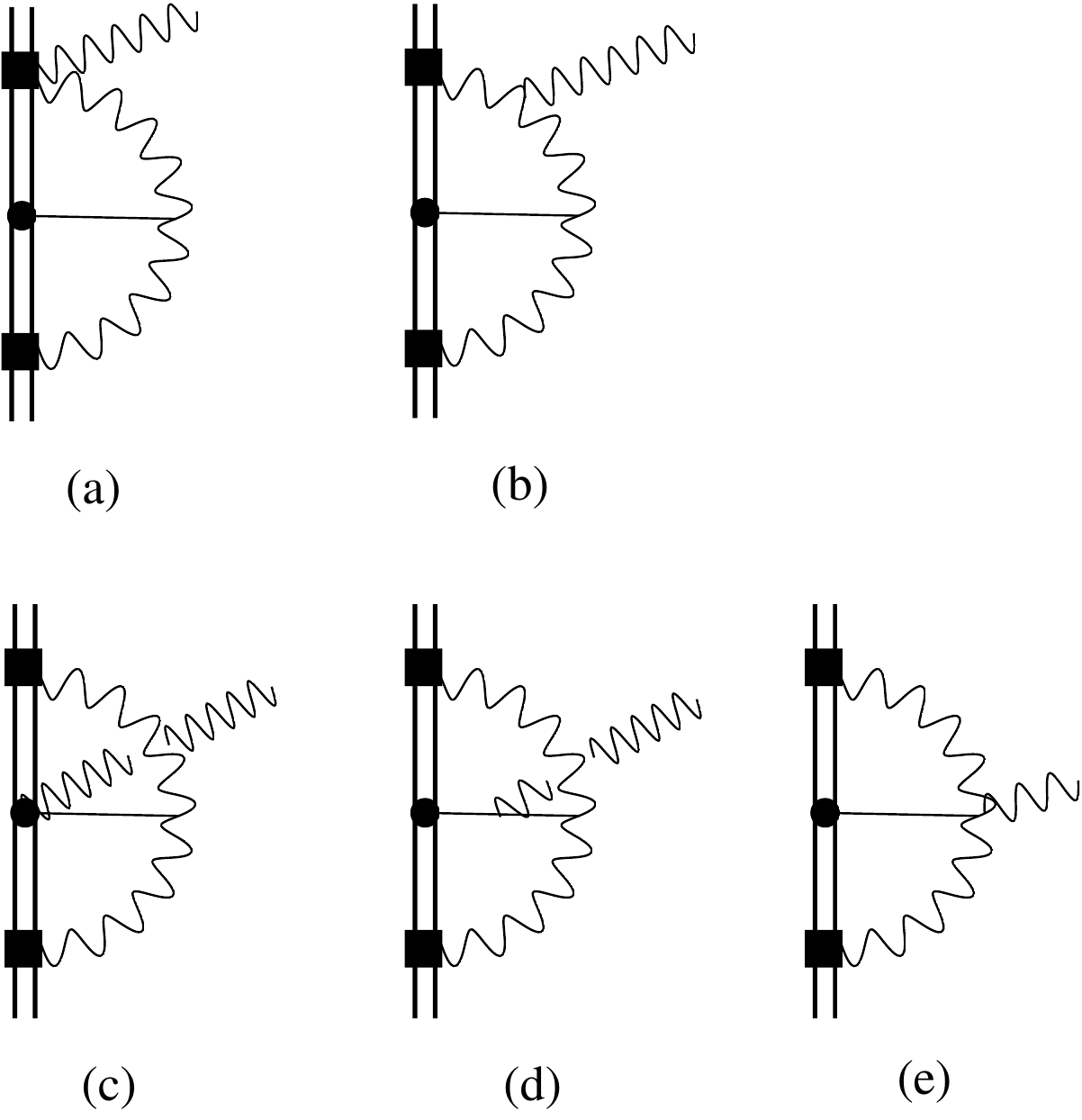}
\caption{The classical RG flow of the composite mass can be found by 
considering the expectation value $\langle T^{\mu\nu}\rangle$ 
\cite{Goldberger:2012kf}. These diagrams are logarithmically UV divergent, 
and yield classical RG running of the mass. This RG flow of the mass can also 
be inferred from the leading non-linear radiation reaction, corresponding to 
graph (c) of figure \ref{radreaction} \cite{Galley:2015kus}.}
\label{rgflow2} 
\end{figure} 

The diagrams in figure \ref{quadrad} are not logarithmically divergent, yet 
the non-linear corrections in figure \ref{rgflow2} yield logarithmic UV 
divergences, which give rise to the mass renormalization, providing a new 
time-dependent RG equation for the mass. By considering the singular pieces 
of the graphs in figure \ref{rgflow2} as $d-4\to0$ the following RG equation 
for the composite mass, $M(t)$, is found:
\be \label{rgm}
\mu\frac{d}{d\mu}\ln \bar{M}=
-2G^2\langle I_{ij}^{(3)}I_{ij}^{(3)}\rangle (\mu), 
\ee
where at this order $M(t)$ is simply replaced by its time average, $\bar{M}$.
The mass is scale dependent due to the radiation backscattering off of the 
static background, as seen in figures \ref{tailmemory}(b), \ref{nlotail}, 
\ref{rgflow2}, and \ref{radreaction}(c). The RG running of the composite mass 
is suppressed with respect to that of the mass quadrupole, $I_{ij}$, so the 
coupled RG equations, equations \eqref{rgm} and \eqref{rgq}, may be solved by 
using the solution in equation \eqref{losol}, assuming $\bar{M}$ is scale 
independent. The solution of the RG equation for the mass then reads as follows:
\be
\frac{\bar{M}(\mu)}{\bar{M}_0}=
\exp\left[\frac{\langle I_{ij}^{(2)}I_{ij}^{(2)}\rangle(\mu_0)
-\langle I_{ij}^{(2)}I_{ij}^{(2)}\rangle(\mu)}{\frac{214}{105}\bar{M}_0^2}
\right],
\ee
where the mass as defined in equation \eqref{sppcomp} is the energy of the 
conservative system, not including energy in gravitational radiation.

The solution to the coupled RG equations for $M$ and $I_{ij}$ provides the 
logarithms, which appear in the low frequency mode, $l=2$, distribution of 
emitted gravitons radiated from the binary. Thus the logarithms in the 
running composite mass and mass quadrupole can be predicted at higher PN 
orders. Similar to this analysis, which is derived from the conservation of 
the expectation value of the stress-energy pseudotensor, $\partial_\mu\langle 
T^{\mu\nu}(x)\rangle=0$, the evolution equation for other multipole moments, 
such as the total angular momentum of the system, can also be set up. An 
alternative analysis can be made by studying the radiation reaction, as in 
graph (c) of figure \ref{radreaction} for the RG flow of the composite mass, 
which was analyzed here using figure \ref{rgflow2}.

\begin{figure}[t]
\centering
\includegraphics[width=0.45\linewidth]{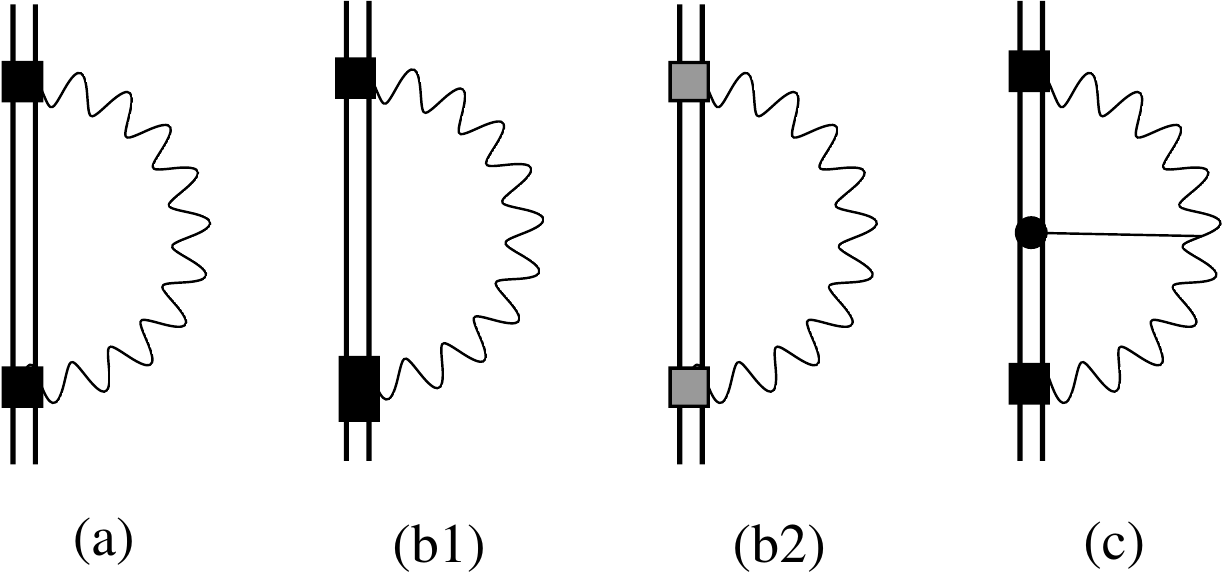}
\caption{Radiation reaction in the effective theory of the composite particle 
for the PN binary inspiral, similar to the self-force in the effective 
point-particle theory for EMRIs \cite{Galley:2008ih}: 
(a) The LO effect at the 2.5PN order \cite{Goldberger:2004jt,Galley:2009px}; 
(b1) and (b2) The NLO effect at the 3.5PN order \cite{Galley:2012qs}. Here 
the black rectangle represents the mass octupole, $I_{ijk}$, and the gray 
squares represent the current quadrupole, $J_{ij}$;
(c) The leading non-linear effect, which enters at the 4PN order, tackled in 
\cite{Foffa:2011np,Galley:2015kus}. As opposed to the former linear effects, 
this has a contribution to the conservative action.} 
\label{radreaction} 
\end{figure}

The radiation reaction in the PN binary inspiral was tackled in an EFT approach 
in \cite{Galley:2009px}, where it is computed by integrating out the 
radiation field modes from the effective theory of the composite particle. In 
this work the LO effect at the 2.5PN order was reproduced (see also in 
\cite{Goldberger:2004jt}), as shown in graph (a) of figure \ref{radreaction}. 
Next, the NLO effect at the 3.5PN order, which is still linear in $G$ as 
can be seen in graphs (b1) and (b2) of figure \ref{radreaction}, was reproduced 
in \cite{Galley:2012qs}\footnote{Related with this progress, the LO finite size 
correction to the radiation reaction in classical electrodynamics was also 
found in \cite{Galley:2010es}; see also \cite{Forgacs:2012qt, 
Galley:2012gv}.}. The leading non-linear tail effect, which yields a 
radiation reaction force at the 4PN order, was first approached in 
\cite{Foffa:2011np}, and further studied in \cite{Galley:2015kus}, using the 
abovementioned classical formulation for dissipative systems from 
\cite{Galley:2012hx,Galley:2014wla}. This non-linear effect, depicted in 
figure \ref{radreaction}(c), contributes a piece, which is non-local in time, 
to the effective action of the composite particle in the point-mass sector 
\cite{Damour:2014jta}, and it contributes to the binding energy logarithmic 
corrections, which were first obtained via the EFT approach in 
\cite{Goldberger:2012kf}\footnote{The analogy of the 
leading non-linear tail effect to the well-known Lamb shift and Bethe 
logarithm from electrodynamics was also discussed in \cite{Porto:2017shd}.}. 
Hence, unlike the linear effects of the radiation reaction in figure 
\ref{radreaction}, the logarithmically divergent part at order $G^2$ from 
graph (c) is a conservative effect, which affects the orbital dynamics of the 
binary. 

Graph (c) of figure \ref{radreaction} is a self-energy graph of the theory, 
i.e.~a renormalization graph of the mass monopole of the effective theory of 
the composite particle, and thus it enables one to infer the RG flow of the 
composite mass \cite{Galley:2015kus}. In fact, the study in 
\cite{Galley:2015kus} indicates that the renormalization of the effective 
theory of the composite particle can be more efficiently approached via an 
analysis of the radiation reaction, rather than via an analysis, which involves 
the expectation value of the stress-energy pseudotensor as in 
\cite{Goldberger:2012kf}. Moreover, the analysis in \cite{Galley:2015kus} in 
terms of the radiation reaction reveals the subtle interplay between the 
radiation (or far) and the potential (or near) zones, where the counterterm 
for the UV divergence of the radiation reaction potential originates from an 
IR singularity in the orbital region. 

Finally, the leading radiation reaction linear in the spins of the binary 
components, which enters at the 4PN order, was reproduced in 
\cite{Maia:2017gxn}, where the LO spin-orbit tail effect was also previously 
reproduced in \cite{Porto:2012as}. The leading radiation reaction quadratic 
in the spins, including the backreaction from spin-induced finite size 
effects, which enters at the 4.5PN order, was computed in 
\cite{Maia:2017yok}.

\subsection{EFTofPNG public code}
\label{eftofpngcode}

\begin{figure}[t]
\centering
\includegraphics[width=0.5\linewidth]{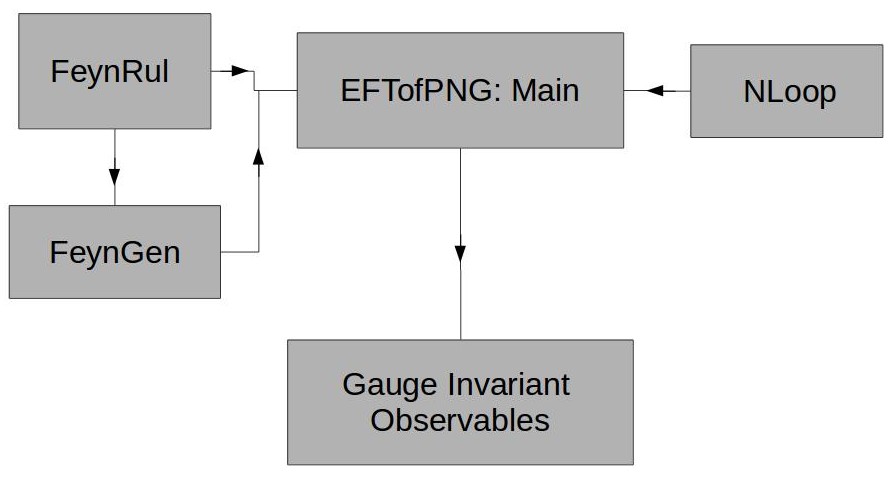}
\caption{A sketch of the comprehensive `EFTofPNG' public package version 
$1.0$, which incorporates the EFT framework for high precision Feynman 
computation in PN gravity, including spin effects \cite{Levi:2017kzq}. The 
code is modular with independent units, and a pipeline unit, where the flow 
among these units is shown here. This is the first public code in PN theory, 
and also serves as a generic high precision Feynman computation tool.}
\label{eftofpng} 
\end{figure}

Eventually, one would like to capitalize on the outcome of the EFT framework 
for the PN binary inspiral to arrive at the practical inclusion of the 
various results in the GW flux and in the gravitational waveform amplitude and 
phase. This was indeed a significant driver in creating `EFTofPNG' --- a public 
Mathematica package, which incorporates the EFT framework for high precision 
Feynman computation in PN gravity \cite{Levi:2017kzq}. In its current first 
version, it has applied to the conservative sector and covers the current state 
of the art in the field. The code was designed to be self-contained and modular 
with independent units, including a pipeline unit, in order to serve diverse 
purposes within the broad community, for which it was made accessible. An 
outline of the EFTofPNG package version $1.0$ is depicted in figure 
\ref{eftofpng}. This is the first comprehensive code in PN theory made public, 
so as to fill in the current analytic gap for the major part of the 
gravitational waveform models, and complement current GW public codes, which 
include numerical codes, codes for BH perturbation theory \cite{BHPToolkit}, 
and codes for data analysis\footnote{We note a recent public 
numerical code, `PRECESSION', which specifically studies the dynamics of 
spinning BHs in the PN regime \cite{Gerosa:2016sys}. However, this code deals 
with the outcome of a given PN dynamics, rather than deriving the PN theory 
like the EFTofPNG code does.}. Moreover, it is expected that the code will be 
continually extended and publicly developed to cover higher-order PN sectors, 
non-conservative sectors, and further GW observables. However, it should 
be stressed that beyond its obvious utility for PN gravity, the EFTofPNG 
is also a useful high precision Feynman computation tool, which handles the 
generic perturbative diagrammatic expansion in a unique and efficient way 
and contains a stand-alone multi-loop computation unit. 

As for GW observables from the conservative sector, 
which are included in the code, we note that finally, neither the effective 
interaction potentials, nor the Hamiltonians, are gauge invariant (GI). However, 
the Hamiltonians are related to the total energy of the binary system, which 
is a global GI quantity, like the total angular momentum of the binary. In 
addition, for the simplified case of binaries with non-spinning components in 
quasi-circular orbits the energy depends only on the magnitude of the orbital 
angular momentum. This relation between binding energy and angular momentum 
is GI, and is thus a useful tool to evaluate different analytic and numerical 
descriptions of the binary dynamics \cite{Damour:2011fu}. There is another 
useful GI relation between the binding energy and the orbital frequency, that 
together with the energy flux of the GWs, can be used to derive the GW phasing. 
All these GI quantities and relations are also included in the EFTofPNG code. 
Finally, a further important useful GI quantity, to be possibly considered in 
the code in the future, is the periastron advance 
\cite{LeTiec:2011bk,Tiec:2013twa,Hinderer:2013uwa}.

\section{Field theory for gravity at all scales}
\label{future}

With significant developments over the last decade, since its initiation, the 
field of EFTs in PN gravity has firmly demonstrated that seemingly disparate 
physical domains, such as QFT and classical gravity, are in fact related. 
This attests to the unity of physics. The EFT framework for PN gravity has 
proved to supply a robust methodology to boost the progress in PN theory, in 
particular for the timely study of GWs from the inspiral of comparable mass 
compact binaries. GW observations are thereby expected to carry insightful 
implications the other way around: For various subjects in theoretical 
physics, from QCD and gravity, to astrophysics and cosmology. As was 
highlighted in section \ref{intro}, PN theory exclusively covers a 
considerable portion of the CBC GW signal, which must be accurately modeled 
in order to extract the utmost possible information from GW data. Numerical 
simulations cannot in essence cover the NR regime, and all the more so when 
the spins of the components of the binary are taken into account and their 
capacity becomes extremely limited.

Concerning the output of the EFT methodology for the binary inspiral in the 
conservative sector, there are two key ingredients of the orbital dynamics, 
which should be produced in a straightforward manner in order to improve the 
accuracy of theoretical GW templates: The EOMs and the Hamiltonians. This 
was attained in the EFT framework, in particular in the realistic yet much 
more challenging case where spins are present \cite{Levi:2015msa}, and is 
incorporated in the pipeline unit of the EFTofPNG code \cite{Levi:2017kzq}. 
It should be noted that standard Hamiltonians still need to be further 
worked out into EOB Hamiltonians in order to be incorporated into the EOB GW 
models. This is currently not implemented up to the level of PN accuracy 
attained in the spin sectors. Further, more generally, due to the complexity 
of spin effects, their incorporation into GW templates falls far behind the 
state-of-the-art results of PN theory. As for the EOMs of the components of 
the binary, progress should also be made in finding solutions to the EOMs to 
high PN order, a challenging task which again lags behind the 
state-of-the-art PN theory and even more so in the spin sectors. 
Interestingly, in \cite{Galley:2016zee}, a dynamical RG method, designed
to calculate such closed form solutions to a desirable accuracy without 
orbit averaging, was put forward. 

Despite the notable progress within the EFT approach to the binary inspiral, 
there is still much room and demand for development in the field. First, 
in general, our conceptual grasp of classical spins in gravity should be 
further improved. In particular, the EFT framework via coset construction in 
\cite{Delacretaz:2014oxa}, suitable for slowly rotating gravitating objects, 
should be scrutinized, which may possibly lead to a unified formulation 
suitable for both slowly and rapidly rotating objects in the appropriate 
limits. In any case, building on the formulation in 
\cite{Delacretaz:2014oxa}, concrete PN predictions for slowly rotating 
objects should be produced. Spin-induced Wilson coefficients, which play a 
role at least as of the 2PN order, should be studied via a formal matching 
procedure. Further, the treatment of the non-conservative sectors within the 
EFT approach seems to be underdeveloped with respect to other methods in PN 
theory, and with respect to the progress accomplished within the EFT approach 
in the conservative sector. In particular, it should be stressed that tail 
and dissipative effects manifest in the late inspiral phase, which 
constitutes the transition into the strong field regime. Finally, public 
codes like the EFTofPNG package should be further developed in order to improve 
in computational efficiency, but more importantly, they should be extended 
along with formal advancement in the field, especially so as to cover the 
non-conservative sectors.

Whether one is strictly interested in to analytically covering the GW CBC signal, 
entering the strong field regime as well, or much more generally looking to obtain 
a complete theory of gravity at all scales, it is clear that field theory 
advances at all levels should be invoked. First, standard working knowledge 
in scattering amplitudes can be imported from gauge theory to gravity 
\cite{Smirnov:2006ry}, and vice versa, to further advance high precision 
computation, as was demonstrated in \cite{Foffa:2016rgu,Damour:2017ced}. 
Furthermore, modern advances in 
scattering amplitudes, such as the BCFW on-shell recursion relations 
\cite{Britto:2004ap} and unitarity methods, were demonstrated to reproduce 
low order results in classical gravity \cite{Neill:2013wsa,Vaidya:2014kza}. 
These advances imply that tree level data encode all multiplicity at the 
integrand level, i.e.~that the integrands of extremely complicated multi-loop 
predictions can be verified systematically and invariantly with compact 
on-shell tree level data. Thus, more generally, it has been shown how to 
extract classical higher-order loop results for gravitational scattering from 
advanced on-shell unitary methods and for higher spin particles in 
\cite{Bjerrum-Bohr:2013bxa,Bjerrum-Bohr:2014zsa,Bjerrum-Bohr:2016hpa,
Cachazo:2017jef,Guevara:2017csg}. In order to actually go into the strong 
field regime of the CBC signal, the gravitational scattering of two BHs, 
which can be deduced from the quantum gravitational scattering amplitude of 
two particles, was mapped into the EOB framework \cite{Damour:2014afa, 
Damour:2016gwp,Vines:2017hyw,Damour:2017zjx}. It was indeed recently 
discussed how to strategically get from scattering amplitudes, using 
generalized unitarity, to the classical scattering of two BHs 
\cite{Bjerrum-Bohr:2018xdl,Cheung:2018wkq}.

Proceeding to candidate theories of quantum gravity, i.e.~extending gravity 
in the UV, intriguing and powerful duality and correspondence relations have 
been discovered in the context of scattering amplitudes, between 
supersymmetry and supergravity theories \cite{Kawai:1985xq, 
Bern:2008qj,Bern:2010ue,Bern:2010yg}. The color-kinematics, or so-called BCJ 
duality, and the related double copy correspondence, state that Yang-Mills 
amplitudes, for example, can be mapped onto their gravity theory counterparts 
by applying a prescribed set of color to kinematics replacement rules 
\cite{Carrasco:2015iwa}. First, this correspondence provides an exciting 
opportunity to reveal more on the underlying origin of these relations in the 
fundamental structure of these QFTs, gauge and gravity theories, and thus 
extend our understanding and reach of both. However, this correspondence 
also offers a compelling novel approach for handling perturbative (and exact) 
computation within classical gravity, and particularly for studying 
GWs. In fact, the exploration of how to formulate the double copy 
correspondence in the classical perturbative context has been already 
initiated in a recent series of works \cite{Luna:2016due,Goldberger:2016iau, 
Luna:2016hge,Cheung:2017ems,Goldberger:2017frp,Adamo:2017nia,Luna:2017dtq, 
Goldberger:2017vcg,Chester:2017vcz,Goldberger:2017ogt,Li:2018qap,
Ilderton:2018lsf,Shen:2018ebu}.

Turning instead to very large scales, we are once again puzzled by the 
observational discovery of the accelerated expansion of the Universe, or the 
threefold equivalence of the IR structure of gauge and gravity theories, 
among soft theorems, memory effects, and asymptotic symmetry. These have led 
us to the realization that our understanding of gravity is not only lacking 
at the quantum regime, but also at the classical IR one. In fact, the 
problems of dealing with gravity in the UV and IR sides may well be tied 
together. Let us then discuss first the triple classical and quantum 
equivalence noted above. Soft theorems are statements about the universal 
characteristic behavior of scattering amplitudes at low energies, and they 
tell us about the IR structure of gauge and gravity theories with massless 
particles (see, e.g.~\cite{Strominger:2017zoo}). In particular, the equivalence 
between soft graviton theorems and gravitational memory effects was recently 
demonstrated in \cite{Strominger:2014pwa,Pasterski:2015tva}. The former 
concerns momentum space poles in scattering amplitudes, while the latter 
concerns a shift in asymptotic data between late and early times, and they 
are simply connected via a Fourier transform. Furthermore, there is going 
development on gravitational memory effects and their observable imprint in 
GWs \cite{Flanagan:2014kfa,Nichols:2017rqr}, their manifestation in 
cosmological GWs \cite{Chu:2015yua,Chu:2016ngc}, and related further 
advancements in soft theorems \cite{Laddha:2018rle,Laddha:2018vbn}. 

However, the cosmological constant problem and dark matter puzzle call for 
alternative theories of gravity at large scales, even if only for the 
classical regime. Indeed, a plethora of alternative theories of gravity in 
the IR has been put forward through the years: Theories with extra fields, 
such as massive gravity (see, e.g.~\cite{Hinterbichler:2011tt,deRham:2014zqa}) 
or Horndeski theories (see, e.g.~\cite{Deffayet:2013lga}); higher derivative 
and non-local theories, e.g.~Galileons (see, e.g.~\cite{Deffayet:2013lga}); 
or higher-dimensional theories, in the context of large extra dimensions in 
particle physics, or inspired by string theory. We also refer the reader to 
the comprehensive reviews in \cite{Clifton:2011jh,Joyce:2014kja}. The theoretical 
and phenomenological study of these alternative theories serves to develop general 
tests of gravity, also based on GR, through the extraction of the latest 
information from GW data (see, e.g.~\cite{Cannella:2009he}). Conversely, the 
ongoing advances in GW and cosmological observations may greatly benefit our 
ability to pin down an ultimate viable theory of gravity (see, 
e.g.~\cite{deRham:2016nuf,Cardoso:2018zhm}). 

Different aspects of such alternative theories have been already studied 
by applying the EFT approach (see, e.g.~\cite{Cannella:2008nr, 
Sanctuary:2008jv}). Preliminary explicit implementations in radiating binary 
systems in Galileon theories have been made in \cite{Chu:2012kz}, and also 
following the EFT framework from \cite{Goldberger:2004jt,Goldberger:2007hy} 
in \cite{deRham:2012fw,deRham:2012fg}, and were recently followed up with 
numerical work in \cite{Dar:2018dra}. Thus, it is desirable to apply the EFT 
framework, replacing GR with other alternative theories of gravity, to 
produce predictions for the possible GW signal from binaries. Furthermore, 
building on the $n$-body case initially explored with the EFT approach in 
\cite{Chu:2008xm}, it may be beneficial for the study of such alternative 
theories to model GWs from $n$-body systems, which may also be facilitated 
using higher multiplicities from scattering amplitudes. 
Although two-body systems are much more abundant in nature, if 
GWs were to be detected, e.g.~from three-body systems, this would 
provide for richer testing of GR, since the two-body sector is a degenerate 
case as far as the non-linearities of GR are concerned. At the practical level 
the EFTofPNG code, currently implemented for binaries in GR, can be easily 
accommodated to handle the more generic case in such modified theories of 
gravity. 

It is noteworthy that the coupling of spin to the various fields in each 
candidate theory of gravity, classical or quantum, plays a crucial role in 
distinguishing among these theories. This further highlights the vital 
importance of gaining a profound understanding of spin, in particular also 
in classical gravity.

To conclude, there are myriad prospects of using modern insights and tools from 
the realm of field theory to study gravity, which will surely be 
taken advantage of in the endeavor to zero in on a complete theory of 
gravity across all scales. Indeed, also in this broader sense, this line of 
research, which has been advanced with the use of EFTs in PN gravity, is certain 
to continue developing in the years to come and lead us towards resolving 
at least some of the formidable fundamental puzzles that we currently face. 

\acknowledgments

I am grateful to John Joseph Carrasco for his significant 
encouragement and support, and for comments on the manuscript.
I would like to thank Alessandra Buonanno, Yi-Zen Chu, Thibault Damour, 
Stefano Foffa, Walter Goldberger, Pierpaolo Mastrolia, Riccardo Sturani, 
Christian Sturm, and Pierre Vanhove for their comments on the manuscript 
and support, and to Barak Kol and Ira Rothstein for support.
I would like to warmly acknowledge my collaborators on this line of research, 
Barak Kol, Michael Smolkin, and Jan Steinhoff. 
Finally, I kindly thank Alexandre Le Tiec for allowing to use the plot in 
figure \ref{methods} in this review.

It is with great pleasure that I extend my gratitude to Feng-Li Lin and the 
Taipei GW group, in particular to Chian-Shu Chen, and Sheng-Lan Ko for the 
kind and warm hospitality, and for their keen interest in understanding the 
subtleties of the topic throughout my lecture series.

The work of ML is supported by the European Research Council under the 
European Union's Horizon 2020 Framework Programme FP8/2014-2020 ``preQFT'' 
grant no.~639729, ``Strategic Predictions for Quantum Field Theories'' 
project.

\bibliographystyle{jhep}
\bibliography{gwbibtex}

\end{document}